\documentclass[journal]{IEEEtran}

\usepackage{amssymb}
\usepackage{amsmath}
\usepackage{bm}
\usepackage{graphicx}
\usepackage{xcolor}
\usepackage{import}
\usepackage[utf8]{inputenc}
\usepackage{listings}
\usepackage[noend]{algpseudocode}
\usepackage[hyphens]{url}
\usepackage{multirow}
\usepackage[center, font={small,it}]{caption}
\usepackage[bookmarks=true]{hyperref} 
\usepackage{array}
\usepackage{rotating}   
\usepackage{makecell}
\usepackage{colortbl}
\usepackage{enumerate}
\usepackage{booktabs}
\usepackage{subfig}
\usepackage{paralist}
\usepackage[ruled,vlined]{algorithm2e}
\usepackage{float}
\usepackage{stfloats}

\newcommand{\subparagraph}{} 

\let\url\nolinkurl

\algrenewcommand\algorithmiccomment[1]{\hfill \textcolor{gray}{$\triangleright$ \textit{#1}}}

\DeclareUnicodeCharacter{00A0}{ }

\newcommand{\specialcell}[2][c]{
	\begin{tabular}[#1]{@{}l@{}}#2\end{tabular}}

\makeatletter
\g@addto@macro\normalsize{
  \setlength\abovedisplayskip{4pt}
  \setlength\belowdisplayskip{4pt}
  \setlength\abovedisplayshortskip{4pt}
  \setlength\belowdisplayshortskip{4pt}
}
\makeatother

\renewcommand{\tablename}{Table}

\renewcommand{\figurename}{Figure}

\renewcommand{\paragraph}[1]{\vspace{0.1cm}\noindent{\bf #1.}}

\newcommand{\mysubsubsection}[1]{\smallskip\subsubsection{#1}}

\ifCLASSINFOpdf
\else
\fi

\begin{document}

\title{The Security Reference Architecture for Blockchains: Towards a Standardized Model\\for Studying Vulnerabilities, Threats, and Defenses\thanks{$\copyright$ 2020 IEEE.  Personal use of this material is permitted.  Permission from IEEE must be obtained for all other uses, in any current or future media, including reprinting/republishing this material for advertising or promotional purposes, creating new collective works, for resale or redistribution to servers or lists, or reuse of any copyrighted component of this work in other works.}}

\author{
	\IEEEauthorblockN{
		Ivan Homoliak\IEEEauthorrefmark{1}\IEEEauthorrefmark{2}
	}
	\and
	\IEEEauthorblockN{
		Sarad Venugopalan\IEEEauthorrefmark{1}
	}
	\and
	\IEEEauthorblockN{
		Dani\"el Reijsbergen\IEEEauthorrefmark{1}\\
	}
	\and
	\IEEEauthorblockN{
		Qingze Hum\IEEEauthorrefmark{1}
	}
	\and
	\IEEEauthorblockN{
		Richard Schumi\IEEEauthorrefmark{1}
	}
	\and
	\IEEEauthorblockN{
		Pawel Szalachowski\IEEEauthorrefmark{1}\\
	}
	\and
	\IEEEauthorblockA{
		\IEEEauthorrefmark{1}Singapore University of Technology and Design
	}
	\and \\
	\IEEEauthorblockA{
		\IEEEauthorrefmark{2}Brno University of Technology
	}
}

\maketitle

\begin{abstract}
Blockchains are distributed systems, in which security is a critical factor for their success.
However, despite their increasing popularity and adoption, there is a lack of standardized models that study blockchain-related security threats. 
To fill this gap, the main focus of our work is to systematize and extend the knowledge about the security and privacy aspects of blockchains and contribute to the standardization of this domain.

We propose the security reference architecture (SRA) for blockchains, which adopts a stacked model (similar to the ISO/OSI) describing the nature and hierarchy of various security and privacy aspects.
The SRA contains four layers: (1) the network layer, (2) the consensus layer, (3) the replicated state machine layer, and (4) the application layer.
At each of these layers, we identify known security threats, their origin, and countermeasures, while we also analyze several cross-layer dependencies. 
Next, to enable better reasoning about security aspects of blockchains by the practitioners, we propose a blockchain-specific version of the threat-risk assessment standard ISO/IEC 15408 by embedding the stacked model into this standard. 
Finally, we provide designers of blockchain platforms and applications with a design methodology following the model of SRA and its hierarchy.

\end{abstract}


\section{Introduction}
\label{sec:intro}
The popularity of blockchain systems has rapidly increased in recent years, mainly due to the decentralization of control that they aim to provide.
Blockchains are full-stack distributed systems in which multiple layers, (sub)systems, and dynamics interact together.
Hence, they should leverage a secure and resilient networking architecture, a robust consensus protocol, and a safe environment for building higher-level applications.
Although most of the deployed blockchains need better scalability and well-aligned incentives to unleash their full potential, their security is undoubtedly a critical factor for their success.
As these systems are actively being developed and deployed, it is often challenging to understand how secure they are, or what security implications are introduced by some specific components they consist of.
Moreover, due to their complexity and novelty (e.g., built-in protocol incentives), their security assessment and analysis requires a more holistic view than in the case of traditional distributed systems.

Although some standardization efforts have already been undertaken, they are either specific to a particular platform~\cite{EEA-standards} or still under development~\cite{iso-security-threats,iso-reference-architecture}.
Hence, there is a lack of platform-agnostic standards in blockchain implementation, interoperability, services, and applications, as well as the analysis of its security threats~\cite{gartner-lack-of-standards,barry-medium}.
All of these areas are challenging, and it might take years until they are standardized and agreed upon across a diverse spectrum of stakeholders. 
In this work, we aim to contribute to the standardization of security threat analysis.
We believe that it is critical to provide blockchain stakeholders (developers, users, standardization bodies, regulators, etc.) with a comprehensive systematization of knowledge about the security and privacy aspects of today's blockchain systems.

In this work, we aim to achieve this goal, with a particular focus on system design and architectural aspects.
We do not limit our work to an enumeration of security issues, but additionally, discuss the origins of those issues while listing possible countermeasures and mitigation techniques together with their potential implications.
As our main contribution, we propose the security reference architecture (SRA) for blockchains, which is based on models that demonstrate the stacked hierarchy of different threat categories (similar to the ISO/OSI hierarchy~\cite{zimmermann1980osi}) and is inspired by security modeling performed in the cloud computing~\cite{liu2011nist,xiao2013cloud}.
As our next contribution, we enrich the threat-risk assessment standard ISO/IEC 15408~\cite{cc2017} to fit the blockchain infrastructure.
We achieve this by embedding the stacked model into this standard.

This paper is based on our previous work outlining the security reference architecture~\cite{homoliak-sra-short}. 
We substantially modify and extend it by the following:
\begin{itemize}
    \item We provide a theoretical background related to the security reference architecture and the environment of blockchains, their types, failure models, consensus protocols, design goals, and means to achieve these goals.
    
    \item For each layer, we model particular attacks or their categories through vulnerability/threat/defense graphs, while we provide reasoning about several important relations and causalities in these graphs.
    
    \item We modify and significantly extend the application layer, where we propose a novel functionality-oriented categorization, as opposed to the application-domain-oriented categorizations presented in related work~\cite{casino2018systematic,zheng2018blockchain}.
    
    \item We extend and revise the consensus layer by mining-pool-specific attacks, time-spoofing attacks, and we provide a more fine-grained categorization.     
    
    \item For each layer, we present an incident table that lists and briefly describes examples of attacks that have occurred worldwide: either caused by malicious parties or by researchers who demonstrated their practical feasibility.
    
    \item In the lessons learned, we describe the hierarchy of security dependencies among particular categories, based on which, we propose a methodology for designers of blockchain platforms and applications.

\end{itemize}

\smallskip
The rest of the paper is organized as follows:
\begin{inparaitem}
    \item[] We describe the scope and methodology of our research as well as quantitative analysis of the included literature in \autoref{sec:methodology}.
    \item[] In \autoref{sec:background}, we summarize the background on blockchain systems.
    \item[] Next, in \autoref{sec:model} we introduce the security reference
    architecture whose layers are discussed in the follow-up sections.
    \item[] In detail, \autoref{sec:network} deals with the security and privacy of the network layer,
    \item[] \autoref{sec:consensus} focuses on the consensus layer,
    \item[] \autoref{sec:smart_contracts} overviews the replicated state
    machine layer,   
    \item[] and \autoref{sec:apps} with \autoref{sec:apps-applications} describe applications built on top of the blockchain.
    \item[] In \autoref{sec:lessons}, we elaborate on lessons learned and propose a methodology for designers of blockchain-based solutions.
    \item[] We discuss the limitations of our work in \autoref{sec:discussion} and we compare our research to related work in \autoref{sec:related}.
    \item[] Finally, we conclude the paper in \autoref{sec:conclusion}.
\end{inparaitem}
 
\section{Methodology \& Scope}
\label{sec:methodology}
In contrast to conventional survey approaches, such as grounded theory for rigorous literature review~\cite{wolfswinkel2013using}, we do not sample included research from existing databases queried with specific terms.
Instead, we study and analyze security-oriented literature for vulnerabilities and threats related to the blockchain infrastructure.
The literature that we select as the input mainly covers existing blockchain-oriented surveys as well as various conferences and journals in security and distributed computing.
Moreover, we include other materials published in preprints, whitepapers, products' documentation, and forums, which are also related.

We aim to consolidate the literature, categorize found vulnerabilities and threats according to their origin, and as a result, we create four main categories (also referred to as layers).
At the level of particular main categories, we apply sub-categorization that is based on the existing knowledge and operation principles specific to such subcategories, especially concerning the security implications.
If some subcategories impose equivalent security implications, we merge them into a single subcategory.
See the road-map of all the categories in \autoref{fig:overview}.
Our next aim is to indicate and explain the co-occurrences or relations of multiple threats, either at the same main category or across more categories.

\begin{figure}[t]
	\begin{center}		
		\vspace{-0.4cm}	
		\subfloat[][All references]{\includegraphics[width=0.23\textwidth]{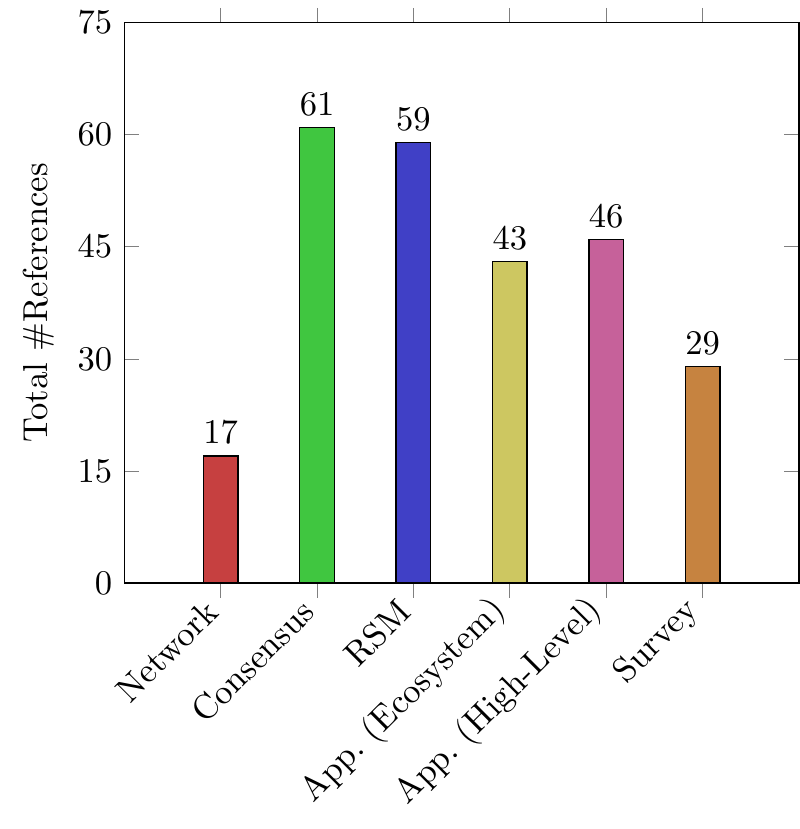}}
		\subfloat[][Academic references]{\includegraphics[width=0.23\textwidth]{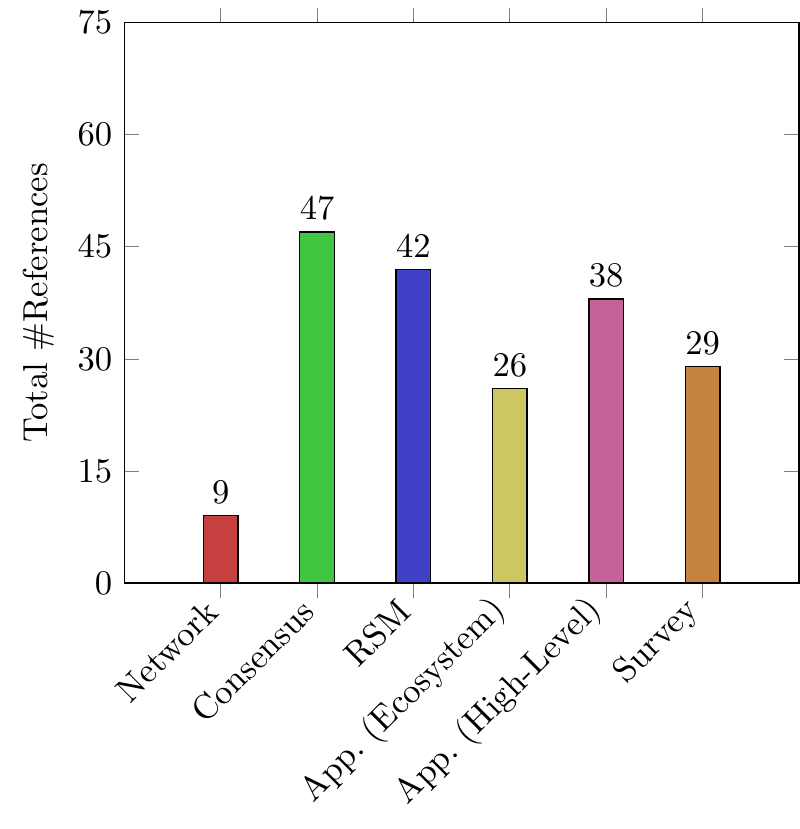}}
		\caption{Blockchain-specific references per category.}
		\label{fig:category_totals}		
	\end{center}	
\end{figure}

\begin{figure}[t]
	\vspace{-0.6cm}
	\begin{center}			
		\subfloat[][$\#$References per year]{\includegraphics[width=0.24\textwidth]{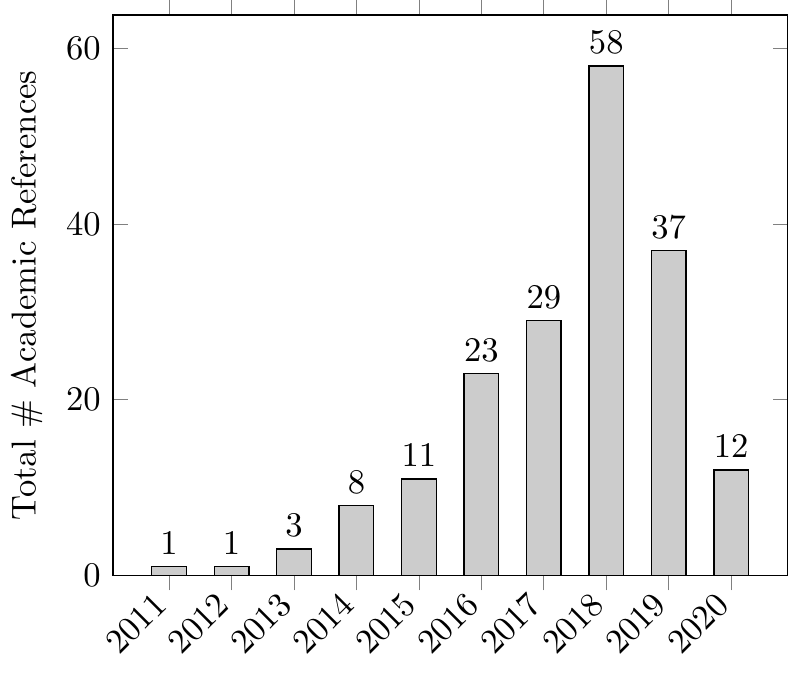}}
		\subfloat[][$\#$References per year/category]{\includegraphics[width=0.23\textwidth]{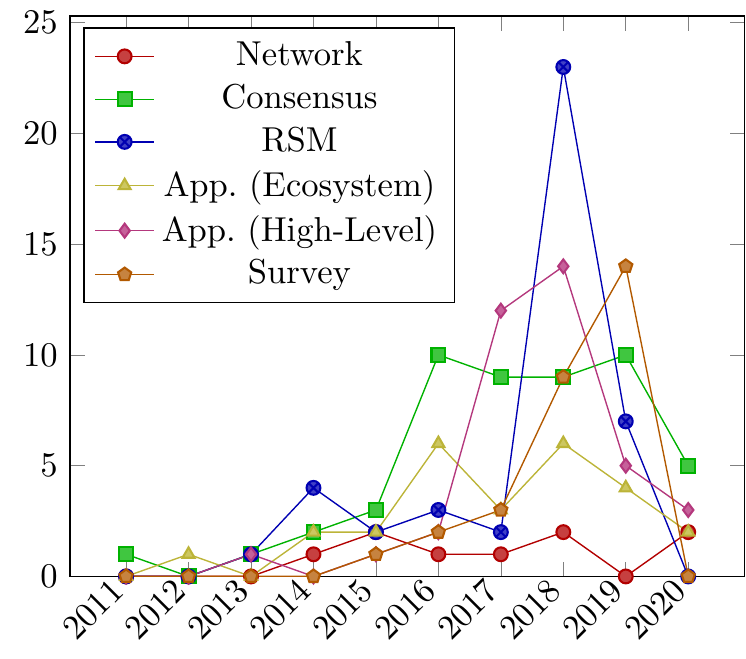}}
		\caption{Blockchain-specific academic references over time.}
		\label{fig:papers_over_time}		
	\end{center}	
\end{figure}

The scope of our work mainly covers aspects related to the blockchain core elements, while we mention common operational security issues (e.g., key management in the blockchain ecosystem) and countermeasures only tangentially if required.
Similarly, we do not pursue threats that emerge outside of the blockchain infrastructure and outside of the extra infrastructure required for certain blockchain-based applications.
Out-of-scope examples are remote exploitation of arbitrary devices (e.g.,  covert mining/crypto-jacking~\cite{tahir2017mining}) and issues related to using blockchain explorers (e.g., spoofing attacks, availability issues).
Examples of vulnerabilities that we assume only tangentially are semantic bugs and programming errors in the infrastructure of the blockchains -- we assume that core blockchain infrastructure is implemented correctly, uses secure cryptographic primitives, and provides expected functionality. 

\subsection{Quantitative Analysis}
For a quantitative summary of the literature, we considered only the references from the main sections that correspond to particular layers of the proposed stacked model (i.e., \autoref{sec:network} to \autoref{sec:apps-applications}) and related work (\autoref{sec:related}). 
We excluded references from other sections and the references on the examples of incidents (Appendix~\ref{appendix:incident-tables}).
Each assumed reference was labeled with three flags to indicate:
\begin{itemize}
 \item The category to which it belongs. Note that some papers belong to multiple categories. 
 \item Whether it is specifically related to blockchains. 
 \item Whether it was written by academics (i.e., such that their affiliation was listed in the document). 
 Note that the positive value of this flag covers also the papers that have not been peer-reviewed but which appeared on public repositories (e.g., arXiv and Cryptology ePrint Archive).
\end{itemize} 
In sum, we collected 276 references, of which 247 were blockchain-specific, 211 were written by academics, and 180 met both criteria.
In \autoref{fig:category_totals}, we show the number of references in each of the main categories for (a) all papers and (b) the academic papers. 
A total of 11 papers were found to belong to two categories.
We observe that most of the references are presented at the application layer, which we conjecture that occurred because each application running on the blockchain might introduce new attack vectors. 

The evolution of the number of references over time is depicted in \autoref{fig:papers_over_time}, where we observe that the number of academic papers about blockchains and security has increased steadily, although 2018 was the year with the highest number of references selected for the current work.\footnote{Note that we started to work on this survey in early 2019. } 
Of the four layers discussed in this paper, the network layer has the lowest number of blockchain-focused papers. 
A possible reason for it could be that among the remaining layers, the consensus and replicated state machine (RSM) layers are more specific to blockchains, while the application layer is popular among practitioners building on top of other layers.
Finally, we observe the number of surveys is steadily growing (culminating in 2019), which indicates increasing interest in this domain.

\section{Blockchains at a Glance}
\label{sec:background}
The blockchain is a data structure representing an append-only distributed ledger
that consists of entries (a.k.a., transactions) aggregated within ordered blocks.
The order of the blocks is agreed upon by mutually untrusting participants running a consensus protocol -- these participants are also referred to as nodes.
The blockchain is resistant against modifications by design since blocks are linked using a cryptographic hash function, and each new block has to be agreed upon by nodes running a consensus protocol.

A transaction is an elementary data entry that may contain arbitrary data, e.g., an order to transfer native cryptocurrency (i.e., crypto-tokens), a piece of application code (i.e., smart contract), the execution orders of such application code, etc. 
Transactions sent to a blockchain are validated by all nodes that maintain a
replicated state of the blockchain.

Blockchains were initially introduced as a means of coping with the centralization of monetary assets management, resulting in their most popular application -- a decentralized cryptocurrency with a native crypto-token.
Nevertheless, other blockchain applications have emerged, benefiting from features other than decentralization, e.g.,  privacy, energy efficiency, throughput, etc.
For the review of the inherent and non-inherent features of the blockchains, we refer the reader to Appendix~\ref{sec:features-of-blockchain}.

\begin{figure}[t]
	\centering        
	\includegraphics[width=0.93\columnwidth]{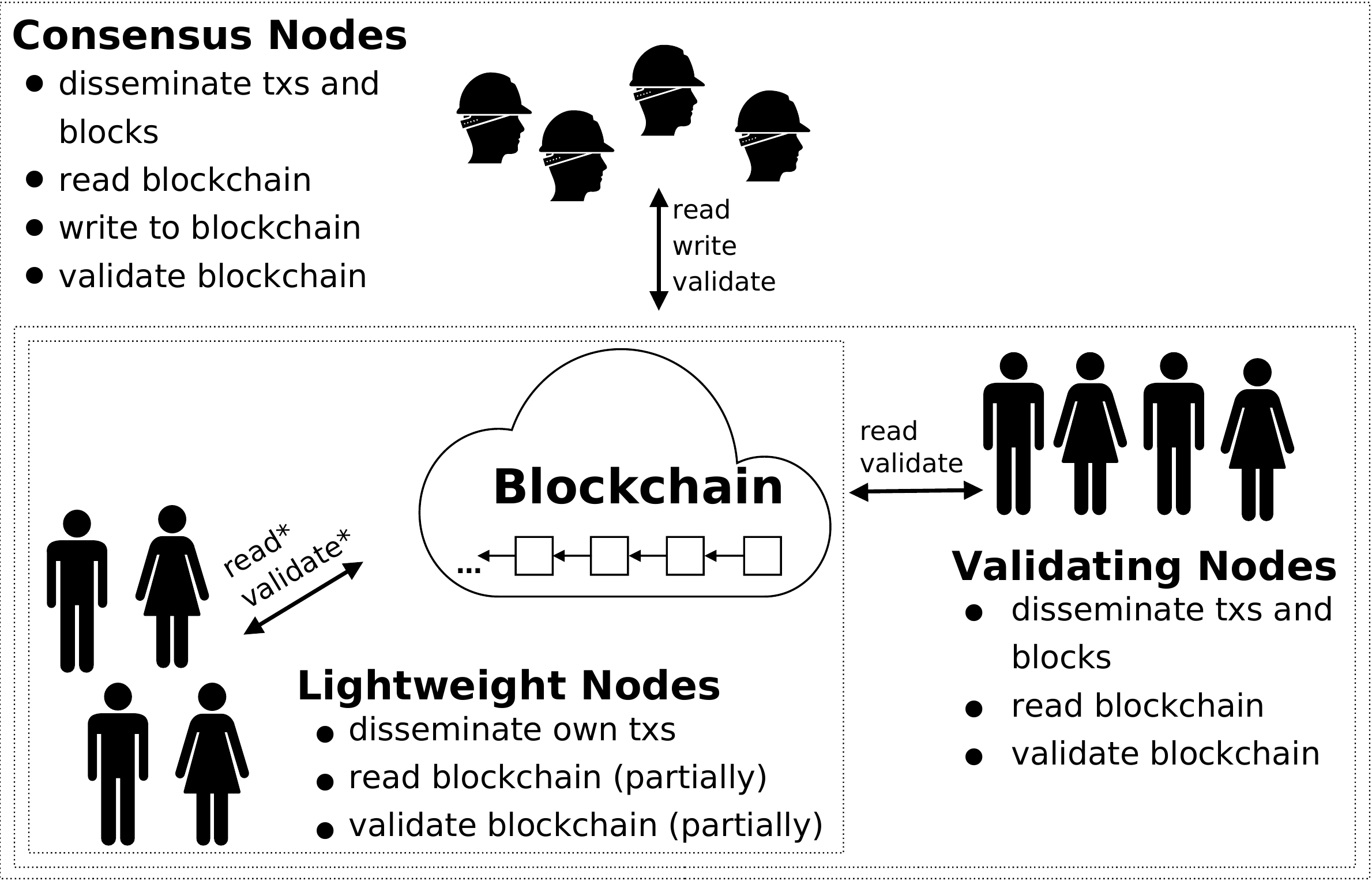}         
	\caption{Involved parties with their interactions and hierarchy.}
	\label{fig:node-types}
\end{figure}

\subsection{Involved Parties}\label{sec:involved-parties}
Blockchains usually involve three native types of parties that can be organized into a hierarchy, according to the actions that they perform (see~\autoref{fig:node-types}):

 \begin{compactdesc}

    \item[\textbf{(1) Consensus nodes}] (a.k.a., \textit{miners} in Proof-of-Resource protocols) actively participate in the underlying consensus protocol. 
    These nodes can read the blockchain and write to it by appending new transactions.
    Additionally, they can validate the blockchain and thus check whether writes of other consensus nodes are correct. 
    Consensus nodes can prevent malicious behaviors (e.g., by not appending invalid transactions, or ignoring an incorrect chain).
     
    \item[\textbf{(2) Validating nodes}] read the entire blockchain, validate it, and disseminate transactions. 
    Unlike consensus nodes, validating nodes cannot write to the blockchain, and thus they cannot prevent malicious behaviors. 
    On the other hand, they can detect malicious behavior since they possess copies of the entire blockchain.
    
    \item[\textbf{(3) Lightweight nodes}] (a.k.a., clients or Simplified Payment Verification (SPV) clients) benefit from most of the blockchain functionalities, but they are equipped only with limited information about the blockchain.  
    These nodes can read only fragments of the blockchain (usually block headers) and validate only a small number of transactions that concern them, while they rely on consensus and validating nodes. 
    Therefore, they can detect only a limited set of attacks, pertaining to their own transactions. 
 \end{compactdesc}

	\paragraph{Additional Involved Parties}
	Note that besides native types of involved parties, many applications using or running on the blockchain introduce their own (centralized) components.

\subsection{Types of Blockchains}\label{sec:types-of-blockchains}
Based on how a new node enters a consensus protocol, we distinguish the following blockchain types:
 \begin{compactitem}
    \item[\textbf{Permissionless}] blockchains allow anyone to join the consensus  protocol without permission.
    To prevent Sybil attacks, this type of blockchains usually requires consensus nodes to establish their identities by running a Proof-of-Resource protocol,  where the consensus power of a node is proportional to its resources allocated. 
    \item[\textbf{Permissioned}] blockchains require a consensus node to obtain permission to join the consensus protocol from a centralized or federated authority(ies), while nodes usually have equal consensus power (i.e., one vote per node).
    \item[\textbf{Semi-Permissionless}] blockchains require a consensus node to obtain some form of permission (i.e., stake) before joining the protocol; however, such permission can be given by any consensus node.
    The consensus power of a node is proportional to the stake that it has. 
 \end{compactitem}

\subsection{Design Goals of Consensus Protocols}\label{sec:backgroun-design-goal}

\subsubsection{Standard Design Goals -- Liveness and Safety}\label{sec:standard-design-goals}\hfill\\
\textbf{Liveness} ensures that all valid transactions are eventually processed -- i.e.,  if a transaction is received by a single honest node, it will eventually be delivered to all honest nodes.
\textbf{Safety} ensures that if an honest node accepts (or rejects) a transaction, then all other honest nodes make the same decision.
Usually, consensus protocols satisfy safety and liveness only under certain assumptions: the minimal fraction of honest consensus power or the maximal fraction of adversarial consensus power.
With regard to safety, literature often uses the term \textit{finality} and \textit{time to finality}. 
\textbf{Finality} represents the sequence of the blocks from the genesis block up to the block $B$, where it can be assumed that this sequence of blocks is infeasible to overturn.
To reach finality up to the block $B$, several successive blocks need to be appended after $B$ -- the number of such blocks is referred to as \textit{the number of confirmations}. 

\mysubsubsection{Specific Design Goals}\hfill\\
As a result of this study, we learned that standard design goals of the consensus protocol should be amended by specific design goals related to the type of the blockchain.
In permissionless type, \textit{elimination of Sybil entities}, \textit{a fresh and fair leader/committee election}, and \textit{non-interactive verification of the consensus result} is required to meet. 
In contrast, the (semi)-permissionless types do not require the elimination of Sybil entities (see details in \autoref{sec:specific-design-goals}).

\mysubsubsection{Means to Achieve Design Goals}
\hfill\\
\paragraph{Simulation of the Verifiable Random Function (VRF)}\label{sec:background-VRF}
To ensure a fresh and fair leader/committee election, all consensus nodes should contribute to the pseudo-randomness generation that determines the fresh result of the election. 
This can be captured by the concept of the VRF~\cite{micali1999verifiable}, which ensures the unpredictability and fairness of the election process.
Therefore, the leader/committee election process can be viewed as a simulation of VRF~\cite{wang2018survey}.
Due to the properties of VRF, the correctness of the election result can be verified non-interactively after the election took place.

\paragraph{Incentive and Rewarding Schemes}\label{sec:incentives}
An important aspect for protocol designers is to include a rewarding/incentive scheme that motivates consensus nodes to participate honestly in the protocol.
In the context of public (permissionless) blockchains that introduce their native crypto-tokens, this is achieved by block creation rewards as well as transaction fees, and optionally penalties for misbehavior.
Transaction fees and block creation rewards are attributed to the consensus node(s) that create a valid block (e.g., \cite{nakamoto2008bitcoin}), although alternative incentive schemes rewarding more consensus nodes at the same time are also possible (e.g., \cite{strongchain}).
While transaction fees are included in a particular transaction, the block reward is usually part of the first transaction in the block (a.k.a., \textit{coinbase} transaction).

\subsection{Basis of Consensus Protocols}\label{sec:background-basis}
\textit{Lottery} and \textit{voting} are two marginal techniques that deal with the establishment of a consensus~\cite{hyperledger1}.
However, in addition to them, their combinations have become popular.

\paragraph{Lottery-Based Protocols}
These protocols provide consensus by running a lottery that elects a leader/committee, who produces the block.
The advantages of lottery-based approaches are a small network traffic overheads and high scalability since the process is usually non-interactive (e.g.,~\cite{nakamoto2008bitcoin},~\cite{chen2017security},~\cite{kiayias2017ouroboros}). 
However, a disadvantage of this approach is the possibility of multiple ``winners'' being elected, who propose conflicting blocks, which naturally leads to inconsistencies called \textit{forks}.
Forks are resolved by fork-choice rules, which compute the difficulty of each branch and select the one. 
For the \textit{longest chain rule}, the chain with the largest number of blocks is selected in the case of a conflict, while for the \textit{strongest chain rule}, the selection criteria involve the quality of each block in the chain (e.g., \cite{sompolinsky2016spectre,sompolinsky2013accelerating,ethereum-classic-GHOST,zamyatinflux,strongchain}).
Note that the possibility of forks in this category of protocols causes an increase of the time to finality, which in turn might enable some attacks such as double-spending. 

\paragraph{Voting-Based Protocols} 
In this group of protocols, the agreement on transactions is reached through the votes of all participants. 
Examples include Byzantine Fault Tolerant (BFT) protocols -- which require the consensus of a majority quorum (usually $\frac{2}{3}$) of all consensus nodes
(e.g.,  \cite{castro1999practical,aublin2013rbft,buchman2018tendermint,kiayias2018ouroboros,duan2014bchain}).
The advantage of this category is a low-latency finality due to a negligible likelihood of forks. 
The protocols from this group suffer from low scalability, and thus their throughput forms a trade-off with scalability (i.e., the higher the number of nodes, the lower the throughput).

\paragraph{Combinations}
To improve the scalability of voting-based protocols, it is desirable to shrink the number of consensus nodes participating in the voting by a lottery, so that only nodes of such a committee vote for a block (e.g.,~\cite{gilad2017algorand}, \cite{daian2017snow}, \cite{hanke2018dfinity}, \cite{zilliqa2017zilliqa}, \cite{kiayias2018ouroboros}). 
Another option to reduce active voting nodes is to split them into several groups (a.k.a., \textit{shards}) that run a consensus protocol in parallel (e.g., \cite{Kokoris-KogiasJ18-omniledger, ZamaniM018-rapidchain}). 
Such a setting further increases the throughput in contrast to the single-group option, but on the other hand, it requires a mechanism that accomplishes inter-shard transactions.

\subsection{Failure Models in Distributed Consensus Protocols}
The relevant literature mentions two main failure models for consensus protocols~\cite{schneider1990implementing}: 

\begin{compactitem}
\item[\textbf{Fail-Stop Failures:}]
A node either stops its operation or continues to operate, while obviously exposing its faulty behavior to other nodes.
Hence, all other nodes are aware of the faulty state of that node (e.g., tolerated in Paxos~\cite{lamport1998paxos}, Raft~\cite{ongaro2014search}, Viewstamped Replication~\cite{oki1988viewstamped}).

\item[\textbf{Byzantine Failures:}]
In this model, the failed nodes (a.k.a., Byzantine nodes) may perform arbitrary actions, including malicious behavior targeting the consensus protocol and collusions with other Byzantine nodes.
Hence, the Byzantine failure model is of particular interest to security-critical applications, such as blockchains (e.g., Nakamoto's consensus~\cite{nakamoto2008bitcoin}, pure BFT protocols~\cite{castro1999practical}, \cite{aublin2013rbft}, \cite{buchman2018tendermint}, \cite{cachin2002sintra}, and hybrid protocols~\cite{gilad2017algorand,Kokoris-KogiasJ18-omniledger,ZamaniM018-rapidchain}). 
\end{compactitem}

\section{The Security Reference Architecture}
\label{sec:model}
We present two models of the security reference architecture, which facilitate systematic studying of vulnerabilities and threats related to the blockchains and applications running on top of them.
First, we introduce the stacked model, which we then project into the threat-risk assessment model.

\subsection{Stacked Model}\label{sec:stacked-model}
\label{sec:LayeredModel}
\begin{figure}[t]
	\centering
	\includegraphics[width=0.45\textwidth]{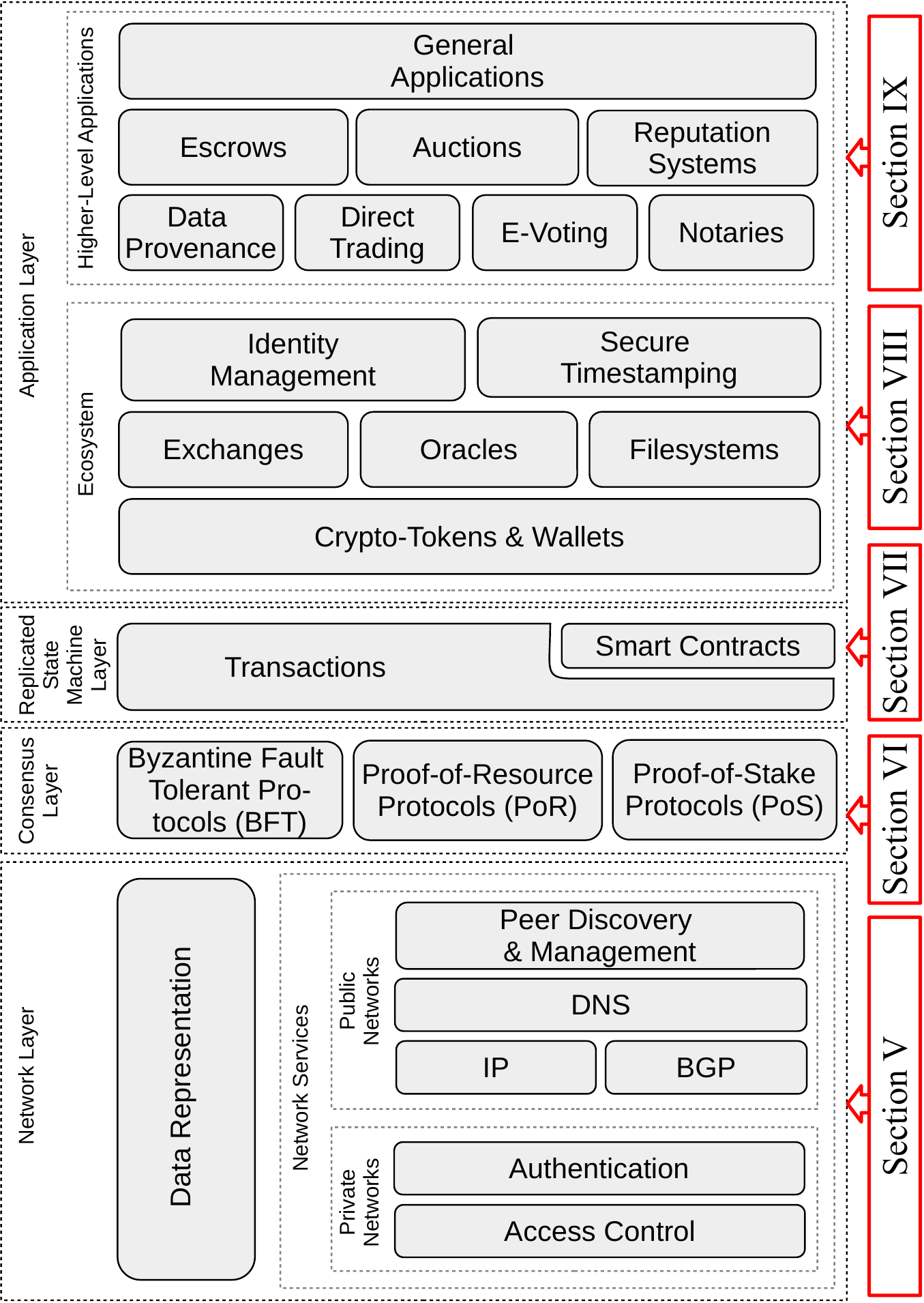}         
	\caption{Stacked model of the security reference architecture.}
	\label{fig:overview}
\end{figure}

To classify the security aspects of blockchains, we
utilize a stacked model consisting of four layers (see \autoref{fig:overview}).  
A similar stacked model was already proposed in the literature~\cite{wang2018survey}, but in contrast to it, we preserve only such a granularity level that enables us to isolate security threats and their nature, which is the key focus of our work.
In the following, we briefly describe each layer.

\begin{compactenum}
	\item[\textbf{(1) The  network layer}] consists of the data
	representation and network services planes. 
	The data representation plane deals with the storage, encoding, and protection of data, while the
	network service plane contains the discovery and communication with protocol peers, addressing, routing, and naming services. 	
	
	\item[\textbf{(2) The consensus layer}] deals with the ordering of transactions, and we divide it into three main categories according to the protocol type: Byzantine Fault Tolerant, 
	Proof-of-Resource,
	 and Proof-of-Stake protocols.

	\item[\textbf{(3) The  replicated state machine (RSM) layer}] deals with the interpretation of transactions, according to which the state of the blockchain is updated.
	In this layer, transactions are categorized into two parts, where the first part deals with the privacy of data in transactions as well as the privacy of the users who created them, and the second part -- smart contracts -- deals with the security and safety aspects of decentralized code execution in this environment.
	
	\item[\textbf{(4) The application layer}] contains the most common end-user functionalities and services.
	We divide this layer into two groups.
	The first group represents the applications that provide common functionalities for most of the higher-level blockchain applications, and it contains the following categories: wallets, exchanges, oracles, filesystems, identity management, and secure timestamping.
	We refer to this group as applications of the blockchain ecosystem.
	The next group of application types resides at a higher level and focuses on providing certain end-user functionality. 
	This group contains categories such as e-voting, notaries, identity management, auctions, escrows, etc.
	We found out that these higher-level applications inherit security aspects from particular categories in the underlying ecosystem group (see \autoref{fig:dependencies}).

\end{compactenum}

\smallskip\noindent
Throughout the paper, we summarize components and categories of particular layers of the reference architecture with their respective security threats, their origin, and corresponding countermeasures and/or mitigation techniques.

\subsection{Threat-Risk Assessment Model}
\begin{figure}[t]
	\begin{center}
		\includegraphics[width=0.47\textwidth]{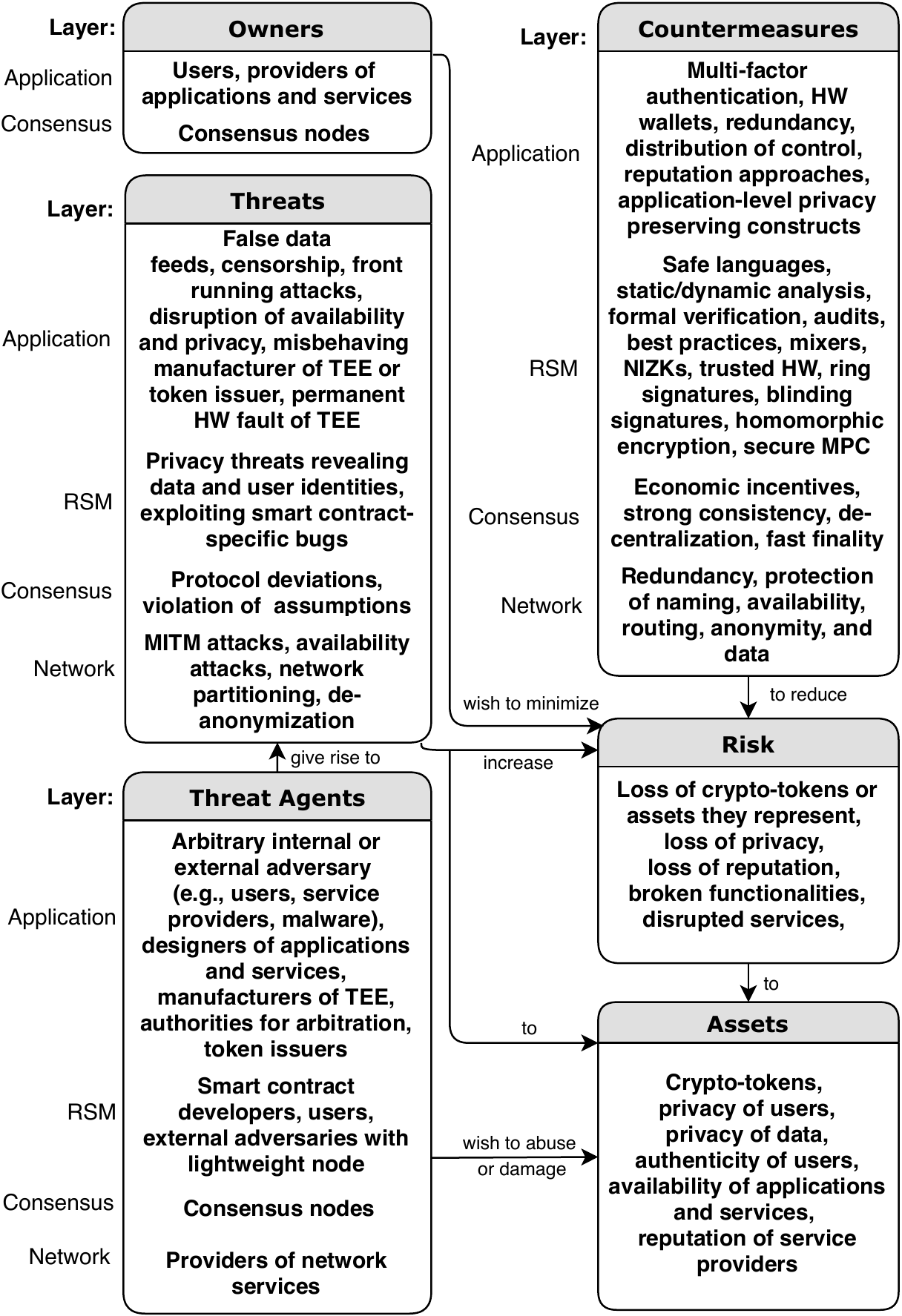}
		\caption{Threat-risk assessment model of the security reference architecture.}
		\label{fig:iso15408}
	\end{center}        
\end{figure}

To better capture the security-related aspects of blockchain systems, we introduce a threat-risk model (see \autoref{fig:iso15408}) that is
based on the template of ISO/IEC 15408~\cite{cc2017} and projection of our stacked model (see \autoref{fig:overview}). 
This model includes the following components and actors: 
 \begin{compactitem}
    \item[\textbf{Owners}] are blockchain users who run any type of node and they exist at the application layer and the consensus layer.
    Owners possess crypto-tokens, and they might use or provide blockchain-based applications and services.
    Additionally, owners involve consensus nodes that earn crypto-tokens from running the consensus protocol.
    \item[\textbf{Assets}] are present at the application layer, and they consist of monetary value (i.e., crypto-tokens or other tokens) as well as the availability of application-layer services and functionalities built on top of blockchains (e.g., notaries, escrows, data provenance, auctions).
	The authenticity of users, the privacy of users, and the privacy of data might also be considered as application-specific assets.
    Furthermore, we include here the reputation of service providers using the blockchain services.
    \item[\textbf{Threat agents}] are spread across all the layers of the stacked model, and they mostly involve malicious users whose intention is to steal assets, break functionalities, or disrupt services.
    However, threat agents might also be inadvertent entities, such as developers of smart contracts who unintentionally create bugs and designers of blockchain applications who make mistakes in the design or  ignore some issues.
    \item[\textbf{Threats}] facilitate various attacks on assets, and they exist at all layers of the stacked model.
    Threats arise from vulnerabilities in the network, smart
    contracts, applications, from consensus protocol deviations, violations of consensus protocol assumptions.
    \item[\textbf{Countermeasures}] protect owners from threats by minimizing the risk of compromising/losing the assets.   
	Alike the threats and threat agents, countermeasures can be applied at each of the layers of our stacked model, and they involve various security/privacy/safety solutions, incentive schemes, reputation techniques, best practices, etc. 
	Nevertheless, we emphasize that their utilization usually imposes some limitations such as higher complexity and additional performance overheads (e.g., resulting in decreased throughput). 

    \item[\textbf{Risks}] are related to the application layer, and they are caused by threats and their agents. 
    Risks may lead to a loss of monetary assets, a loss of privacy, a loss of reputation, service malfunctions, and disruptions of services and applications (i.e., availability issues).
 
 \end{compactitem}

\smallskip\noindent
The owners wish to minimize the risk caused by threats that arise
from threat agents.
Within our stacked model, different threat agents appear at each layer.  
At the \textbf{network layer}, there are service providers including parties
managing IP addresses and DNS names.
The threats at this layer arise from man-in-the-middle (MITM) attacks, 
network partitioning, de-anonymization, and availability attacks. 
Countermeasures contain protection of availability, naming, routing,
anonymity, and data.
At the \textbf{consensus layer}, consensus nodes may be malicious and wish to alter the outcome of the consensus protocol by deviating from it.
Moreover, if they are powerful enough, malicious nodes might violate assumptions of consensus protocols to take over the execution of the protocol or cause its disruption.
The countermeasures include well-designed economic incentives, strong consistency, decentralization, and fast finality solutions.
At the \textbf{RSM layer}, the threat agents may stand for developers who (un)intentionally introduce semantic bugs in smart contracts (intentional bugs represent backdoors) as well as users and external adversaries running lightweight nodes who pose threats due to the exploitation of such bugs. 
Countermeasures include safe languages, static/dynamic analysis, formal verification, audits, best practices, and design patterns.
Other threats of the RSM layer are related to compromising the privacy of data and user identities with mitigation techniques involving mixers, privacy-preserving cryptography constructs (e.g., non-interactive zero-knowledge proofs (NIZKs), ring signatures, blinding signatures, homomorphic encryption) as well as usage of trusted hardware (respecting its assumptions and attacker models declared).
At the \textbf{application layer}, threat agents are broad and involve arbitrary internal or external adversaries such as users, service providers, malware, designers of applications and services, manufactures of trusted execution environments (TEE) for concerned applications (e.g., oracles, auctions), authorities in the case of applications that require them for arbitration (e.g., escrows, auctions) or filtering of users (e.g., e-voting, auctions), token issuers.
The threats on this layer might arise from false data feeds, censorship by application-specific authorities (e.g., auctions, e-voting), front running attacks, disruption of the availability of centralized components, compromising application-level privacy, misbehaving of the token issuer, misbehaving of manufacturer of TEE or permanent hardware (HW) faults in TEE.
Examples of mitigation techniques are multi-factor authentication, HW wallets with displays for signing transactions, redundancy/distributions of some centralized components,  reputation systems, and privacy preserving-constructs as part of the applications themselves. 
We elaborate closer on vulnerabilities, threats, and countermeasures (or mitigation techniques) related to each layer of the stacked model in the following sections. 

\medskip
\paragraph{Involved Parties \& Blockchain's Life-Cycle}
In \autoref{sec:background}, we presented several types of involved parties in the blockchain infrastructure (see \autoref{fig:node-types}). 
We emphasize that these parties are involved in the operational stage of the blockchain's life-cycle.
However, in the design and development stages of the blockchain's life-cycle, programmers and designers should also be considered as potential threat agents who influence the security aspects of the whole blockchain infrastructure (regardless of whether their intention is malicious or not).
This is of great concern especially for applications built on top of blockchains (i.e., at the application layer) since these applications are usually not thoroughly reviewed by the community or public, as it is typical for other (lower) layers.

\section{Network Layer}
\label{sec:network}

\begin{figure}[t]
	\centering
	\includegraphics[width=0.45\textwidth]{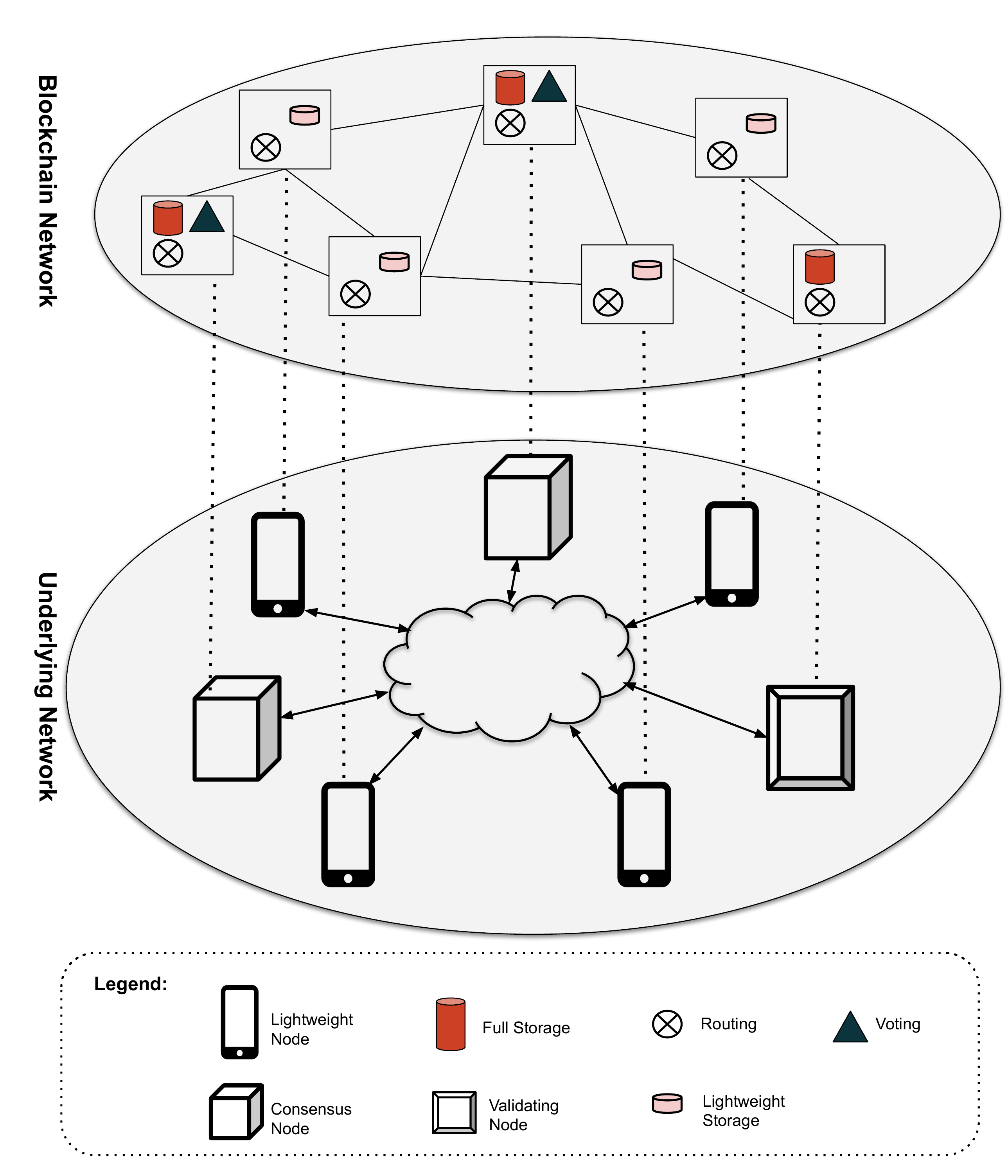} 	
	\caption{Overlaying of blockchains over a private/public network.}
	\label{fig:overlay}
\end{figure}

Blockchains usually introduce peer-to-peer overlay networks built on top of other networks (see \autoref{fig:overlay}). 
Hence, blockchains inherit security and privacy issues from their underlying networks. 
In our model (see \autoref{fig:overview}), we divide the network layer into  \textit{data representation} and \textit{network services} sub-planes. 
The data representation plane is protected by cryptographic primitives that ensure data integrity, user authentication, and optionally confidentiality, privacy, anonymity, non-repudiation, and accountability. 
The main services provided by the network layer are peer management and discovery, which rely on the internals of the underlying network, such as domain name resolution (i.e., DNS) or network routing protocols. 
Based on permission to join the blockchain system, the networks are either private or public. 
In the following, we discuss the pros and cons of private and public networks as well as their security threats and mitigation techniques.

\subsection{\textbf{Private Networks}}
\label{Private Network}
A private network ensures low latency, a centralized administration, privacy, and meeting regulatory obligations (e.g., HIPAA\footnote{Health insurance portability and accountability act, \url{https://hipaa.com/}.} for healthcare data).
The organization owning the network provides access to local participants as well as to external ones when required; hence, systems deploying private networks belong to the group of permissioned private blockchains. 
Private networks inherently provide authentication and access control. 

\mysubsubsection{Pros}
    
    \textit{Access control}
    is achieved by centralized authentication of users. 
    A private network has full control over routing paths and physical resources used, which enables suitable regulation of the network topology with regard to the given requirements.
    \textit{Data privacy} is ensured by permissioned settings.
    \textit{User identities}
     might only be revealed  within a private group of nodes. 
    \textit{Fine-grained authorization controls}
    are applied by the operators of the network resources to implement the security principle of minimal exposure and thus mitigate insider threat attacks on a local network. 
    \textit{Resource availability}
    is easier to manage and foresee, as all network participants and the deployment scenario are known ahead of time.

\mysubsubsection{Cons}
\begin{inparaenum}

	Virtual Private Network (VPN)
    connectivity is required to communicate between private networks spread over different geographical locations. 
    While VPNs are in general secure, they inherit the disadvantages of running a service over the Internet.
    Private networks are suitable only for permissioned blockchains.
\end{inparaenum}

\begin{figure}[t]
	\begin{center}		
		\includegraphics[width=0.82\columnwidth]{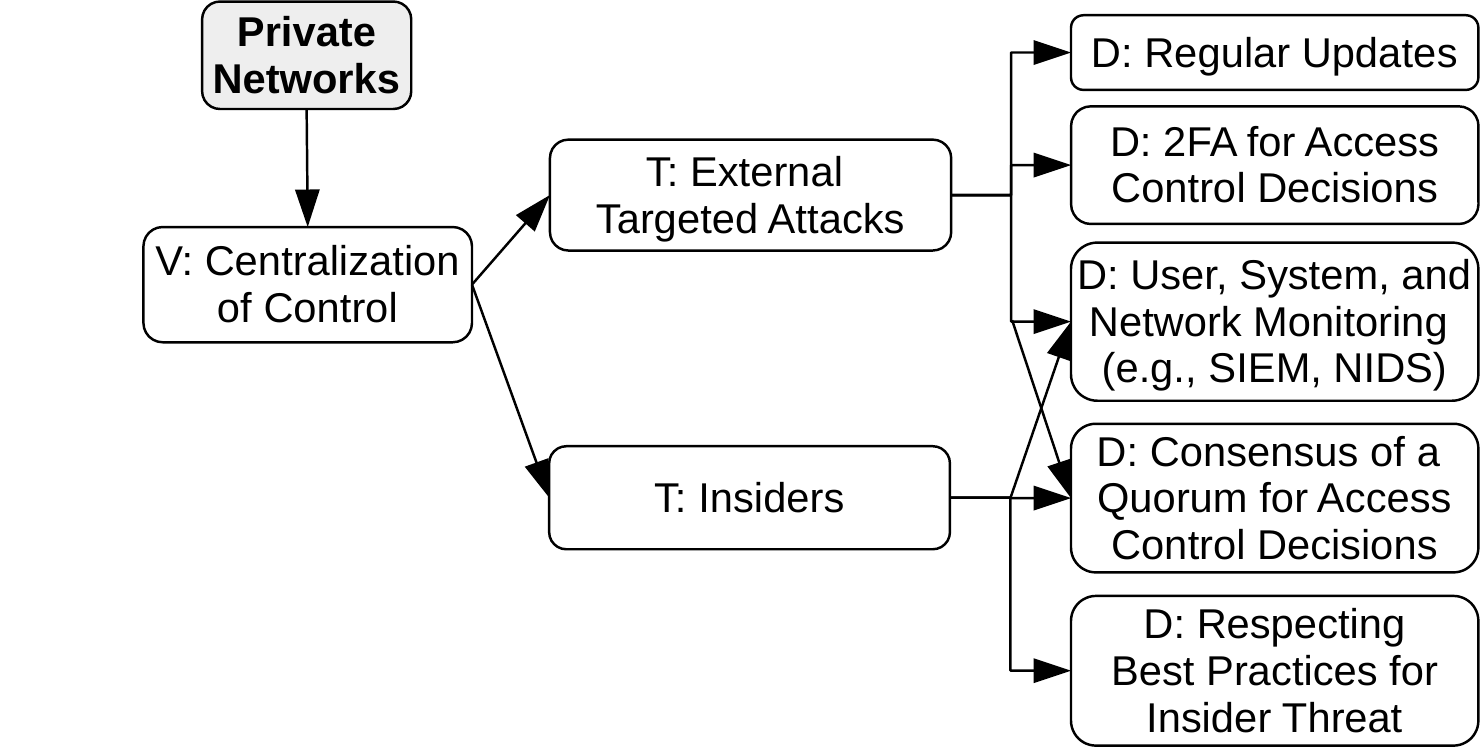} 		
		\caption{Vulnerabilities, threats, and defenses in private networks (network layer).}
		\label{fig:attacks-private-networks}
	\end{center}	
\end{figure}

\mysubsubsection{Security Threats and Countermeasures}
We present a taxonomy of vulnerabilities, threats, and defenses against them in \autoref{fig:attacks-private-networks}.
    We identified \textit{insiders} and \textit{external targeted attacks} as the specific security threats for the (permissioned) blockchains running over the private networks. 
    These threats are possible due to a centralization of access control that might occur in private networks, and thus permissioned blockchains.
    An external attacker might exploit a network or system vulnerability and obtain access to an element responsible for access control to the blockchain.
    In the case of the insider threat, she may already have the necessary privileges or obtain them by exploiting the system, network, or organization vulnerabilities.
    As a result, the insider might add malicious consensus nodes into the network or remove legitimate ones, and thereby increase the adversarial consensus power that is manifested at the consensus layer. 
    In turn, this may lead to plenty of attacks occurring at the consensus layer (see \autoref{sec:consensus}) such as attacks on violation of protocol assumptions.

    \textit{Countermeasures} include regular software updates, monitoring of users, network, and systems (e.g., by SIEM, anti-virus software, intrusion detection systems), prevention techniques that minimize trust and maximize trustworthiness such as two-factor authentication for access control decisions (effective for the external attacker), as well as respecting best practices for mitigation of insider threat~\cite{2016-Collins-CERTv5}.
	Another option of coping with the centralization of access control is to embed decentralized access control into the consensus layer and thus mandating a requirement on reaching a consensus of a quorum of nodes for each access control decision.    
	In contrast to the other mentioned countermeasures, the embedding of access control into the consensus layer is more effective since it eliminates a single-point-of-failure of centralized access control service, and thus it makes an increase of adversarial consensus power more difficult for the attacker -- the attacker has to compromise a high number of independent consensus nodes.

\mysubsubsection{Side Effects of Countermeasures}
Most of the mentioned countermeasures do not cause negative effects on the features of the blockchains under normal circumstances.
The exception might be embedding of access control into the consensus layer under the assumption that the set of consensus nodes in the network is extremely dynamic, and thus nodes are entering and leaving the blockchain very often. 
In such a case, the throughput of the blockchain might be decreased.

\subsection{\textbf{Public Networks / The Internet}}
Public networks provide high decentralization, openness, and low entry barrier, while network latency, privacy, and network control are put aside.
These networks are naturally required by all public (permissionless) blockchain systems.

\mysubsubsection{Pros}
\begin{inparaenum}

    \item[\textit{High availability}] is attractive to multi-homed nodes since they have alternate routes to send and receive messages. 
    Multi-homed nodes 
    may benefit from disseminating 
    blocks across multiple channels, thereby increasing the chance of blocks being appended to the blockchain.
    \item[\textit{High decentralization}] is achieved through geographical dispersion of nodes. 
    Public peer-to-peer (p2p) networks are harder to shut down. 
    \item[\textit{Openness and low entry barriers}] on the Internet are achieved through wide adoption, technology interoperability (e.g., using TCP/IP), economic  (e.g., low cost of broadband connection), and societal (e.g., resistance to regulations) factors. 
\end{inparaenum}

\mysubsubsection{Cons}\label{ssec:InternetCons}

\begin{inparaenum}
\item[\textit{Single-point-of-failure --}] 
DNS with its hierarchy, IP addresses, and autonomous systems (ASes) are managed by  centralized parties -- Internet Corporation for Assigned Names and Numbers (ICANN); in particular, Internet Assigned Numbers Authority (IANA). 
\item[\textit{External adversaries}]
pose a threat to public networks. 
These adversaries can be  classified based on their capabilities to which the  blockchain network may be exposed to~\cite{internetadversary}: 
(1) resources under attacker control (e.g., botnets, DNS and BGP servers),
(2) stolen or masqueraded identities (e.g., IP addresses participating in an eclipse attack or route manipulation),
(3) MITM attacker (i.e., eavesdropping and spoofing),
(4) the exploitation of common network vulnerabilities, 
(5) revealing secrets (e.g., de-anonymizing peers).
\item[\textit{Efficiency}] 
-- although an average Internet bandwidth has been improved in recent years, distribution of powerful infrastructure is not uniform, which results in a different latency among peers, and thus the overall latency of the network is increased; this might result in the loss of created blocks and thus wasting  consensus power.
\end{inparaenum}

\begin{figure}[t]
	\begin{center}		
		\includegraphics[width=0.97\columnwidth]{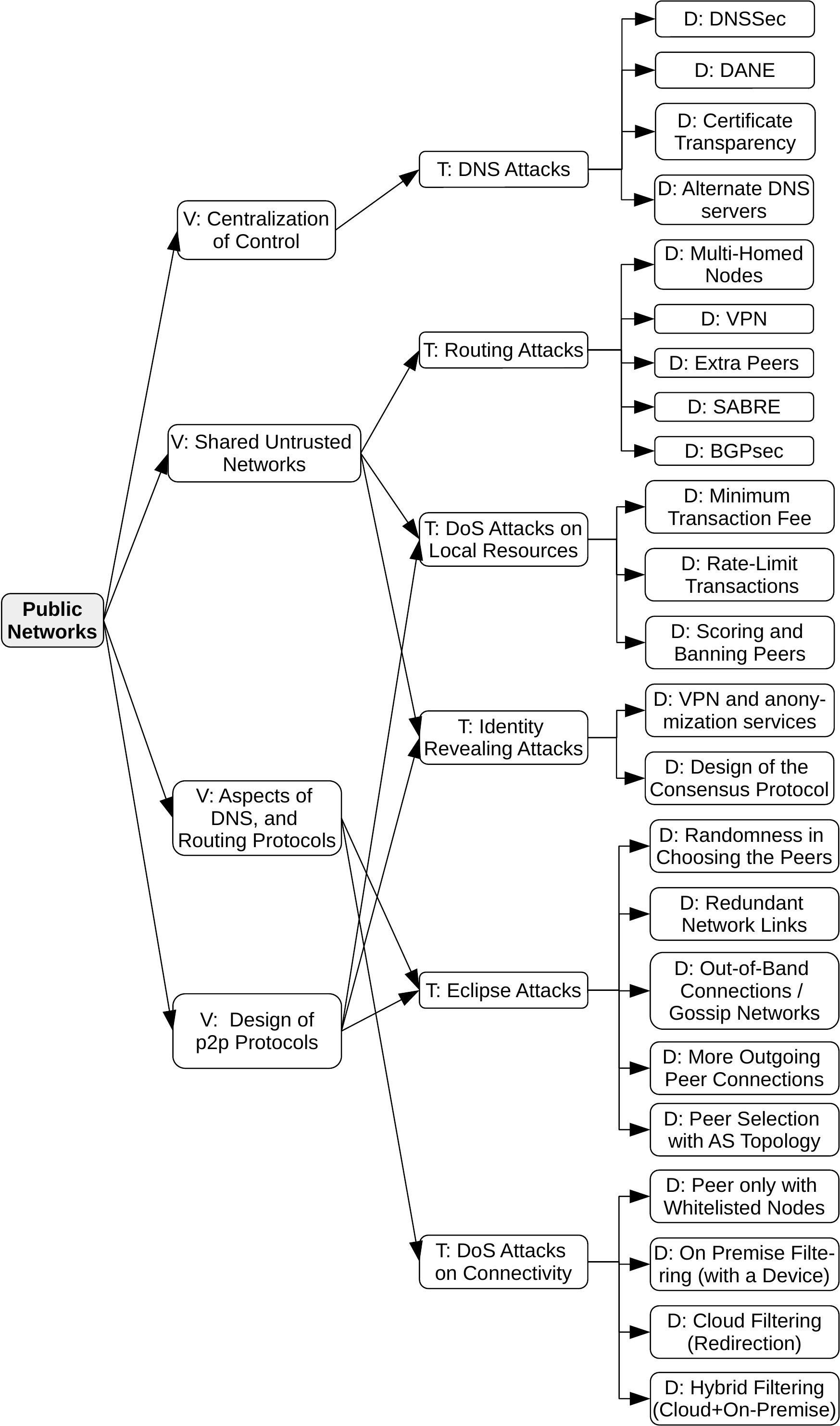} 		
		\caption{Vulnerabilities, threats, and defenses in public networks (network layer).}
		\label{fig:attacks-public-networks}
	\end{center}	
\end{figure}
    
\mysubsubsection{Security Threats and Countermeasures}
\label{ssec:InternetSecurityThreats}
We present a taxonomy of vulnerabilities, threats, and defenses related to public networks in \autoref{fig:attacks-public-networks}, while in \autoref{tab:incidents-network} of Appendix, we list several incidents that occurred in practice.
In the following, we describe these threats as well as possible defense techniques. 
\begin{compactdesc}

\item[\textbf{DNS attacks}] arise from cache poisoning
that mainly affects blockchains employing centralized DNS bootstrapping to retrieve online peers from a hard-coded list of DNS seeders. 
One \textit{countermeasure} is a security extension of DNS, called DNSSEC, which provides authentication and data integrity.
In addition to standard DNS, name resolution can also be made using alternate DNS servers~\cite{Bradshaw2016}.

\item[\textbf{Routing attacks}] are traffic route diversions, hijacking, or DoS attacks.
Besides simple data eavesdropping or modification, these attacks may lead to network partitioning, which in turn raises the risks of 51\% attacks or selfish mining attacks presented at the consensus layer (see \autoref{sec:consensus}). 
Apostolaki  et  al.~\cite{apostolaki2017hijacking}  demonstrated  that the Bitcoin  protocol  is  vulnerable  to  BGP  routing  attacks where  the  attacker  controlling  a  transit autonomous system (AS) can modify inter-domain routes for a few Bitcoin nodes and cause network partitioning.
However, perpetration of this attack reveals the identity of the malicious AS, which might have immediate reputation consequences.

\textit{Countermeasures} are multi-homed nodes (or using VPN) for route diversity, choosing extra peers whose connections do not pass through the same ASes, preference of peers hosted on the same AS within the same /24 prefix (to reduce risk of partitions), and fetching the same block from multiple peers~\cite{apostolaki2017hijacking}. 
Another mitigation is SABRE~\cite{apostolaki2018sabre}, a secure relay network that runs alongside with the Bitcoin network. 
Further,  BGPsec is a security extension for BGP used between neighboring ASes, and it assures route origin and propagation by cryptographic verification.

\item[\textbf{Eclipse attacks}] aim to hijack all connections of a node to its peers in a blockchain network.
Consequently, all traffic received and sent by the node is under the full control of the attacker.
Eclipse attacks arise from threats on DNS and routing in the network, and they may be a result of vulnerabilities in p2p protocols~\cite{heilman2015eclipse,wust2016ethereum,marcus2018low}.
Eclipse attacks increase chances of selfish mining and double-spending attacks (see \autoref{sec:consensus}) -- the eclipsed victims may unknowingly vote for an attacker's chain, and thus cause a network partitioning. 
Erebus~\cite{tran2020stealthier} is a stealthier attack causing network partitioning as compared to Apostolaki et al.~\cite{apostolaki2017hijacking}.
However, Erebus is not a routing attack since it does not involve BGP prefix hijacking (which is easy to detect) and has a very small network traffic footprint.
In Erebus, the attacker controls a large number of shadow IP addresses and influences the victim's peer selection mechanism to pick all outgoing connections with shadow IP addresses.
This is achieved through slowly flooding the victim's peer tables by incoming connections from the attacker-controlled shadow IP addresses.

\textit{Countermeasures:}
Improving randomness in choosing peers was proposed in the work of Heilman et al.~\cite{heilman2015eclipse} by several rules that manage the peer table.
Another mitigation strategy against eclipse attacks is to use redundant network links or out-of-band connections to verify transactions (e.g., by a blockchain explorer).
Eclipse attacks can also be detected by employing out-of-band gossip networks (e.g., web-servers)~\cite{alangot2020decentralized} that communicate with lightweight clients to exchange their views on the blockchain (i.e., block headers) along with native web traffic. 
Erebus attacks can be made much harder by decreasing the size of the peer tables, increasing the number of peers, preferring the peers that provide fresh data, and incorporation the topology of ASes into the peer selection process.
Also, note that countermeasures for DNS and routing attacks are applicable here as well.

    \item[\textbf{DoS attacks on connectivity}] of consensus nodes may result in a loss of consensus power, thus preventing consensus nodes from being rewarded~\cite{minerddos}.
    For validating nodes, this attack leads to a disruption of some blockchain-dependent services~\cite{fullnodeddos}. 
    \textit{Countermeasures:}
    One mitigation is to peer only with white-listed nodes.
    Methods to prevent volumetric DDoS include on-premise filtering (i.e., with an extra network device), cloud filtering (i.e., redirection of traffic through a cloud when DDoS is detected or through a cloud DDoS mitigation service), or hybrid filtering~\cite{hybriddos}. 

    \item[\textbf{DoS attacks on resources}] such as memory and storage, may reduce the peering and consensus capabilities~\cite{hdddos} of nodes.
    An example attack is flooding the network with low fee transactions (a.k.a., penny-flooding), which may cause memory pool depletion, resulting in a system crash. 
    A possible mitigation is raising the minimum transaction fee and the rate-limit to the number of transactions.
    Several mitigating techniques were applied to Bitcoin~\cite{weaknessesdos} nodes including scoring DoS attacks and banning misbehaving peers.
    DoS attacks on connectivity may also target (additional) centralized elements of blockchain infrastructure, such as servers communicating with hosted wallets (see \autoref{sec:wallets}), which in turn might lead to application layer attacks targeting clients of the wallets.
    
    \item[\textbf{Identity revealing attacks}] are conducted by linking the IP address of a node with an  identity propagated in transactions~\cite{BiryukovP14,millertopology2015}. 
    Traffic analysis using Sybil listeners can reveal the linkage of node IP addresses and their transactions~\cite{chainalysis2015}.
    \textit{Countermeasures} include using VPNs or anonymization services, such as Tor.
    See \autoref{sec:privacy-threats} for further identity and privacy-protecting mechanisms at the RSM layer.        

\end{compactdesc}    

\mysubsubsection{Side Effects of Countermeasures}
The anonymization services cause deterioration of connectivity for consensus nodes, and these nodes might not distribute the created block on time and thus lose their reward. 
On the other hand, the slow connectivity of anonymization services can be acceptable for validating nodes and clients transmitting messages related to the creation and validation of their transactions.
The trade-off for connectivity and anonymity can be provided by VPN services, which are fast but they are usually operated by a centralized party, and thus the risk of de-anonymization is higher than in the case of anonymization services.
Increasing the number of outgoing peer connections as protection against eclipse attacks~\cite{tran2020stealthier} can increase the volume of data that needs to be transferred, which might negatively impact the throughput of blockchains.
Furthermore, fetching the same blocks from multiple peers~\cite{apostolaki2017hijacking} may contribute to the network congestion under certain configurations of blockchains focusing on a high throughput -- this might lead to a slow down in keeping touch with the tip of the blockchain and thereby impact a response time of an application running on top of it.
The response time of the application might be also impacted by cloud and hybrid filtering that relay some incoming traffic through 3rd party services.

\section{Consensus Layer}
\label{sec:consensus}
The consensus layer of the stacked model deals with the ordering of transactions, while the interpretation of them is left for the RSM layer (see \autoref{sec:smart_contracts}).
The consensus layer includes three main categories of consensus protocols concerning different principles of operation and thus their security aspects.
First, we focus on the security aspects that are generic to all categories, and then we detail each category. 

\subsection{Generic Attacks}\label{sec:general-consensus-attacks}

We present a taxonomy of the generic threats to all types of consensus protocols, their origins, and defenses against them in \autoref{fig:attacks-consensus-generic}.
These threats originate mainly from \textit{violation of protocol assumptions} but also due to \textit{a long time to the finality} of some consensus protocols.
In the following, we describe these threats as well as possible defense techniques.

\mysubsubsection{Security Threats and Mitigations}\label{sec:generic-threats}
\begin{compactitem}
    \item[\textbf{Adversarial Centralization of Consensus Power.}] 
    In these attacks, a design assumption about the decentralized distribution of consensus power is violated.
    Examples of this category are \textit{51\% attacks} for PoR and PoS protocols as well as
    $\frac{1}{3}$ of \textit{Byzantine nodes} for BFT protocols (and their combinations).
    In a 51\% attack, the majority of the consensus power is held by the adversary, thus also the result of the protocol is under her control.
    In \textit{Byzantine attacks}, a quorum of $\frac{1}{3}$ adversarial consensus nodes might cause the protocol to be disrupted or even halted.
    As a design-oriented countermeasure, it is important to promote decentralization by incentive schemes that reward honest participation and discourage~\cite{miller2015nonoutsourceable} or punish~\cite{buterin2017casper,daian2017snow} protocol violations. 
    Another mitigation that makes these attacks more expensive is a statistical analysis of sudden anomalies in the history of the consensus power distribution among nodes, which can be embedded in the fork-choice rule of the consensus protocol~\cite{yang2019effective}.

    \item [\textbf{Breaking Network Assumptions.}]
	Protocols assuming synchronous or partially synchronous network delivery would inevitably fail when this assumption does not hold. 
	For instance, this assumption can be violated in BFT protocols by \textit{an unpredictable network scheduler}, as demonstrated on PBFT protocol~\cite{miller2016honey}.
    This fact motivates asynchronous BFT protocols that can be based on threshold-based cryptography, which enables reliable and consistent broadcast~\cite{cachin2002sintra}~\cite{miller2016honey}.
    
    \begin{figure}[t]
    	\begin{center}		
    		\includegraphics[width=0.98\columnwidth]{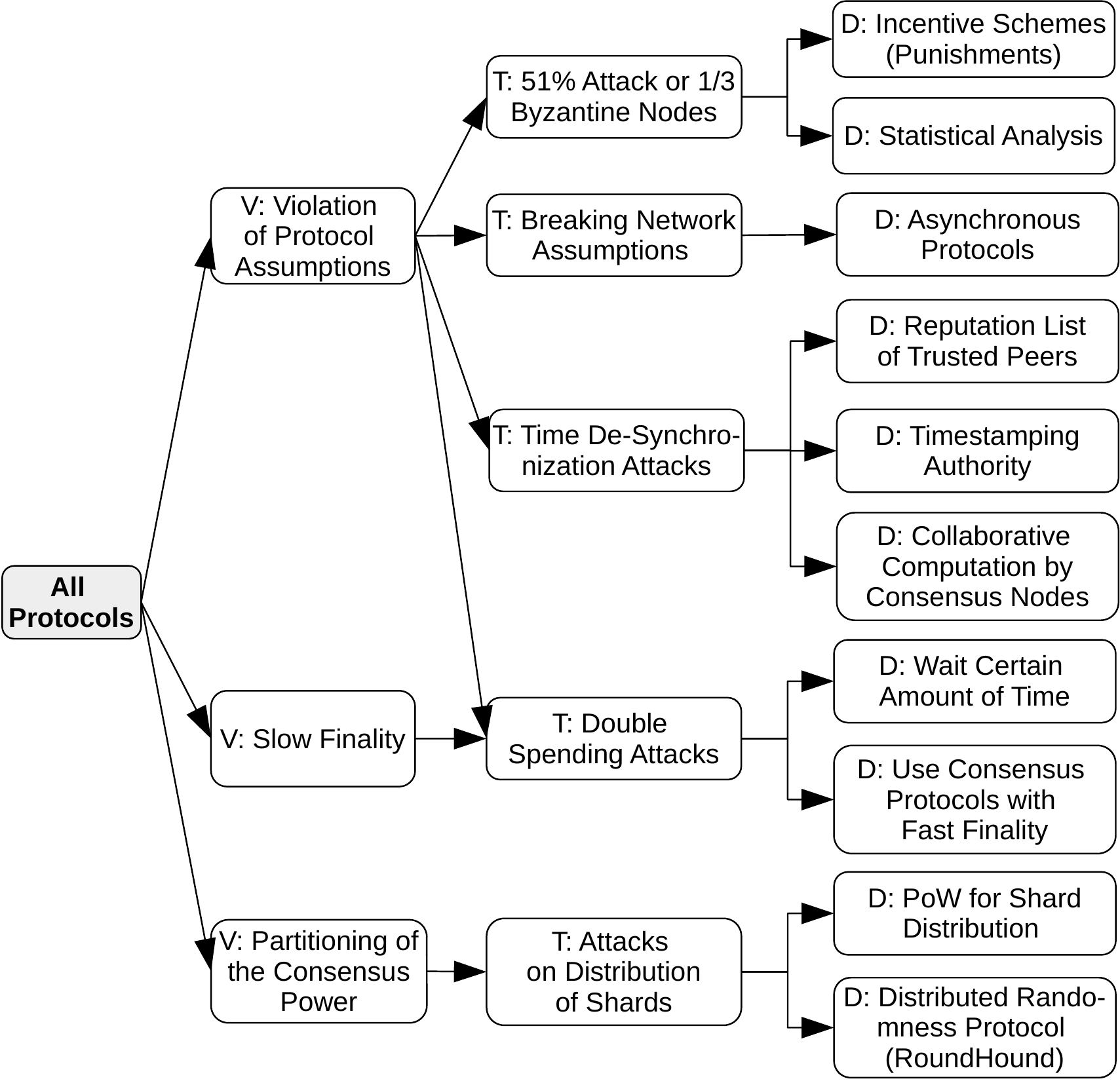} 		
    		\caption{Generic threats and defenses of the consensus layer.}
    		\label{fig:attacks-consensus-generic}
    	\end{center}	
    \end{figure}

    \item[\textbf{Time De-Synchronization Attacks.}]
    Usually, besides system time, nodes in PoW and PoS maintain network time that is computed as the median value of the time obtained from the peers.
    Such a time is often put into the block header, while nodes, upon receiving a block, validate whether it fits freshness constraints.
    An attacker can exploit this approach by connecting a significant number of nodes and propagate inaccurate timestamps, which can slow down or speed up the victim node's network time~\cite{timejacking}.  
    When such a desynchronized node creates a block, this block can be discarded by a network due to freshness constraints. 
    To avoid de-synchronization attacks, a node can build a reputation list of trusted peers or employ a timestamping authority~\cite{szalachowski2018short}. 
    Another option to improve the accuracy of block timestamps is to compute them collaboratively by consensus nodes~\cite{strongchain}.

	\item[\textbf{Double-Spending Attack.}]
	This attack is possible due to the creation of two or more conflicting blocks with the same height, resulting in inconsistencies called \textit{forks}.
	Thus, some crypto-tokens might be temporarily spent in both conflicting blocks, while only a single block is later included in the honest chain. 
	A double-spending attack mainly affects consensus protocols with slow finality.
	This attack usually occurs as a consequence of 51\% attacks.\footnote{Note that in the case of PoR protocols, this attack may also occur as a consequence of the selfish mining attack.}
	To prevent this attack, it is recommended to wait a certain amount of time (i.e., time to the finality) until a block ``is settled'' or utilize consensus protocols with fast time to the finality, such as BFT protocols and their combinations.

	\item[\textbf{Attacks on Shards.}]
	Sharding means that consensus nodes are distributed among subgroups (i.e., shards) such that each node only validates the transactions in its group. 
	Shards operate in parallel and can achieve higher scalability and throughput since each shard has a throughput similar to an entire non-sharded blockchain.
	On the other hand, sharding has the potential to harm security because each shard has a lower number of participating nodes than the entire blockchain, which means that it may be easier for the attacker to compromise a single shard than the entire blockchain~\cite{mango2018shard,bjelic2018shard}. 
	The main mitigation technique is to achieve a truly random distribution of nodes among the shards, and thus minimize the potential for adversaries to bias the randomness used for shard distribution. 	
	For example, Elastico~\cite{luu2016elastico} uses PoW to distribute nodes among shards, whereas Omniledger~\cite{Kokoris-KogiasJ18-omniledger} uses a bias-resistant distributed randomness protocol (e.g., RandHound~\cite{syta2017scalable}). 
	Poor design of sharding protocols may also lead to vulnerabilities such as replay attacks~\cite{sonnino2019replay}.

\end{compactitem}

\mysubsubsection{Side Effects of Countermeasures}
Some generic countermeasures presented in the above text might impact the features of the blockchains.
For example, the application layer countermeasure that uses the timestamping authority brings centralization issues, which might impact the availability of the service and enable misbehaving of the authority that might provide imprecise time. 
To cope with this issue, instead of using the application layer countermeasure, the computation of block timestamps can be embedded into the consensus protocol, e.g., by ensuring that multiple consensus nodes simultaneously contribute to global time using partial solutions~\cite{strongchain}. 

Enriching a fork-choice rule by statistical analysis of the history~\cite{yang2019effective} might protect against sudden changes in the distribution of the overall consensus power (e.g., rent-and-attack in PoW protocols), but not against a slow gradual increase of the consensus power by a coalition.
Furthermore, this mitigation technique incentivizes consensus nodes to use stable identifications (i.e., the same key pairs) to increase the strength of the honest chain and thus the chances that it becomes the main chain after a temporary fork.
This might impact the privacy and security of consensus nodes, who are disincentivized to rotate keys. 
Note that some privacy issues can be resolved at the RSM layer (see \autoref{sec:smart_contracts}).

Using a BFT consensus protocol helps to significantly decrease the likelihood of double-spending attacks, but on the other hand, it worsens the scalability of the blockchain and thus throughput, especially in the case of a high number of consensus nodes.
These issues can be resolved by combining a BFT voting protocol with lottery-based protocols that reduce the size of the actively communicating nodes to a small committee (e.g.,~\cite{gilad2017algorand},~\cite{daian2017snow},~\cite{hanke2018dfinity},~\cite{zilliqa2017zilliqa},~\cite{kiayias2018ouroboros}).
The scalability of BFT protocols can be also improved by using threshold-based signature aggregation with a gossip-based communication pattern (e.g.,~\cite{long2019scalable,silva2020comparison}).

Distributed randomness protocols and PoW for distribution of shards bring additional overheads, which may reduce the throughput of the blockchain; however, this reduction is negligible in contrast to the throughput improvement due to sharding.

\begin{figure*}[t]
	\begin{center}		
		\includegraphics[width=0.62\textwidth]{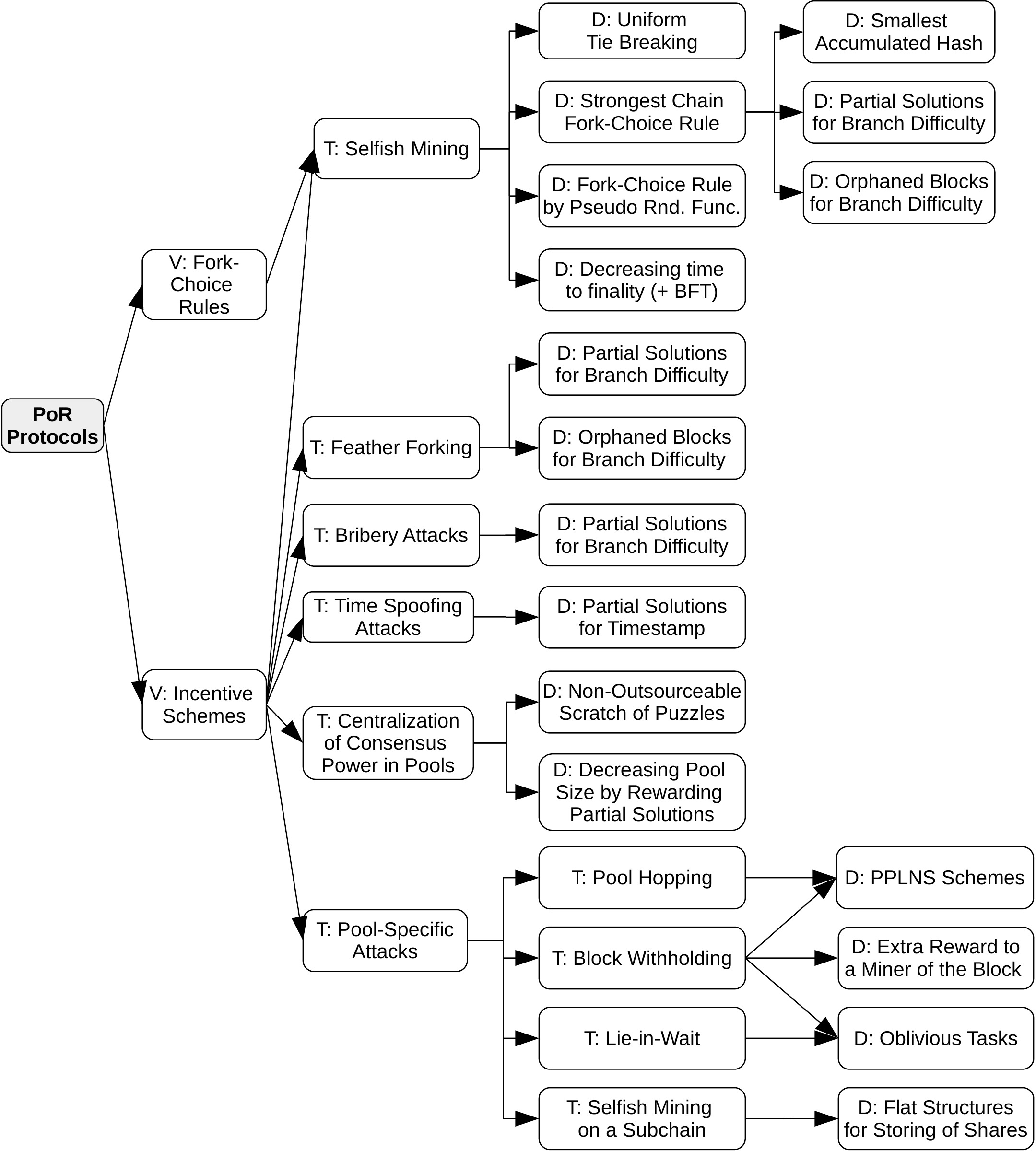} 		
		\caption{Vulnerabilities, threats, and defenses of PoR protocols (consensus layer).}
		\label{fig:attacks-consensus-PoR}
		\vspace{-0.3cm}
	\end{center}	
\end{figure*}

\subsection{Proof-of-Resource Protocols (PoR)}\label{sec:PoR}
Protocols from this category require nodes to prove the spending of a scarce resource in a lottery-based fashion~\cite{hyperledger1}. 
Scarce resources may stand for:
    \textit{(1) Computation}
    that is represented by Proof-of-Work (PoW) protocols (e.g., Bitcoin, Ethereum).
    \textit{(2) Storage} used in the setting of Proof-of-Space protocols~\cite{Dziembowski_proofsof} (e.g., Spacecoin~\cite{spacecoin}, SpaceMint~\cite{honsi2017spacemint}).
    \textit{(3) Crypto-tokens} spent for 
     Proof-of-Burn protocols~\cite{karantias2019proof} (e.g., Slimcoin~\cite{slimcoin}). 
    \textit{(4) Combinations and modification} of the previous types, such as storage and computation, called Proof-of-Retrievability (e.g., Permacoin~\cite{miller2014permacoin}) and  
    storage over time, which is represented by Proof-of-Space protocols (e.g., Filecoin~\cite{filecoin}).
    Another hybrid example of this category is a combination of PoR with elapsed time, such as in PeerCoin~\cite{peercoin}.
    However, it is a philosophical question whether to consider elapsed time as a resource that is spent or as a stake that is invested -- note that literature often categorizes PeerCoin as the first instance of a (hybrid) PoS protocol, hence we incline towards the second option.
    
PoR protocols belong to the first generation of consensus protocols, and they are mostly based on Nakamoto Consensus~\cite{nakamoto2008bitcoin} that utilizes PoW, inheriting its pros (e.g., high scalability) and cons (e.g., low throughput). 
For the detailed analysis of several PoW designs, we refer the reader to~\cite{2019-pow-evaluation}.

\mysubsubsection{Pros}
In PoR protocols, malicious overriding of the history of the blockchain (or part of it) requires spending at least the same amount of resources as was spent for its creation.
This is in contrast to the principles of PoS protocols, where a big enough coalition  may override the history at almost no cost.

\mysubsubsection{Cons} PoR protocols imply high operational costs.
Moreover, these protocols provide only probabilistic finality, which enables attacks forking the last few blocks of the chain.  

\subsubsection{Security Threats and Mitigations}
We present a taxonomy of the attacks related to PoR protocols, their origins, and defenses against them in \autoref{fig:attacks-consensus-PoR}, while  we list several real-world incidents in \autoref{tab:incidents-cons} of Appendix.
In the following, we describe these attacks as well as possible defense techniques.
\begin{compactitem}
    \item [\textbf{Selfish Mining:}] 
    In selfish mining~\cite{eyal2018majority},\footnote{Note that selfish mining is theoretically possible even in PoS protocols~\cite{brown2019formal}, but requiring them to have predictable randomness for the leader election, which is usually a design-oriented vulnerability (see~\autoref{sec:backgroun-design-goal}). However, in PoR, selfish mining is possible even with unpredictable randomness for the leader election.} an adversary attempts to privately build a secret chain and reveal it to the public only when an honest chain is ``catching up'' with the secret one. 
    The longest-chain rule causes honest miners to adopt the attacker's chain and invalidate the honest chain, thus wasting their consensus power.
    This attack is more efficient when the consensus power of a selfish miner reaches some threshold (e.g., 30\%).    
     The selfish mining strategy was later generalized~\cite{sapirshtein2016optimal} and extended to other variants that increase the profit of the attacker~\cite{nayak2016stubborn}.
    \textit{Countermeasures:} 
    For the case of the longest-chain rule, the first introduced mitigation is uniform tie-breaking~\cite{eyal2018majority}, which tells consensus nodes to choose the chain to extend uniformly at random, regardless of which one they received first. 
    However, this technique is less effective when assuming network delays~\cite{sapirshtein2016optimal}.     
    As the longest-chain rule enables this attack, it is recommended to use other fork-choice rules that also account for the quality of solutions and make the decision deterministic, as opposed to a uniform tie-breaking.
    An example of such a rule is to select the block based on the smallest hash value of the header.
    Another example is to include partial solutions~\cite{zamyatinflux, pass2017fruitchains,strongchain}
    or solutions representing full (orphaned) blocks~\cite{sompolinsky2013accelerating,zhang2017publish} in the computation of a chain quality.
    These partial or orphaned solutions can be incentivized by rewards to further improve decentralization.
    Another option for a deterministic fork-choice rule is using a pseudo-random function~\cite{kogias2016byzcoin}, which moreover provides unpredictability,\footnote{As opposed to uniform-tie breaking that provides only unpredictability, this solution additionally brings determinism.} hence an attacker cannot determine his chances to win a tie.     
    Finally, PoW protocols can be combined with BFT protocols, where PoW is used only for joining the protocol and BFT for consensus itself (e.g.,~\cite{Kokoris-KogiasJ18-omniledger,ZamaniM018-rapidchain,kogias2016byzcoin,zilliqa2017zilliqa,luu2016elastico}).     	
    
    \item [\textbf{Feather Forking:}]
    In this attack~\cite{feather-forks}, the adversary creates incentives for rational miners to collectively censor certain transactions.
    Before a mining round begins, an adversary announces that he will not extend the block containing blacklisted transactions, and thus will attempt to extend a forked chain.
    Although this strategy is not profitable for the adversary and the success rate is dependent on his consensus power, rational nodes prefer to join the censorship to avoid the potential loss.
    \textit{Countermeasures:} design-oriented protection is to minimize the chance of the attacker being successful, which can be done by including (and rewarding) partial solutions~\cite{zamyatinflux, rizun2016subchains,pass2017fruitchains,strongchain}
    or full orphaned blocks~\cite{sompolinsky2013accelerating,zhang2017publish} into branch difficulty computation.
    
    \item [\textbf{Bribery Attacks:}]
    Whereas feather forking involves adversaries who try to influence the behavior of miners by threatening to hurt their profits, bribery attacks involve the offering of direct rewards to miners. 
   	For example, consensus nodes could be bribed to enable double-spending attacks~\cite{bonneau2016buy} or to reorder transactions within a block, and thus enable the transaction front-running by other means than natural priority gas auctions (PGAs)~\cite{daian2019flash}.\footnote{In PGAs, users (arbitrage bots) compete with each other to be the first to interact with a smart contract (e.g., due to profit from intra-chain exchanges -- see \autoref{sec:exchanges}).}
   
   	\textit{Countermeasures:} assuming that the miners who accept bribes constitute a minority, a possible mitigation technique is to utilize partial solutions~\cite{strongchain,zamyatinflux, rizun2016subchains,pass2017fruitchains}, which reduce the likelihood that the double-spending attacks succeed. 
   	Regarding ``bribed'' transaction front-running, the misbehavior happens entirely off-chain since miners have complete control over the transaction ordering process, so on-chain mitigation is challenging.
   	Moreover, the likelihood of this attack is directly proportional to the consensus power of the bribed miner; hence, this attack is more feasible for mining pools.   
   	However, since no evidence confirming collusion between mining pools and bots has been found yet, mining pools are likely to be discouraged from accepting bribes due to the fear of consequences (e.g., a decrease in the market value of the crypto-tokens after these attacks are publicly disclosed). 
    
    \item [\textbf{Time-Spoofing Attacks:}]
    Time-spoofing attacks target a time-based difficulty computation algorithm in a PoR protocol with the intention to decrease the difficulty of the puzzle and thus minimize the effort for obtaining the same reward.
    In particular, the attacker is a consensus node that mines blocks with delayed timestamps, which indicates that a puzzle is too hard to meet block creation rate, and therefore difficulty needs to be decreased.
    \textit{Countermeasures:}
    A solution that improves the accuracy of the timestamps may utilize partial solutions found by all nodes into an averaged timestamp computation~\cite{strongchain}.
    Note that the impact of time spoofing attack might be significant also at the application layer, especially in use cases that rely on timestamp accuracy (see \autoref{sec:secure-timestamping}).
    
    \item [\textbf{Pool Specific Attacks:}]
    Since PoR protocols are usually based on a lottery having a single winner~\cite{nakamoto2008bitcoin}, rewards for participation impose a high payout variance for solo miners (i.e., once in a few years).
    As a consequence, mining pools emerged and caused centralization of the mining power, which may result in selfish mining, double-spending, or 51\% attacks.
    \textit{Countermeasures:}
    Non-outsourceable scratch-off puzzles~\cite{miller2015nonoutsourceable} avoid the creation of pools but require each consensus node to meet high demands on connectivity and storage, as opposed to centralized pools, where only a pool operator needs to meet these demands.
    If pools are acceptable, their size can be controlled by protocols that reward partial solutions~\cite{zamyatinflux, rizun2016subchains, pass2017fruitchains,strongchain} 
    and thus minimize payout variance.
    For a detailed analysis of rewarding schemes in pools, we refer the reader to~\cite{rosenfeld2011analysis}.
    In the following, we describe several types of pool-specific attacks.

    \paragraph{a) Pool Hopping} 
    The individual contribution of miners in a pool is proved by broadcasting partial solutions, called \textit{shares}. 
    If pay-per-share (PPS) rewarding is employed (i.e., pool operator instantly rewards miners showing shares), an attacker may jump into another pool after his mining time in a victim pool reaches a certain threshold~\cite{poolhop} since mining at the early stages of a round is statistically more profitable than mining at the end of the round. 
    As a countermeasure pay-per-last-N-shares (PPLNS) scheme and its variants~\cite{schrijvers2016incentive} can be used. 
    PPLNS removes the  concept of rounds and instead of immediate payments, it employs deterred payments after N shares are submitted by a miner.
    
    \paragraph{b) Block Withholding} 
    An attacker may try to sabotage a victim pool -- after mining a block in a victim pool, the attacker discards this block and continues mining at another pool~\cite{rosenfeld2011analysis}.
    Such withholding does not mean a direct gain for the attacker, but she may do a secret agreement with concurrent pool(s) that may reward the attacker for showing a withheld block~\cite{bag2017bitcoin} (a.k.a., sponsored block withholding).
    Mitigation for this kind of attack is using the PPLNS scheme, giving an extra reward to the miner of the block~\cite{bag2016yet}, precluding miners from distinguishing between a share and a full solution (i.e., oblivious tasks). 
    
    \paragraph{c) Lie-in-Wait} 
    If the miner finds a block in a victim pool, she does not immediately submit it to the pool operator, but instead focuses all her available mining power on the victim pool to increase her relative shares within a pool; after some time attacker releases the formerly found block.
    A countermeasure for this attack is an oblivious task~\cite{bag2017bitcoin}. 
    
    \paragraph{d) Selfish Mining on a Subchain}
     Decentralized mining pools, such as p2pool~\cite{p2pool}, achieve decentralization by updating an intermediary coinbase transaction with mined shares.
     To preserve consistency with the previous versions of the coinbase transaction within a mining round, its history is kept in a subchain.
     However, a  chaining data structure enables selfish mining on a subchain, besides the fact that it implies a high stale rate of shares in a subchain.\footnote{Note that the same applies for Flux~\cite{zamyatinflux} and Subchains~\cite{rizun2016subchains} that maintain a subchain but at the level of the whole network (as opposed to p2pool).}
     A possible countermeasure is to use flat data structures for aggregation of shares, such as the Merkle tree or hash of a set~\cite{strongchain}.
         
\end{compactitem}

\mysubsubsection{Side Effects of Countermeasures}
Partial solutions for difficulty computation might cause additional network overheads and thus decrease the throughput of the protocol. 
However, this is not the case for sufficiently long rounds of PoW protocols, such as in StrongChain~\cite{strongchain}. 
On the other hand, rewarding partial solutions, apart from mitigating some threats, helps to promote decentralization --  payout variance is decreased and thus mining pools are not needed in some cases, and in other cases, they can be much smaller.

\subsection{Byzantine Fault Tolerant (BFT) Protocols}\label{sec:BFT}
BFT protocols represent voting-based~\cite{hyperledger1} consensus protocols that utilize Byzantine agreement and a state machine replication~\cite{schneider1990implementing}.
These protocols assume a fully connected topology, broadcasting messages, and a master-replicas hierarchy.
Synchronous examples of this category are PBFT~\cite{castro1999practical}, RBFT~\cite{aublin2013rbft}, eventually synchronous examples are BFT-SMaRt~\cite{bessani2014state}, Tendermint~\cite{buchman2018tendermint}, Byzantine Paxos~\cite{cachin2009ByzantinePaxos}, BChain~\cite{duan2014bchain},
and asynchronous examples are SINTRA~\cite{cachin2002sintra} and HoneyBadgerBFT~\cite{miller2016honey}.
For more details, we refer the reader to a review of BFT protocols and their practical applications in both permissioned and permissionless blockchains~\cite{cachin2017blockchain}.

\mysubsubsection{Pros}
BFT protocols provide high throughput and fast finality. 
Another advantage of BFT protocols is that they can be combined with PoS or PoR protocols to achieve reasonable scalability and retain their other properties.
This is in line with a lottery approach~\cite{hyperledger1} for selecting a portion of all nodes, referred to as a committee, which further runs BFT consensus or its part (e.g., Algorand~\cite{gilad2017algorand}, Zilliqa~\cite{zilliqa2017zilliqa}, DFINITY~\cite{hanke2018dfinity}).

\mysubsubsection{Cons}
The main con of traditional BFT protocols~\cite{cachin2009ByzantinePaxos,castro1999practical} is low scalability caused by a high communication complexity (i.e., $\Theta(n^2)$). 
Since these protocols can work efficiently only with a limited number of consensus nodes, they can be used in their pure form only in permissioned blockchains.

\mysubsubsection{Improvements}
The issues with scalability vs. throughput of BFT protocols can be partially addressed by applying threshold signatures (e.g., HotStuff~\cite{yin2019hotstuff}, ByzCoin~\cite{kogias2016byzcoin}, Theta Blockchain~\cite{long2019scalable}) as well as by optimizing communication patterns from original broadcast to gossiping~\cite{long2019scalable} or communication trees~\cite{kogias2016byzcoin}. 
On top of that, consensus nodes might be partitioned into shards that process transactions in parallel (e.g., Omniledger~\cite{Kokoris-KogiasJ18-omniledger}, RapidChain~\cite{ZamaniM018-rapidchain}).

\begin{figure}[t]
	\begin{center}		
		\includegraphics[width=0.98\columnwidth]{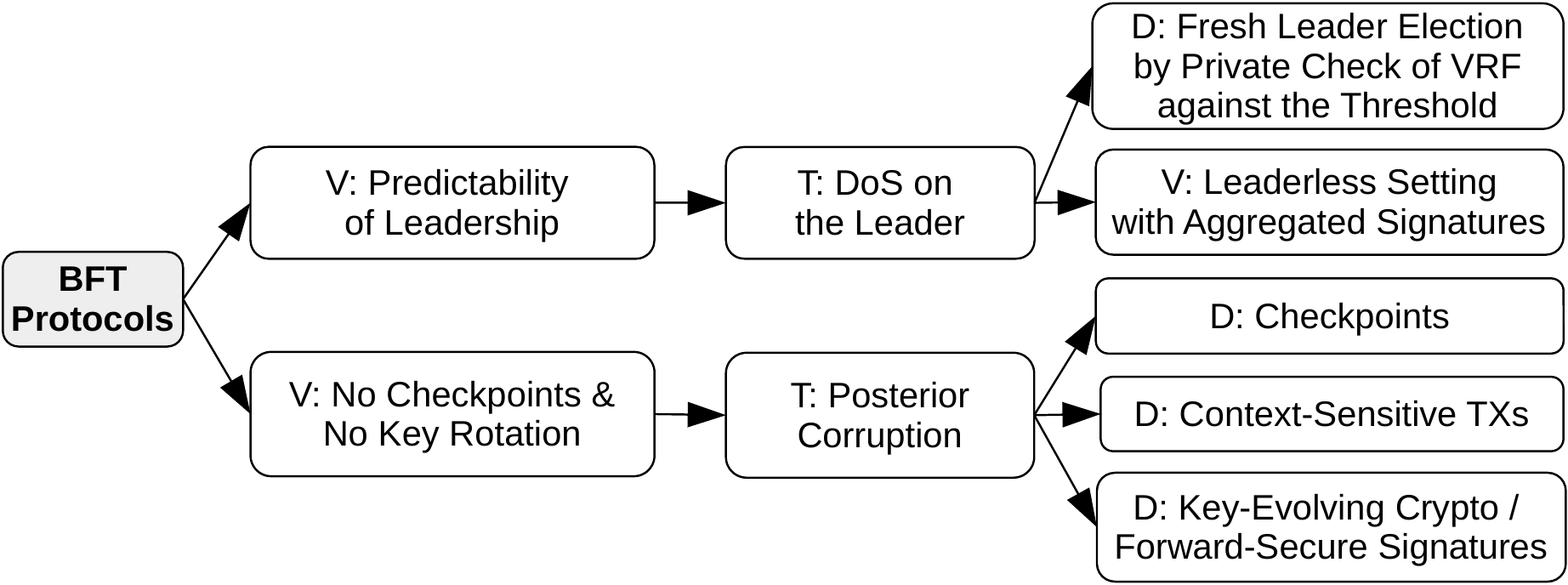} 		
		\caption{Vulnerabilities, threats, and defenses of BFT protocols (consensus layer).}
		\label{fig:attacks-consensus-BFT}
	\end{center}	
\end{figure}

\mysubsubsection{Security Threats and Mitigations}
We present a taxonomy of the attacks related to BFT protocols, their origins, and defenses against them in \autoref{fig:attacks-consensus-BFT}.
In the following, we describe these attacks as well as possible defense techniques.

\begin{compactitem}
	
	\item [\textbf{Denial of Service on a Leader:}]
	Since BFT protocols are mostly intended for (private) permissioned blockchains that are run by trusted participants, they do not assume the  existence of malicious nodes whose goal is to sabotage the protocol.		
	However, assuming such an adversary, a leader of the round might be DoS-ed since her leadership is known before the round starts, which might result in a restart of the round. 
	\textit{Countermeasures:}
	To prevent this attack, a node can privately determine whether it is a potential leader by using Verifiable Random Function (VRF)~\cite{gilad2017algorand}, and immediately release a block candidate; hence, after publishing this data, it is too late for a DoS attack on the node. 
	Another option for coping with this attack can be implemented by aggregating threshold signatures in a leaderless setting~\cite{long2019scalable}. 
	
	\item [\textbf{Posterior Corruption:}]
	Posterior corruption is a specific instance of a violation of protocol assumptions (see 
	\autoref{sec:general-consensus-attacks}) in which the adversarial consensus power reaches $\frac{2}{3}$.
	In posterior corruption, the adversary has to steal private keys of $\frac{2}{3}$ possibly ``retired'' consensus nodes and then rerun the consensus protocol, rewriting the history of the blockchain.
	Although this attack is mainly discussed in the context of PoS protocols (see \autoref{sec:PoS}), we note that in PoS the attacker's motivation is typically financial and exploits an incentive scheme of (semi-)permissionless blockchains. 
	On the other hand, the attacker's motivation might be different in BFT protocols in permissioned settings (e.g., governance or sabotage) since many such BFT protocols do not contain an incentive scheme.  	 
	To mitigate posterior corruption, key-evolving cryptography~\cite{franklin2006survey} and forward-secure digital signatures (e.g., d-ary certificate trees~\cite{bellare1999forward}) can be employed, requiring users to evolve their private keys and erase already used keys.	
	Another option is to employ irreversible checkpoints after a fixed number of blocks or context-sensitive transactions, which put the hash of a recent valid block into a transaction itself~\cite{gavzi2018stake}.
	
\end{compactitem}

\mysubsubsection{Side Effects of Countermeasures and Improvements}
A problem of some threshold signatures (e.g., BLS signatures) is a lack of forward secrecy, which might enable posterior corruption attacks. 
Forward secrecy might be provided by schemes such as $d$-ary certificate trees~\cite{bellare1999forward}.
On the other hand, a disadvantage of some forward-secure digital signatures (e.g.,~\cite{bellare1999forward}) is an extra overhead required for their verification and updating, which negatively influences the throughput of the blockchains.
This problem was addressed in Pixel signatures~\cite{drijvers2020pixel}, in which the authors proposed aggregated signatures supporting forward secrecy and demonstrated bandwidth and storage savings in contrast to $d$-ary certificate trees~\cite{bellare1999forward}.

Pruning the number of nodes that run BFT into committees~\cite{gilad2017algorand} reduces the security level of BFT and provides only probabilistic security guarantees depending on the committee size. 
This phenomenon might be also seen as a deterioration of decentralization.

\subsection{Proof-of-Stake Protocols (PoS)}\label{sec:PoS}
Similar to the PoR category, PoS protocols are based on the lottery approach~\cite{hyperledger1}.
However, in contrast to PoR, no scarce resource is spent; instead, the nodes are required ``to prove investment'' of crypto-tokens to participate in a protocol, and thus eventually earn interest from the invested amount.
The concept of PoS was for the first time proposed in the Bitcointalk forum~\cite{PoS-bitcointalk}.
The first technical realization of PoS is Peercoin~\cite{peercoin}, which is a combination with PoW -- each node has its particular difficulty for PoW, which is based on the age of the coins a node owns. 
Although there exist a couple of pure PoS protocols (e.g., Chains of Activity~\cite{bentov2016cryptocurrencies}, Ouroboros~\cite{kiayias2017ouroboros}), the trend is to combine them in a hybrid setting with PoR (e.g., Proof-of-Activity~\cite{bentov2014proof}, Peercoin~\cite{peercoin}, Snow White~\cite{bentov2016snow}) or BFT protocols (e.g., Algorand~\cite{gilad2017algorand}, Theta Blockchain~\cite{long2019scalable}).
In particular, a combination of PoS with BFT represents a promising approach, which takes advantage of both lottery and voting (i.e., scalability and throughput), where no resources are wasted.

\mysubsubsection{Pros}
The main feature of PoS protocols, as compared to PoR, is their energy efficiency.
Although some PoS protocols are often combined with a PoR technique (e.g.,~\cite{bentov2016snow,peercoin}), the overall energy spent is much smaller than in the case of pure PoR protocols. 

\mysubsubsection{Cons}
The introduction of PoS protocols has brought PoS specific issues and attacks, while these protocols are, at the time of writing, still not formally proven to be secure. 
Next, PoS protocols are semi-permissionless -- a node needs to first obtain a stake from any of the existing nodes to join the protocol.

\mysubsubsection{Security Threats and Mitigations}\label{sec:PoS-threats}
We present a taxonomy of the attacks related to PoS protocols, their origins, and defenses in \autoref{fig:attacks-consensus-PoS}.
In the following, we describe these attacks as well as possible defense techniques.

\begin{compactitem}    
    \item [\textbf{Nothing-at-Stake:}]
    Since generating a block in PoS does not cost any energy, a node can extend two or more conflicting blocks without risking its stake, and hence increase its chance to be rewarded. 
    Such behavior increases the number of forks and thus time to finality.
    \textit{Countermeasures:}
    Deposit-based solutions (e.g.,~\cite{buterin2017casper}) require nodes to make a deposit during some fixed period/round and checkpoint-based solutions (e.g.,~\cite{buterin2017casper,peercoin,daian2017snow,karakostas2020securing})  employ ``state freezing'' at periodic snapshots of the blockchain, while the blockchain can be reversed maximally up to the recent checkpoint.
    Another option is to punish a node that signs two conflicting blocks by embedding cryptographic solutions~\cite{li2017securing} that enable anybody to reveal the identity and a private key of such a node.
    Another countermeasure is to use backward penalization of nodes that produced two or more conflicting chains~\cite{daian2017snow,buterin2017casper}.
    Finally, PoS protocols can be combined with BFT approaches, and thus the probability of forks is negligible (e.g.,~\cite{gilad2017algorand}). 
     
    \item [\textbf{Grinding Attack:}]
    If the leader or committee producing a block is determined before the round starts, then the attacker can bias this process to increase her chances of being selected in the future.
    For example, if a PoS protocol takes only a hash of the previous block for the election process, the leader of a  block may bias a hash value by suitably adjusting the content of the block in a few attempts.
    \textit{Countermeasures:}
    The grinding attack can be prevented by performing a fresh leader election by an interaction of consensus nodes within some committee (e.g., the secure multiparty coin-flipping protocol~\cite{kiayias2017ouroboros}) or by privately checking whether the VRF output is below a certain stake-specific threshold (e.g.,~\cite{gilad2017algorand}).
    The input of the VRF is the user's private key and the randomness unambiguously bound to the previous block; hence each consensus node might compute the only VRF output during each round.

\begin{figure}[t]
	\begin{center}		
		\includegraphics[width=0.98\columnwidth]{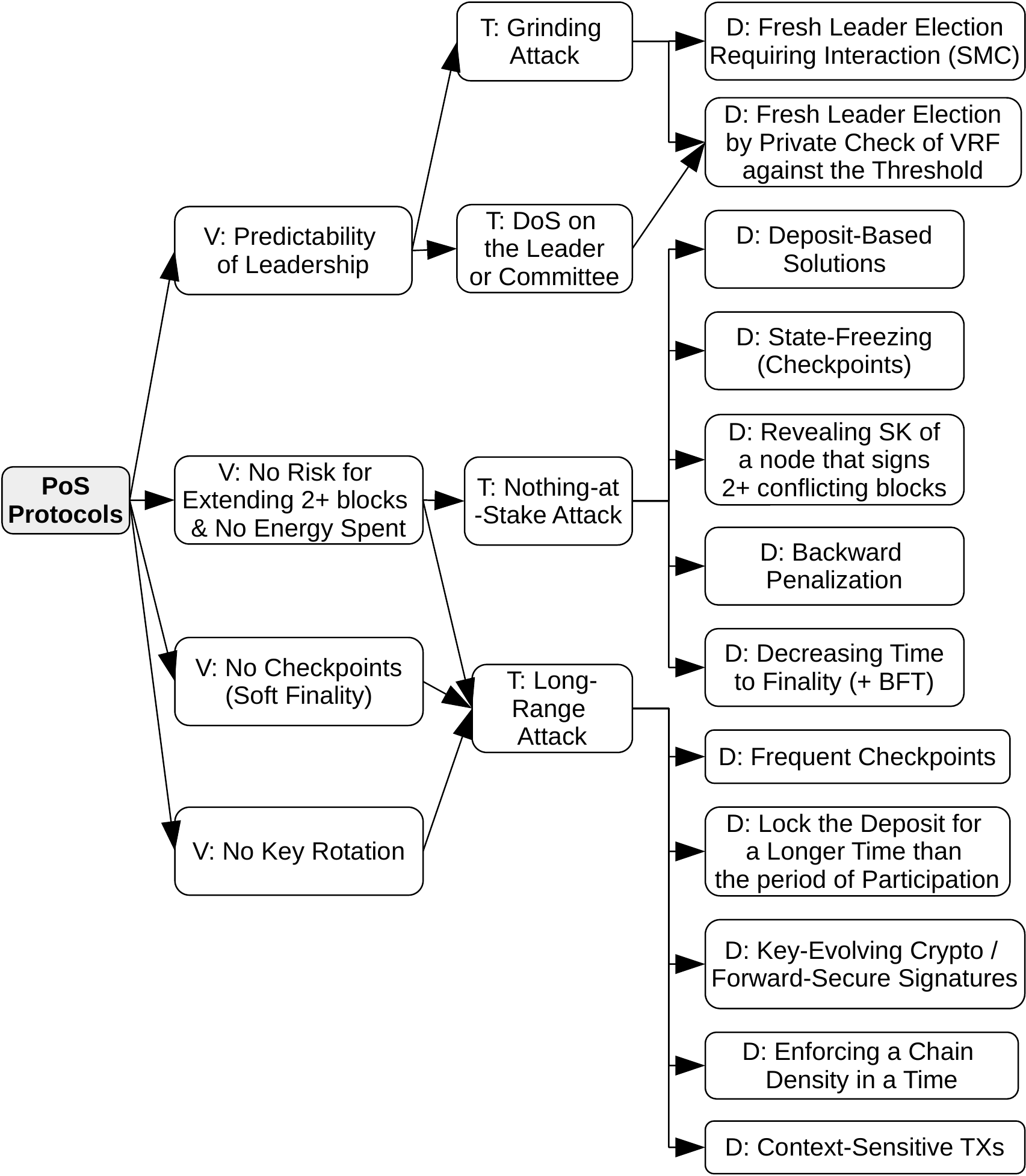} 		
		\caption{Vulnerabilities, threats, and defenses of PoS protocols (consensus layer).}
		\label{fig:attacks-consensus-PoS}
	\end{center}	
\end{figure}

    \item [\textbf{Denial of Service on a Leader/Committee:}]
    Alike in BFT protocols (see	\autoref{sec:BFT}), if a leader or a committee is publicly determined before the round starts~\cite{kiayias2017ouroboros}, then the adversary may conduct a DoS attack against them and thus cause a restart of the round -- this might be repeated until the adversary's desired nodes are elected.
    \textit{Countermeasures:}
    A prevention technique was proposed in Algorand~\cite{gilad2017algorand} -- a node privately determines whether it is a potential leader (or committee member), and immediately releases a block candidate (or a vote) -- hence, after publishing this data, it is too late for a DoS attack.
    The concept of the VRF was also utilized in other protocols (e.g.,~\cite{david2018ouroboros,hanke2018dfinity}).

    \item [\textbf{Long-Range Attack:}]
    In this attack~\cite{buterin2014long} (a.k.a., posterior corruption~\cite{daian2017snow}), an adversary can ``bribe'' previously influential consensus nodes to sell their private keys or steal the private keys by other means. 
    Since consensus nodes may exchange their crypto-tokens for fiat money anytime, selling their keys imposes no expenses and risk.
    If the attacker accumulates keys with enough stake in the past, he may rerun the consensus protocol and rewrite the history of the blockchain. 
    A variant of long-range attack that considers only transaction-fee-based rewarding and infrequent or no check-points is denoted as a \textit{stake-bleeding} attack~\cite{gavzi2018stake}.
    \textit{Countermeasures:}
    One mitigation is to lock the deposit for a longer time than the period of participation in the consensus~\cite{bano2017consensus}.
    The next mitigation technique is frequent periodic checkpointing, which causes the irreversibility of the blockchain with respect to the last checkpoint.
    Another option is to apply key-evolving cryptography~\cite{franklin2006survey} and forward-secure digital signatures~\cite{bellare1999forward}, which require users to evolve their private keys, while already used keys are erased~\cite{david2018ouroboros}.
    Hence, signatures cannot be forged in the case of compromise.
	The third mitigation technique is enforcing a chain density in a time-domain~\cite{gavzi2018stake} for the protocols where the expected number of participants in each round is known (e.g.,~\cite{kiayias2017ouroboros}).
    The last mitigation technique is context-sensitive transactions, which put the hash of a recent valid block into a transaction itself~\cite{gavzi2018stake}.

\end{compactitem}

\mysubsubsection{Side Effects of Countermeasures}
Some countermeasures for threats in PoS protocols might impact the features of the blockchains.
We note that secure multiparty coin-flipping protocol brings requirements on additional interactions among the consensus nodes, and thus it deteriorates the throughput of the protocol.
On the other hand, the throughput does not deteriorate when the leader and the committee are elected non-interactively by VRF.
 
\section{Replicated State Machine Layer}
\label{sec:smart_contracts}
The Replicated State Machine (RSM) layer is responsible for the interpretation and execution of transactions that are already ordered by the consensus layer.
Concerning security threats for this layer are related to the privacy of users, privacy and confidentiality of data, and smart contract-specific bugs. 
We split the security threats of the RSM layer into two parts: standard transactions and smart contracts.

\subsection{Transaction Protection}
Transactions containing plain-text data are digitally signed by private keys of users, enabling anybody to verify the validity of transactions with the corresponding public keys.     
However, such an approach provides only pseudonymous identities that can be traced to real IP addresses (and sometimes to identities) by a network-eavesdropping adversary, and moreover, it does not ensure the confidentiality of data~\cite{feng2019}.
Therefore, several blockchain-embedded mechanisms for the privacy of data and user identities were proposed in the literature, which we further elaborate on.
Note that some privacy-preserving techniques can be applied also on the application layer of our stacked model but imposing higher programming overheads and costs (e.g., see \autoref{sec:e-voting}, \autoref{sec:escrows}, and \autoref{sec:auctions}). 
This is common in the case of blockchain platforms that do not support them natively.

\begin{figure}[t]
	\begin{center}		
		\includegraphics[width=0.98\columnwidth]{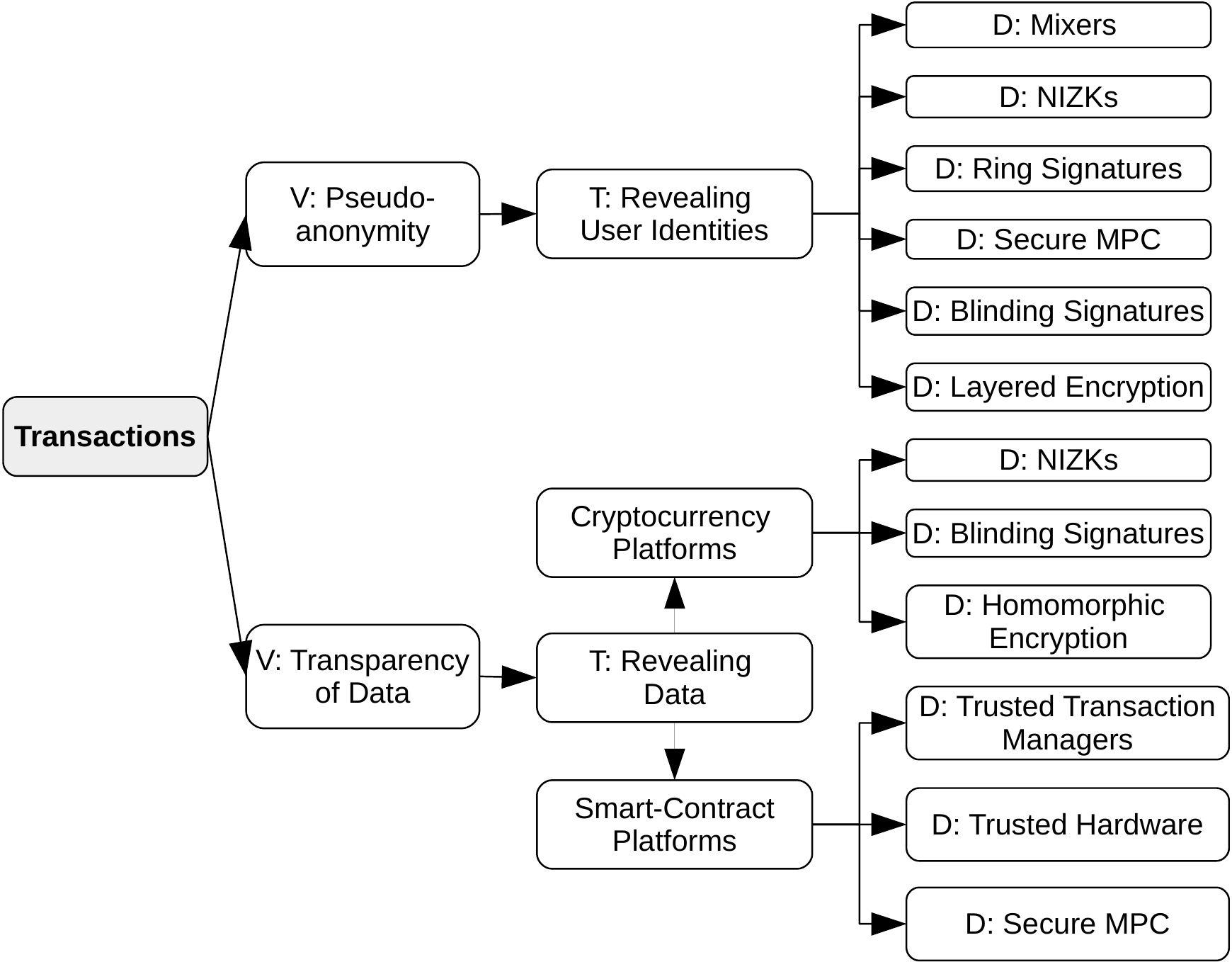} 		
		\caption{Vulnerabilities, threats, and defenses of privacy threats (RSM layer).}
		\label{fig:attacks-RSM-TXs}
	\end{center}	
\end{figure}

\mysubsubsection{Security Threats and Countermeasures}\label{sec:privacy-threats}
We present a taxonomy of vulnerabilities, threats, and defenses related to the privacy of transaction data and user identities in \autoref{fig:attacks-RSM-TXs}.

\begin{compactdesc}  
    
    \item[\textbf{Privacy Threats to User Identity.}]
    In many blockchains, user identities can be linked with their transactions by various deanonymization techniques, such as network flow analysis, address clustering, or transaction fingerprinting~\cite{feng2019,biryukov2014deanonymisation,pustogarov2015deanonymisation}.
     Moreover, blockchains designed with anonymity and privacy features (e.g., Zcash, Monero) are also vulnerable to a few attack strategies~\cite{kappos2018empirical,moser2018empirical}.
    \textit{Countermeasures:}
    Various means are used for obfuscating user identities, including centralized~\cite{bonneau2014mixcoin,valenta2015} and decentralized~\cite{maxwell2013coinjoin,ruffing2014coinshuffle,ziegeldorf2018} mixing services, ring signatures~\cite{noether2015ring}, and non-interactive zero-knowledge proofs (NIZKs)~\cite{miers2013zerocoin,sasson2014zerocash}.
    Some mixers enable internal linkability by involved parties~\cite{maxwell2013coinjoin} or linkability by the mixers~\cite{bonneau2014mixcoin}, which are also potential threats.
    Unlinkability for all parties can be achieved by multiparty computation (MPC)~\cite{ziegeldorf2018}, blinding signatures~\cite{valenta2015}, or layered encryption~\cite{ruffing2014coinshuffle}. 
    Ring signatures provide unlinkability to users in a signing group~\cite{noether2015ring}, enabling only the verification of correctness of a signature, without revealing an identity of a signer.

    \item[\textbf{Privacy of data.}]
    Blind signatures~\cite{heilman2016blindly} and NIZKs such as zk-SNARKs~\cite{sasson2014zerocash} or (shorter) Bulletproofs~\cite{bunz2018bulletproofs} can be used for the preservation of data privacy. 
    Another method is homomorphic encryption, 
     which enables the computation of certain operations over encrypted messages (e.g., ElGamal encryption provides additive homomorphism).
    Privacy and confidentiality for smart contract platforms can be achieved through trusted transaction managers~\cite{kosba2016hawk} utilizing zk-SNARKs, trusted hardware~\cite{cheng2018ekiden}, and secure multiparty computations~\cite{zyskind2015enigma} embedded into these platforms.
    Privacy of data can be achieved even on blockchain platforms without embedded support of privacy-preserving constructs. 
	For example, Zether~\cite{zether} is built on top of the public smart contract platform Ethereum, and it provides a confidential payment mechanism that embeds the balance of users into (secret) exponents of ElGamal encryption.
	Other similar examples that deal with the privacy of data at the application layer of our stacked model are presented in \autoref{sec:e-voting}, \autoref{sec:auctions}, \autoref{sec:escrows}.
\end{compactdesc} 

\mysubsubsection{Side Effects of Countermeasures}
Since protocols of mixing services usually contain a few rounds (that may involve the creation of several transactions), all mixing services slow down the transaction throughput.
The next blockchain feature that is influenced by some mixers is decentralization.
As a consequence, centralized mixing services may misbehave and reveal linkable information of transactions, they can be DoS-ed, or they can steal the funds. 
Accountability for the misbehavior of centralized mixing service is provided in MixCoin~\cite{bonneau2014mixcoin} and Blindcoin~\cite{valenta2015} while the latter additionally provides internal unlinkability by a mixing service.
In contrast to centralized mixers, decentralized mixers remove a trusted third party (i.e., no theft is possible) and provide stronger guarantees for unlinkability of transactions, e.g., CoinShuffle~\cite{ruffing2014coinshuffle} and CoinParty~\cite{ziegeldorf2018} require at least two and $\frac{2}{3} m$ honest participants, respectively, to provide full unlinkability.
Decentralization and availability are also impacted in solutions that utilize  trusted hardware~\cite{kosba2016hawk,cheng2018ekiden}.

The throughput of blockchains is also impacted in cryptographic countermeasures such as NIZKs, ring signatures, and blinding signatures.
Ring signatures cause the large transaction size, which is linear with the number of participants in the anonymity set.
The size of the ring signatures was optimized by cryptographic accumulators in~\cite{sun2017ringct}, which in turn enabled the improvement of the throughput. 
NIZKs utilized in ZeroCoin~\cite{miers2013zerocoin} produce large proofs as well. 
The proof size (and thus throughput) was further optimized by zk-SNARKs in ZeroCash~\cite{sasson2014zerocash}.
However, the disadvantage of zk-SNARKs is the requirement for a trusted setup.
This requirement is eliminated in Bulletproofs~\cite{bunz2018bulletproofs}, which further decrease the size of proofs in transactions and thus improves on throughput.

\subsection{Smart Contracts}\label{sec:smart-contracts}
Smart contracts introduced to automate legal contracts, 
now serve as a method for building decentralized applications on blockchains. 
They are usually written in a blockchain-specific programming language that may be Turing-complete (i.e., contain arbitrary programming logic) or only serve for limited purposes.
In the following, we describe these two contrasting types of smart contract languages and their security aspects. 

\begin{figure}[t]
	\begin{center}		
		\includegraphics[width=0.98\columnwidth]{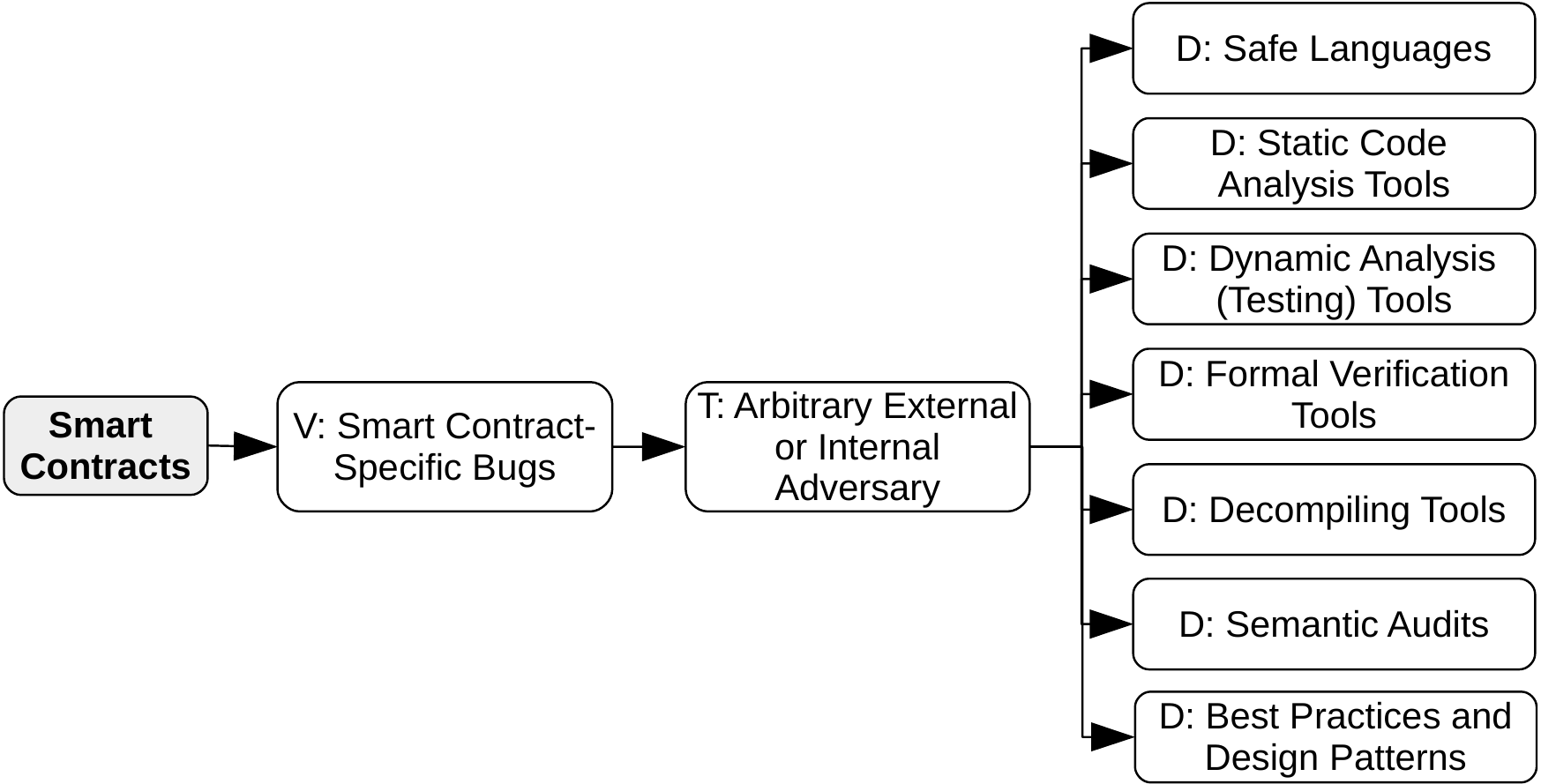} 		
		\caption{Vulnerabilities, threats, and defenses of smart contract platforms (RSM layer).		
		}
		
		\label{fig:attacks-RSM-sc}
	\end{center}	
\end{figure}

\mysubsubsection{Security Threats and Countermeasures}    
We present a taxonomy of vulnerabilities, threats, and defenses inherent to smart contract platforms in \autoref{fig:attacks-RSM-sc}.
\begin{compactdesc}
	\item[\textbf{Turing-Complete Languages.}] 
	An important aspect of this smart contract language category is a large attack surface due to the possibility of arbitrary programming logic.
	Examples of this category are Serpent, Vyper, Yul, Flint, LLL, and Solidity, while as of now Solidity is the most popular and widely-used one.
	\textit{Serpent}\footnote{\url{https://github.com/ethereum/wiki/wiki/Serpent-\%5BDEPRECATED\%5D}} is a high-level language that was designed to be simple and similar to the Python language. 
	However, Serpent was designed in an untyped fashion, lacking out-of-bound access checks of arrays and accepting invalid code by compilers~\cite{serpentAudit}, which opened the door for plenty of vulnerabilities. 
	Hence, Serpent showed to be an unsuccessful attempt to simplify the coding phase. 
	\textit{Vyper}\footnote{\url{https://vyper.readthedocs.io/en/v0.1.0-beta.9/}} is an experimental language designed to ease the audit of smart contracts and increase security -- it contains strong typing, bounds checks, and overflows. 
	\textit{Yul}\footnote{\url{https://solidity.readthedocs.io/en/v0.5.10/yul.html}}
	is a typed  intermediate language for Ethereum, which can be compiled to bytecode for the EVM 1.0, EVM 1.5 and eWASM platforms.
	Snippets of Yul code can be inserted as an inline assembly within Solidity code to perform optimizations that are applicable for these three platforms.
	\textit{Flint}~\cite{flint} is a type-safe language for Ethereum smart contracts. The major focus of this language is its robustness, and it also provides some 
	special features such as caller protection, which can help to produce robust contracts.
	\textit{Lisp Like Language (LLL)}\footnote{\url{https://lll-docs.readthedocs.io/}} is a low-level language that is similar to Assembler. 
	It aims to be simple and to support the creation of clean code; for example, it removes the need to code the stack and jump management. 
	Moreover, it enables a focus on the resource-constrained nature of Ethereum and allows optimized use of the resources.
	\textit{Solidity}\footnote{\url{https://solidity.readthedocs.io/}} is an object-oriented statically-typed language that is primarily used by the Ethereum platform. 
	Contracts written in Solidity can contain various types of vulnerabilities~\cite{atzei2017survey,SoliditySecurityList,historicalVulnerabilities}, which resulted in many incidents in the past.
	\autoref{tab:smartcontractincidents} of Appendix outlines the most prominent incidents and the associated vulnerabilities.
	In addition, the table classifies vulnerabilities according to an existing smart contract weakness classification (SWC) registry~\cite{SmartContractSecurity}.\footnote{Note that for some vulnerabilities there are no publicly available references on incidents supporting the existence of vulnerabilities.} 
	
	\smallskip
	\textit{Countermeasures:}
	Mitigation techniques for such vulnerabilities are static or dynamic analysis (testing) tools, formal verification tools, security audits, as well as respecting best practices and using known design patterns~\cite{SmartContractSecurity,trailofbits}.     
	The literature contains various smart contract analysis tools for the detection of vulnerabilities~\cite{parizi2018empirical,SCSecurityTools}. 
	In the following, we give an overview of them:
	
	\begin{compactitem}
		\item 
		\textit{\underline{Static analysis tools}} such as linters, try to find vulnerabilities by inspecting the source code. 
		For example, SmartCheck~\cite{tikhomirov2018smartcheck}, Solhint,\footnote{\url{https://github.com/protofire/solhint}}  Solium~\cite{solium}, and Slither\footnote{\url{https://github.com/crytic/slither}} belong to this category. Another example, sCompile~\cite{sCompile}, works statically, but it also includes a dynamic component.
		
		\item 
		\textit{\underline{Dynamic analysis tools}} seek vulnerabilities while executing the code of smart contracts.
		For example, simple forms of dynamic analysis are unit testing with hand-crafted tests or replay testing~\cite{ContractVis}, where existing executions (or manually captured ones) are used to check if the same results can be reproduced.
		A more automated form of testing is fuzzing \cite{sutton2007fuzzing}, which  generates unexpected, undefined, random, or invalid inputs to trigger a crash or reveal defects and vulnerabilities. 
		Fuzzers, like ContractFuzzer~\cite{jiang2018contractfuzzer},
		Echidna,\footnote{\url{https://github.com/crytic/echidna/}} and Harvey \cite{Harvey} can be used for automated smart contract testing as well.
		Another technique of dynamic analysis is symbolic execution~\cite{DBLP:journals/cacm/King76}, where a program is executed with symbolic values (i.e., logical expressions) that make it possible to explore all reachable paths of a program. 
		For this technique, there are tools like Securify~\cite{tsankov2018securify}, Manticore~\cite{manticore}, Oyente~\cite{luu2016making}, 
		and Osiris~\cite{osiris}.
		
		\item
		\textit{\underline{Formal verification tools}} usually apply an abstract model or a semantic definition to check for the security and/or correctness properties of smart contracts. 
		One type of this category is represented by semantic-based approaches that work with a semantic language specification defining the expected behavior of smart contracts. 
		Examples of this type are FSolidM~\cite{DBLP:journals/corr/abs-1711-09327}, and  Kevm~\cite{hildenbrandt2017kevm}. 	    
		Other types of formal verification tools are semantic-based approaches that work with a behavioral model~\cite{DBLP:conf/ntms/AbdellatifB18}.
		Several formal verification methods~\cite{DBLP:conf/ntms/MurrayA19,DBLP:conf/icsca/BaiCDH18,DBLP:conf/ithings/NehaiPD18} apply abstract models (e.g.,  finite state machines) that define the expected states and outputs of smart contracts for a given input. 
		The underlying models are used for checking the functional correctness or the presence of vulnerabilities (e.g., Zeus~\cite{zeus}). 
		Additionally, such models can be applied to proving certain security properties~\cite{DBLP:conf/post/GrishchenkoMS18}.

		\item \textit{\underline{Decompiling tools.}}
		The source code of contracts is often not public in contrast to their bytecode.
		For this reason, bytecode decompilers, like Erays~\cite{zhou2018erays}, Eveem,\footnote{\url{https://eveem.org/}} or Porosity~\cite{suiche2017porosity} can be used to (partially) reconstruct the source code of a contract.
		Additionally, there exist various static bytecode analyzers, like  Maian~\cite{nikolic2018finding}, 
		MadMax~\cite{MadMax}, Vandal~\cite{Vandal}, 
		and automated exploit generators, like Teether~\cite{krupp2018teether} that can be utilized to find vulnerabilities in the bytecode. 
		
	\end{compactitem}
	
	\smallskip
	
	\item[\textbf{Turing-Incomplete Languages.}]
	The main pro of this category is its design-oriented goal of a small attack surface and the emphasis on safety, which is achieved at the cost of limited expressiveness.
	Examples of this category are Pact, Scilla, Bitcoin Script, Ivy, and Simplicity.
	\textit{Pact}~\cite{popejoy2016pact} is a declarative language intended for the Kadena blockchain and provides type inference and module-guarded tables to prevent direct access to the module. 
	Pact is equipped with the ability to express and check properties of its programs, also leveraging satisfiability modulo theories (SMT) solvers.
	\textit{Scilla}~\cite{sergey2018scilla} is designed to achieve expressiveness and tractability while enabling formal reasoning about contract behavior.
	Every computation utilizes an automata-based model, and computations are realized as standalone atomic transitions that strictly terminate.
	Scilla enables external calls only in the last instruction of a contract, which simplifies proving safety and thus mitigates a few vulnerabilities.
	\textit{Bitcoin Script}~\cite{DBLP:conf/esorics/KlompB18} is a stack-based language for the Bitcoin platform.
	It has limited complexity and processing requirements, and its main purpose is transaction processing.
	\textit{Ivy} is a high-level declarative predicate language for the 
	Bitcoin platform. 
	It can be compiled to a Bitcoin script and its main advantage is its comprehensibility, which enables fast writing and an easy understanding of the code.
	\textit{Simplicity}~\cite{simplicity} is a typed functional language that works with combinators. 
	It is equipped with (formal) denotational and operational semantics, which facilitate the estimation of the required computing resources.
\end{compactdesc}

\mysubsubsection{Side Effects of Countermeasures}
Since all of the smart contract related countermeasures are performed before the deployment of smart contracts as part of the development stage of the blockchain-based applications, they do not negatively impact any blockchain features.
Note that only countermeasures related to the operational stage of blockchain-based applications might influence blockchain features.

\section{Application Layer: Ecosystem Applications}
\label{sec:apps}
\begin{figure}[t]
	\centering
	\includegraphics[width=0.47\textwidth]{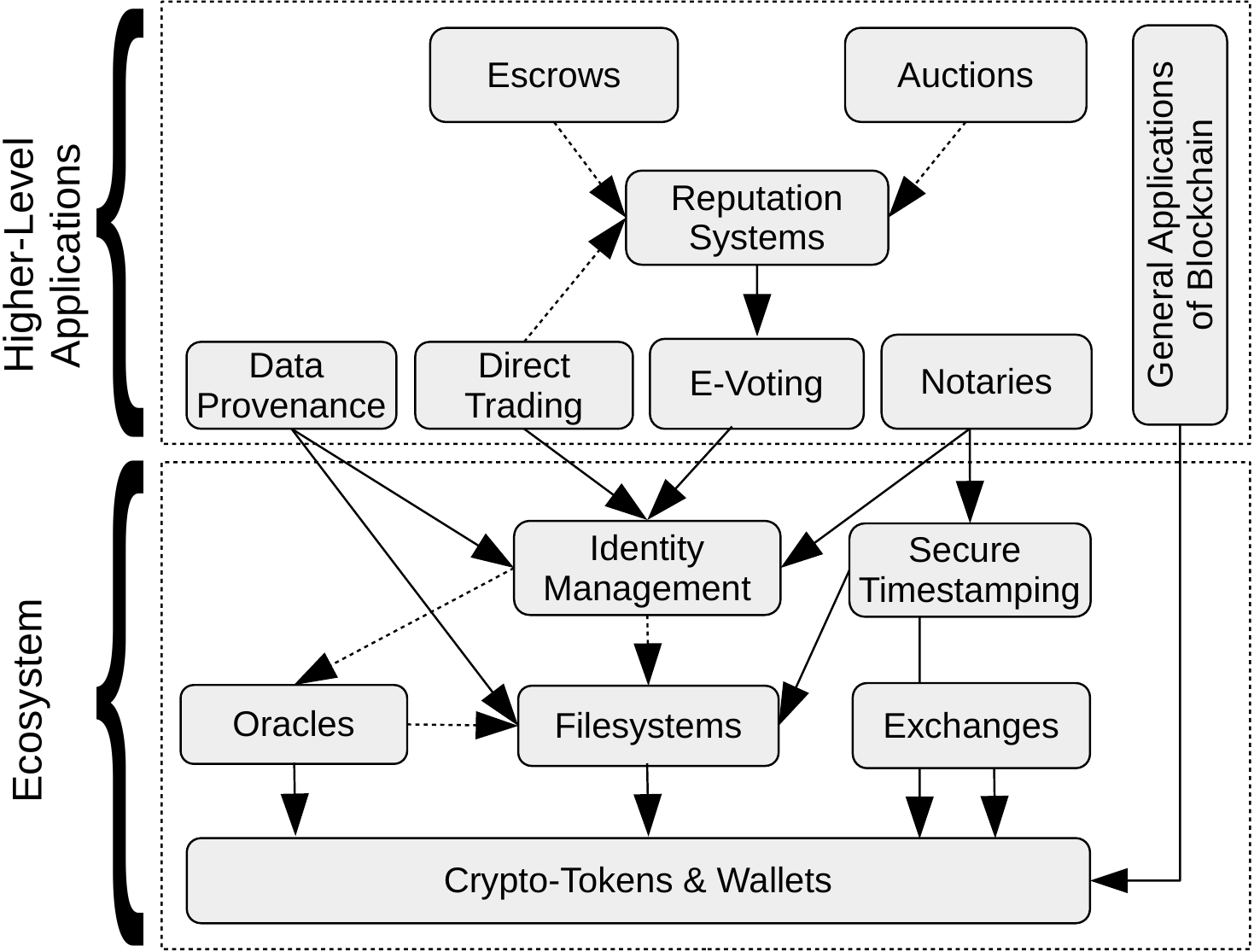}
	\caption{Hierarchy in inheritance of security aspects across categories of the application layer. Dotted arrows represent application-specific and optional dependencies.}\label{fig:dependencies}   
\end{figure}

We present a functionality-oriented categorization of the applications running on or utilizing the blockchain in \autoref{fig:dependencies}, where we depict hierarchy in the inheritance of security aspects among particular categories.
In this categorization, we divide the applications into categories according to the main functionality/goal that is to be achieved by using the blockchain.
Security threats of this layer are mostly specific to particular types of applications.
Nevertheless, there are a few application-level categories that are often utilized by other higher-level applications. 
In the current section, we isolate such categories into a dedicated application-level group denoted as an \textit{ecosystem}, while we describe the rest of the applications in \autoref{sec:apps-applications}.
The group of ecosystem applications contains five categories: 
(1) \textbf{crypto-tokens and wallets}, (2) \textbf{exchanges}, (3) \textbf{oracles}, (4) \textbf{filesystems}, (5) \textbf{identity management}, and (6) \textbf{secure-timestamping}.
We accompany the application layer with several incidents in \autoref{tab:incidents-app} of Appendix.

\subsection{Crypto-Tokens \& Wallets}\label{sec:wallets}
Besides blockchains that provide cryptocurrencies with native crypto-tokens, there are blockchain applications that use crypto-tokens to provide owners with rights against a third party (i.e., counter-party tokens) or with the possibility of transferring asset ownership (i.e.,  ownership/colored tokens)~\cite{BCPframework}. 
All types of tokens require the protection of private keys and secrets linked with user accounts.
For this purpose, two main categories of wallets have emerged: \textit{self-sovereign wallets} (a.k.a., non-custodial) and \textit{hosted wallets}~\cite{eskandari2018first,2015-Bitcoin-SOK}.
All crypto-tokens are exposed to  technical and regulatory risks, while non-native tokens are also exposed to legal risks~\cite{BCPframework}.

\paragraph{Self-Sovereign Wallets}
Users of self-sovereign wallets locally store their private keys and directly interact with the blockchain platform using these keys, while they verify the inclusion of their transactions by SPV client software. 
The instances of these wallets differ in several points. 
One of them is the manner in which the keys are isolated -- there are software wallets that store the keys within the user PC (e.g., Bitcoin Core,\footnote{\url{https://bitcoin.org/en/download}} MyEtherWallet\footnote{\url{https://www.myetherwallet.com/}}) 
as well as hardware wallets that store keys in sealed storage, while they expose only signing functionality (e.g., Trezor,\footnote{\url{https://trezor.io/}}  Le\-dger\footnote{\url{https://www.ledger.com/products/ledger-nano-s}}).
The next type of wallets enables functionality and security customization through a smart contract (e.g., TrezorMultisig2of3,\footnote{\url{https://github.com/unchained-capital/ethereum-multisig}} Ethereum MultiSigWallet,\footnote{\url{https://github.com/ConsenSys/MultiSigWallet}} SmartOTPs~\cite{homoliak2018air}).

\begin{figure*}[th]
	\begin{center}		
		\includegraphics[width=0.65\textwidth]{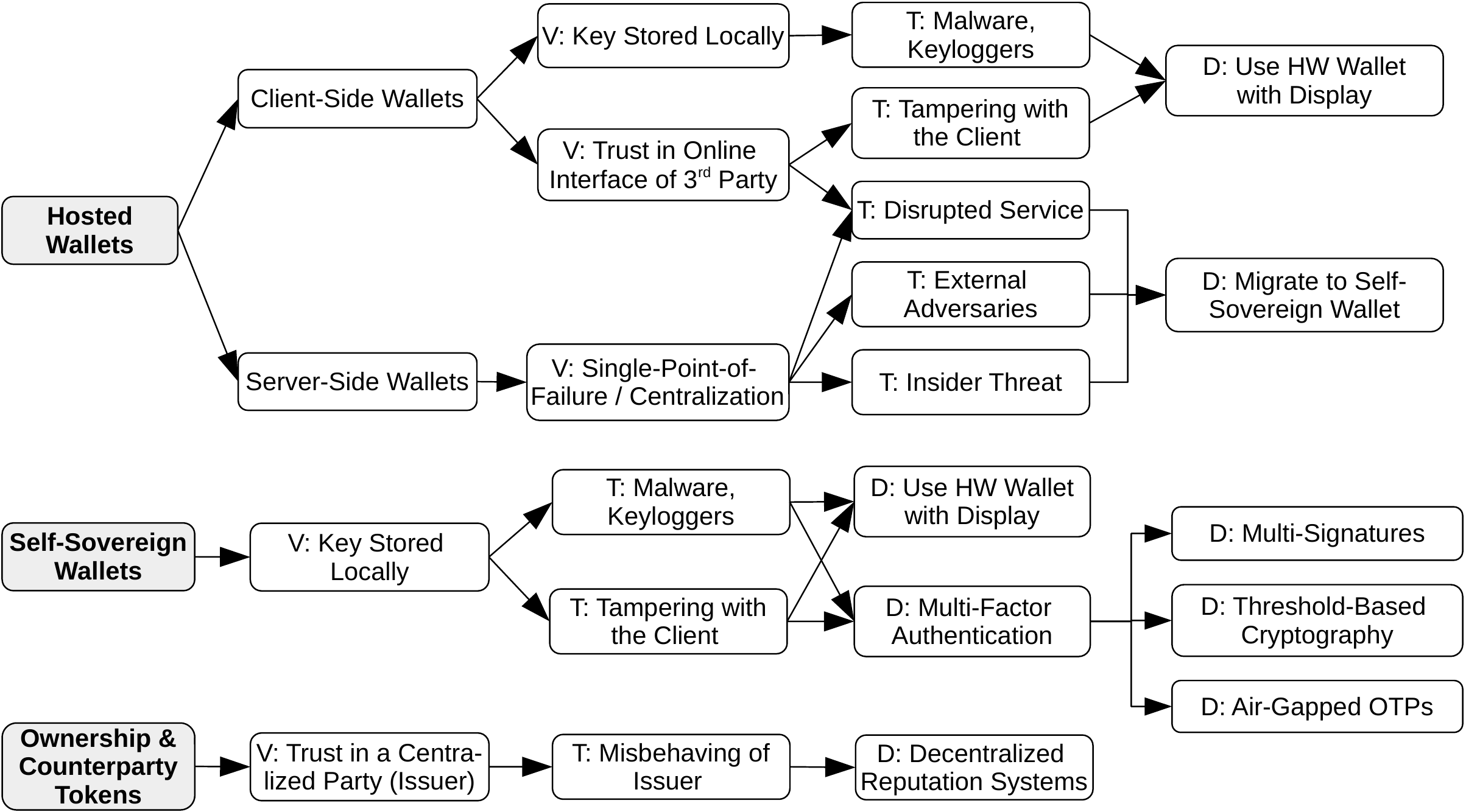} 		
		\caption{Vulnerabilities, threats, and defenses of the crypto-token $\&$ wallets category (application layer).}
		\label{fig:attacks-APP-wallets}
		\vspace{-0.4cm}
	\end{center}	
\end{figure*}

\paragraph{Hosted Wallets}
Hosted wallets require a centralized party, which provides an interface for interaction with the wallet and the blockchain.
If a hosted wallet has full control over private keys, it is
referred to as a \textit{server-side wallet} (e.g., Coinbase\footnote{\url{https://www.coinbase.com/}}), 
while in the case when keys are stored in the user's browser, a wallet is referred to as a \textit{client-side wallet} (e.g., Blockchain Wallet\footnote{\url{https://blockchain.info/wallet/}}). 
We refer the reader to works~\cite{eskandari2018first,homoliak2018air} for a security overview of miscellaneous wallet solutions. 

\mysubsubsection{Security Threats and Mitigations}
We present a taxonomy of vulnerabilities, threats, and defenses related to the crypto-token wallets category in \autoref{fig:attacks-APP-wallets}.
Server-side wallets pose a single-point-of-failure, which can be exploited  by external or internal adversaries, and moreover, it can be subjected to availability attacks such as DoS.
Since server-side wallets have been the target in several security incidents~\cite{2018-coindesk-bithumb}, \cite{2014-Mt-Gox}, \cite{2016-Bitfinex-hack}, their popularity has declined in favor of client-side wallets.
Client-side wallets do not expose private keys to a centralized party but store it locally. 
Nevertheless, they still trust in the online interface provided by such a party, and thus their availability is dependent on this party.
Other threats with client-side wallets are client tampering and malware/keyloggers, which focus on deceiving the user while signing a transaction or stealing the key.
Possible mitigations of these attacks include hardware wallets that display details of transactions to the user, while the user confirms signing by a button (e.g., Trezor,  Le\-dger).

In contrast, self-sovereign wallets do not trust in a third party nor rely on its availability.
However, these wallets are susceptible to key theft (i.e., malware~\cite{2015-CCSM-SecureWorks}, keyloggers~\cite{2015-Bitcoin-SOK,2017-keylogger-bc-malware}).
One protection is to use a hardware wallet with a display as described above.
Another option is to protect self-sovereign wallets by multi-factor/(-step) authentication using multi-signatures (e.g., TrezorMultisig2of3, Ethereum MultiSigWallet), threshold-based cryptography~\cite{goldfeder2015securing}, or air-gapped OTPs~\cite{homoliak2018air}.
In the case of counter-party and ownership tokens presented at the application layer of existing public blockchains, we emphasize the additional vulnerability caused by trusting in the centralized party that issues such tokens -- the provided counter-party and ownership rights are only virtual, which imposes a significant risk.
While this risk cannot be eliminated in this application scenario,  a possible mitigation technique for preventing fraudulent issuers is to use decentralized reputation-based systems (see \autoref{sec:reputation-systems}) and notaries (see \autoref{sec:notaries}) that might be built on top of them.

\mysubsubsection{Side Effects and Implications of Countermeasures}
A disadvantage of some multi-factor authentication solutions is that they require the execution of smart contracts, which increases the costs and might slow down the throughput of the system.
In contrast, threshold-based cryptography constructs save these costs since they produce only a single signature, which appears as it was made by a single party.
However, threshold-based cryptography requires off-chain computation, in which the duration of execution is dependent on the number of co-signing parties.

Although self-sovereign wallets provide higher security in contrast to client-side and server-side hosted wallets, they impose overheads for storing the non-negligible part of the blockchain (i.e., headers) to validate the inclusion of signed transactions within their SPV client software -- this is especially a concern when users have multiple SPV clients for multiple blockchains.
BTC Relay is an early optimization attempt that reduces storage requirement by outsourcing the validation of transactions from the source blockchain to the target blockchain in exchange for a small fee paid to the relay nodes that submit headers to the target blockchain. 
However, the costs of BTC Relay were too high and its users tend to rather store headers of source blockchain on their own.
Zk-relay~\cite{westerkamp2020zkrelay} optimizes these costs by using zk-SNARKs with batched block validation, achieving an improvement by a factor of 187.

\subsection{Exchanges}\label{sec:exchanges}
If the user wishes to exchange crypto-tokens, she might either directly find a counter-party wishing to exchange the opposite pair or approach an exchange that might be centralized or decentralized (DEX).
In the case of centralized exchanges, the security threats and implications are due to centralization, and the only countermeasure is to use decentralized exchange solutions that we further focus on.

\paragraph{Direct Cross-Chain Exchange with Atomic Swap}
Atomic swaps\footnote{\url{https://en.bitcoinwiki.org/wiki/Atomic_Swap}} assume two parties owning crypto-tokens in two different blockchains, and these parties wish to execute exchange atomically, i.e., either both of the parties receive the agreed amount or neither of them. 
The atomic swap protocol enables conditional redemption of the funds in the first blockchain upon revealing the hash pre-image (i.e., secret) that redeems the funds on the second blockchain.
This protocol is based on two Hashed Time-Lock Contracts (HTLC) that are deployed by both parties in both blockchains, and it requires 4 transactions (see details in Appendix~\ref{appendix:atomic-swaps}).

\paragraph{Cross-Chain DEX}
Although atomic swaps are, in theory, sufficient means for the execution of fair cross-chain exchange, the situation is more complicated in practice.
In particular, there might not exist a contra-party exchanging the opposite pair or the user might not be aware of it.
This motivates DEXes, which facilitate the process of maintaining and matching the existing orders, act as a contra-party or intermediary, while guaranteeing fairness (e.g., Komodo\footnote{\url{https://komodoplatform.com/atomic-swap-technology/}}). 
The users match the orders, reward DEX, and afterward perform an atomic swap on their own.
If users wish to trade more obscure crypto-tokens, for which there is no matching counter-order, DEX may serve as a counter-party and do the atomic swap with the user.
Moreover, if the users wish to trade colored tokens for native crypto-tokens of different blockchains (e.g., $\mathbb{A}$ sells an asset for BTC and $\mathbb{B}$ buys it for ETH), 
DEX might serve as an intermediary who executes the three-way atomic swap~\cite{herlihy2018atomic} (see details in Appendix \ref{appendix:atomic-swaps}).

\begin{figure*}[ht]
	\begin{center}
		\vspace{-0.4cm}			
		\includegraphics[width=0.70\textwidth]{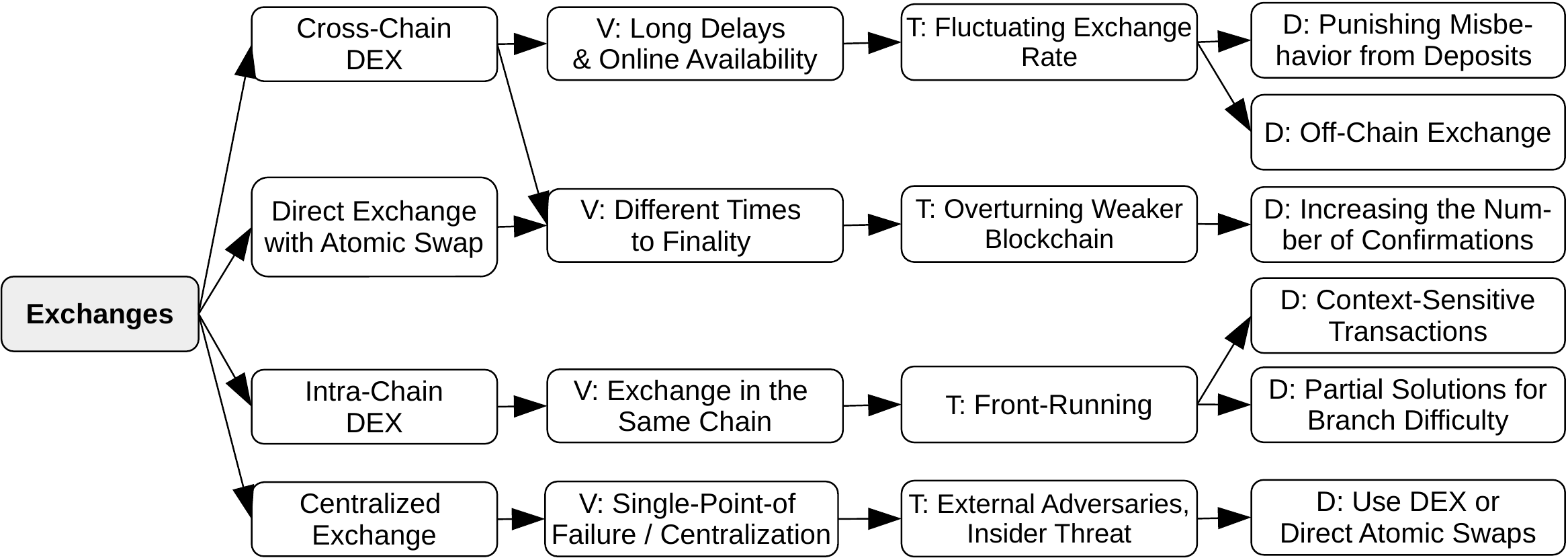} 		
		\caption{Vulnerabilities, threats, and defenses of the exchanges category (application layer).}
		\label{fig:attacks-APP-exchanges}
	\end{center}	
	\vspace{-0.4cm}
\end{figure*}

\paragraph{Intra-Chain DEX}
Some intra-chain DEX designs (e.g., Maker Market, EtherOpt, and Intrinsically Tradeable Tokens) require parties to post buy\&sell offers on the blockchain, while smart contracts perform matches and execute trades. 
However, each placing of an order or its modification requires a payment for the inclusion of a transaction. 
Therefore, designs with off-chain order matching became more popular; in these designs, only trades are executed on-chain, while orders and their matching is performed off-chain.
An example is 0x~\cite{warren20170x} protocol handling DEX of ERC20 tokens (e.g., applied in EtherDelta\footnote{\url{https://etherdelta.com/}}). 
The next intra-chain exchange design is known as the \textit{automated market maker} (AAM)~\cite{berg2009hanson}.
AAM is applicable within a smart contract-based DEX that contains deposited reserves of traded ERC20 tokens; examples are Euler~\cite{euler-exchange}, Bancor~\cite{bancor-exchange}, and Uniswap.\footnote{\url{https://uniswap.io/}}
AAM provides high liquidity since users are not required to match their orders, and they can directly do the exchange with the smart contract.

\paragraph{Cross-Chain Communication}
The concept of cross-chain exchange can be further generalized into cross-chain communication (CCC), which deals with the interoperability of applications running on different blockchains. 
The security aspects of CCC are very similar to the exchanges, and we refer the interested reader to the work of Zamyatin et al.~\cite{zamyatinsok}.

\mysubsubsection{Security Threats and Mitigations}
We present an overview of vulnerabilities, threats, and defenses related to the exchanges category in \autoref{fig:attacks-APP-exchanges}.

In \textbf{centralized exchanges}, threats are caused by external and internal adversaries, and they are identical to those of server-side hosted wallets (see \autoref{sec:wallets}) since server-side wallets always provide exchange services.
So far centralized exchanges posed the most attractive target for adversaries that have caused huge financial losses~\cite{Oosthoek-centralized-exch}.
There are many operation security (OPSEC) countermeasures such as multi-factor authentication, split of the funds to hot and cold wallets, however, none of them eliminates the single-point-of-failure coming from the centralization.
Therefore, effective mitigation from the user point of view is to use decentralized exchange solutions such as DEXes and atomic swaps; however, they also contain some vulnerabilities.

Different blockchains of \textbf{cross-chain} decentralized (direct and DEX-based) exchanges might have a different time to finality, and thus the likelihood that one blockchain will be overturned is higher than in the case of the other one. 
Therefore, the number of required confirmations might be agreed upon by involved parties beforehand. 
However, this results in longer delays for the execution of the protocol and the need for both parties to be online.
In some cases, such long delays might cause fluctuation in the exchange rate, making the exchange not attractive at a later time.
As a mitigation technique, off-chain exchanges (within side-chains) might be used, where each blockchain is updated only with the final transaction.
Off-chain real-time exchanges might also be achieved with the use of TEE~\cite{bentov2017tesseract} (see below). 
Another mitigation for intentional delaying of the exchange by any party is to use deposit-based bonds, which will be restored only when a particular party acts timely.

Since \textbf{intra-chain} exchanges are executed in the single blockchain, they give rise to the transaction front-running, in which the adversary (a.k.a, arbitrage bots) front-run user trades by transactions containing higher fees and by optimizing network latency~\cite{daian2019flash}.
Moreover, such adversaries might compete with each other by ``bidding'' a higher transaction fee~\cite{daian2019flash}, which in turn targets the ordering mechanism of the consensus layer, where miners might tend to overturn or fork the blockchain while including only transactions with the highest fees.
A mitigation technique to these threats is represented by context-sensitive transactions~\cite{gavzi2018stake}, which do not allow overturning of the blockchain, only its extension.
The same effect can be achieved by partial solutions included in branch difficulty computation~\cite{strongchain,zamyatinflux, rizun2016subchains,pass2017fruitchains}.
Note that context-sensitive transactions and partial solutions are a means of the consensus layer.

\mysubsubsection{Side Effects and Implications of Countermeasures}
We assume centralized exchanges as the baseline that requires only two transactions for settlement of cross-chain exchange or one transaction for intra-chain exchange.
Note that such settlement is done only sporadically since centralized exchanges inherently work off-chain in real-time.
In contrast, direct atomic swaps require 4 transactions while three-way atomic swaps require 6 transactions.
Therefore, using these constructs negatively affects the throughput of blockchains. 
Nevertheless, the performance can be improved by off-chain execution of atomic swaps, which provides almost immediate response, e.g., off-chain swaps are possible in two-parties payment channels as well as their extension to multiple parties known as the lightning network~\cite{poon2016bitcoin}. 
Such off-chain solutions greatly improve the scalability, since parties can
transact directly and involve blockchain consensus only when they wish to settle their balances (e.g., once per day or week). 
However, to avoid misbehavior in which a ``stale'' balance is settled on-chain, these systems require that the parties constantly monitor the blockchain state. 
Such an always-online assumption can be relaxed by employing watching
services~\cite{dcwc18,mccorry2018pisa,liu2020fail} (a.k.a., watchtowers), which, however, incur extra costs.

TEE can be also utilized as an off-chain means that improves the throughput of exchange service.
For example, Tesseract~\cite{bentov2017tesseract} is a real-time exchange that leverages TEE for communication with users and a target blockchain.
To enter the system, users submit time-locked refill transactions paying to Tesseract's controlled address in the target blockchain, and then Tesseract uses its SPV client to verify their inclusion. 
Existing users submit bid\&sell requests to Tesseract, which performs matching and executes trades within TEE.
When users want to sync with the on-chain state, they ask Tesseract to generate settlement transactions.
Since decentralization would be impacted by using a single service of Tesseract, which might censor user requests, the authors incorporate the Paxos consensus protocol among multiple mutually untrusted Tesseract nodes; this also increases the fault-tolerance and avoid funds of users becoming stuck in contrast to one instance of Tesseract.
We note that censorship evidence (not resistance) could be alternatively ensured in Tesseract by smart contract-based censorship resolution~\cite{szalachowski2018blockchain}, which, however, implies some extra costs for smart contract execution.
Finally, TEE-based solutions might be vulnerable to attacks on trusted hardware. 
As a mitigation technique to reliance on a trusted manufacturer of TEE, it is possible to use a quorum of several redundant TEEs from multiple manufacturers.
Furthermore, it is important to note that the assumption about the code executed in TEE is its bug-freeness, and thus one might not use return-oriented programming or other techniques to ex-filtrate sealed secrets or private parameters; this is an out-of-the-scope attacker model for TEE.

\subsection{Oracles}
\label{sec:oracles}
Oracles (a.k.a., \textit{data feeds}) are trusted entities that provide plausible data that reflects the state of the world beyond the blockchain.
The authors of~\cite{GuarnizoS19} and~\cite{ellisdecentralized}  define a few security properties of oracles in smart contract platforms: 
\begin{figure*}[t]
	\begin{center}		
		\includegraphics[width=0.72\textwidth]{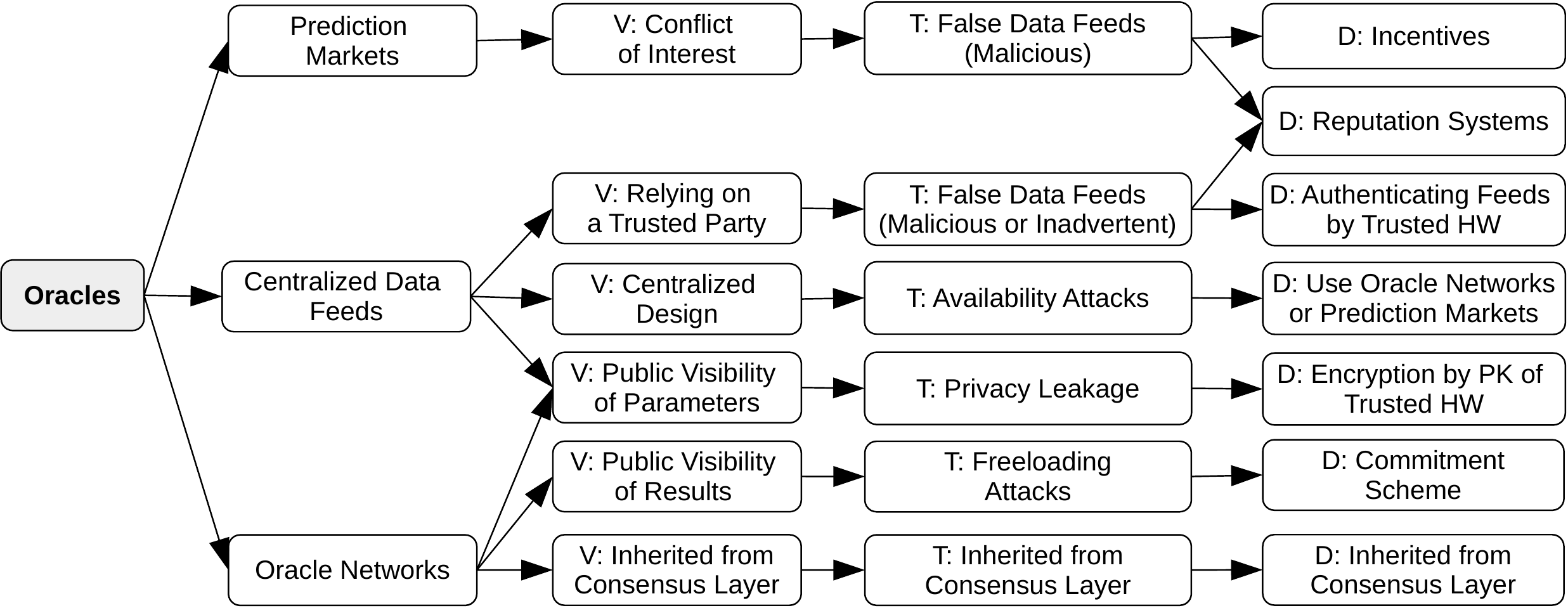} 		
		\caption{Vulnerabilities, threats, and defenses of the oracles category (application layer).}
		\label{fig:attacks-APP-oracles}
		\vspace{-0.4cm}
	\end{center}	
\end{figure*}
\begin{compactitem}
	\item[\textbf{Authenticity:}] Data are authentic if they are produced by content providers agreed by the consumers of the data.
	
	\item[\textbf{Integrity:}] Provided data should not be modified nor deleted after creation. 
	Therefore, content providers should guarantee the correctness of the newly created data and publicly prove their consistency with the past.	
	
	\item[\textbf{Confidentiality:}] Sometimes, input parameters may contain confidential or private data. 
	Therefore, an oracle should support such parameters and their handling.
	
	\item[\textbf{Availability:}] Since the execution of dependent smart contracts relies on data feeds delivered by oracles, they need to provide high availability. 
	
\end{compactitem}

\smallskip\noindent
We categorize existing oracles according to security implications into three categories.
\textit{Prediction markets} (e.g., Augur~\cite{peterson2015augur}, Gnosis\footnote{\url{https://gnosis-pm-contracts.readthedocs.io/en/latest/}}) were created for the purpose of trading the outcome of events -- individuals are incentivized to accurately wager on outcomes serving as data feeds, while outcomes are provided by either a centralized reporter or a quorum of reporters.
\textit{Centralized data feeds} provide arbitrary data from a single centralized source, and they build on existing blockchain platforms (e.g., Oraclize,\footnote{https://www.oraclize.it/} Town Crier~\cite{zhang2016town}, PDFS~\cite{GuarnizoS19}).
Finally, \textit{oracle networks} internally run a consensus protocol for decentralized agreement on data (e.g., ChainLink~\cite{ellisdecentralized}, Witnet~\cite{de2017witnet}).

\paragraph{Prediction Markets}
Augur~\cite{peterson2015augur} is a solution designed as the Ethereum smart contract, and it uses its own reputation token.
The user creating the market specifies a designated reporter, who delivers the result of the event after it happens.
However, the reporter might not report the result or report incorrect results.
When the reporter does not report within the specified time frame, Augur shifts the role of a designated reporter on a first-come-first-serve basis.
After reporting the outcome, the Augur users have a specific time frame to run a decentralized dispute resolution, and thus obtain a different outcome of the event in the case of misreporting.
Augur incentivizes its users to report correct outcomes and file disputes only in justified cases by rewards and deposit-based bonds.
Another example of prediction markets is Apollo from Gnosis, which had originally a centralized data source facilitated by Gnosis but was replaced by a decentralized one in version 2.0. 

\paragraph{Centralized Data Feeds}
Oraclize enriches the data provided to smart contracts by authenticity proofs  that are built upon various technologies such as auditable virtual machines 
and trusted computing.
Since authenticity proofs can be large, Oraclize can store these proofs in the distributed file system IPFS\footnote{\url{https://ipfs.io/}} instead of directly providing them to the smart contracts.
Town Crier~\cite{zhang2016town} is an approach that provides authenticated data feeds to smart contracts by bridging them with public webs through a TEE component. 
A linkage of TEE with a smart contract is made by storing a public key (PK) of TEE at the smart contract of Town Crier. 
It relies on the X.509 public key infrastructure (PKI), due to which, the provided data are provably authenticated.
PDFS~\cite{GuarnizoS19} allows content providers to link their web resources with corresponding smart contracts in the blockchain.
In PDFS, the data of content providers are managed in an auditable manner, enabling publicly-verifiable data transparency and consistency of data with the past.
Besides content providers, PDFS introduces two entities that arrange a smart-contract-based agreement by specifying a particular content provider required for the execution of the code in the agreement. 
To ensure that updates are consistent with the past, the authors apply a history tree data structure~\cite{crosby2009efficient}.
The authors additionally support the means to publicly prove censorship by a content provider.

\paragraph{Oracle Networks}
ChainLink~\cite{ellisdecentralized} builds on top of existing blockchains with support for smart contracts, and it distributes the provisioning of data feeds to multiple oracle providers.
In detail, ChainLink maintains oracle providers and their reputation, who are selected by a smart contract based on their reputation to form an aggregated final result.
When building the final result, ChainLink discards outliers and utilizes the BFT protocol to reach a consensus on a final aggregated value.
Witnet~\cite{de2017witnet} is an approach similar to ChainLink but in contrast to ChainLink, Witnet runs its own oracle network with a native token.
Witnesses (i.e., content providers) earn reputation points when the content that they deliver matches with the majority's content, and they lose reputation points otherwise.
The reputation points serve as a stake in the consensus protocol of the oracle network, hence the higher a node's reputation, the higher the chance that it produces a block.
Since more witnesses might become block producers, Witnet allows multiple chains in parallel, forming a DAG.

\mysubsubsection{Security Threats and Mitigations}
We present a taxonomy of vulnerabilities, threats, and defenses related to the oracle category in \autoref{fig:attacks-APP-oracles}.
In the following, we describe these threats as well as possible defense techniques.

\textbf{Prediction markets} may suffer from conflict-of-interest since the creator of the market specifies a data reporter, who might also participate in the market and later report false data for her convenience.
Since a prediction market might yield significant financial value for the users with a ``correct'' guess, the malicious reporter might bribe other reporters of dispute resolution round and still be profitable.
Therefore, the incentive protocols of prediction markets must count on this situation and incorporate feasible rewards for honest reporters.
Another mitigation for similar attacks is to keep reporters accountable and maintain their reputation in a decentralized fashion (\autoref{sec:reputation-systems}), which involves identity management and verification (see \autoref{sec:identity-management}).

\textbf{Centralized data feeds} rely on a trusted party~\cite{GuarnizoS19,oraclize} that may misbehave or accidentally produce wrong data.
For both cases, decentralized identity management can bring accountability while reputation systems further build on it, which disincentivizes malicious behaviors. 
Another option to cope with possible misbehavior of a trusted party of oracle service is to embed the logic of oracle service into a TEE component~\cite{zhang2016town}, whose code is publicly attested.
TEE component can interact with the external world using X.509 public key infrastructure (PKI), due to which obtained data are provably authenticated.
Since some requests of data feeds might contain private parameters,\footnote{All transactions of permissionless blockchains are visible to public.} they can be encrypted by a PK of the TEE and further processed within TEE that communicates with its remote data provider through an encrypted channel, while  communication is facilitated by oracle's service operator, as demonstrated in Town Crier~\cite{zhang2016town}.
Centralized data feeds might be subject to attacks on availability, leading to interrupted service.
A mitigation technique is to use solutions with a higher redundancy, such as oracle networks.

\textbf{Oracle networks} eliminate trust in a single party by running a consensus protocol either natively~\cite{de2017witnet} or utilizing an existing smart contract platform to facilitate the service and its consensus~\cite{ellisdecentralized}. 
Running the native consensus protocol of an oracle network imposes the security threats related to the consensus layer (see \autoref{sec:consensus}).
Specific threats to oracle networks are freeloading attacks, in which an oracle provider might copy a publicly visible value provided by other oracles without any effort. 
The authors of ChainLink~\cite{ellisdecentralized} propose the usage of a commitment scheme to cope with this attack.

\mysubsubsection{Side Effects and Implications of Countermeasures}
The data provision time of prediction markets may be too long for many applications, and they are convenient to use only for specific use cases that are limited to provided data events.
The data provision time is prolonged even more in the case of disputes, whose resolution may require several days or weeks.
In contrast to limited data of prediction markets, centralized data feeds enrich the data domain and significantly shorten the provisioning time.

In the case of oracle providers that offer authenticated data feeds using trusted hardware~\cite{oraclize,zhang2016town}, a vulnerability in trusted hardware (caused by a manufacturer) may result in the entire data feed being compromised.
As a mitigation technique to reliance on a trusted manufacturer of TEE, it is possible to use a quorum of several redundant TEEs from multiple manufacturers.

Since ChainLink~\cite{ellisdecentralized} is an oracle network running over the public smart contract platform, it imposes significant execution costs.
To reduce the on-chain cost of BFT execution, the authors discuss the use of threshold-based signatures for collective off-chain signing of the final value; however, freeloading attacks remain unresolved. 
When freeloading attacks are resolved by a commitment scheme, it negatively impacts the provisioning delay and costs for data providers, who have to submit another transaction with a commitment.

\begin{figure*}[th]
	\begin{center}		
		\vspace{-0.4cm}
		\includegraphics[width=0.72\textwidth]{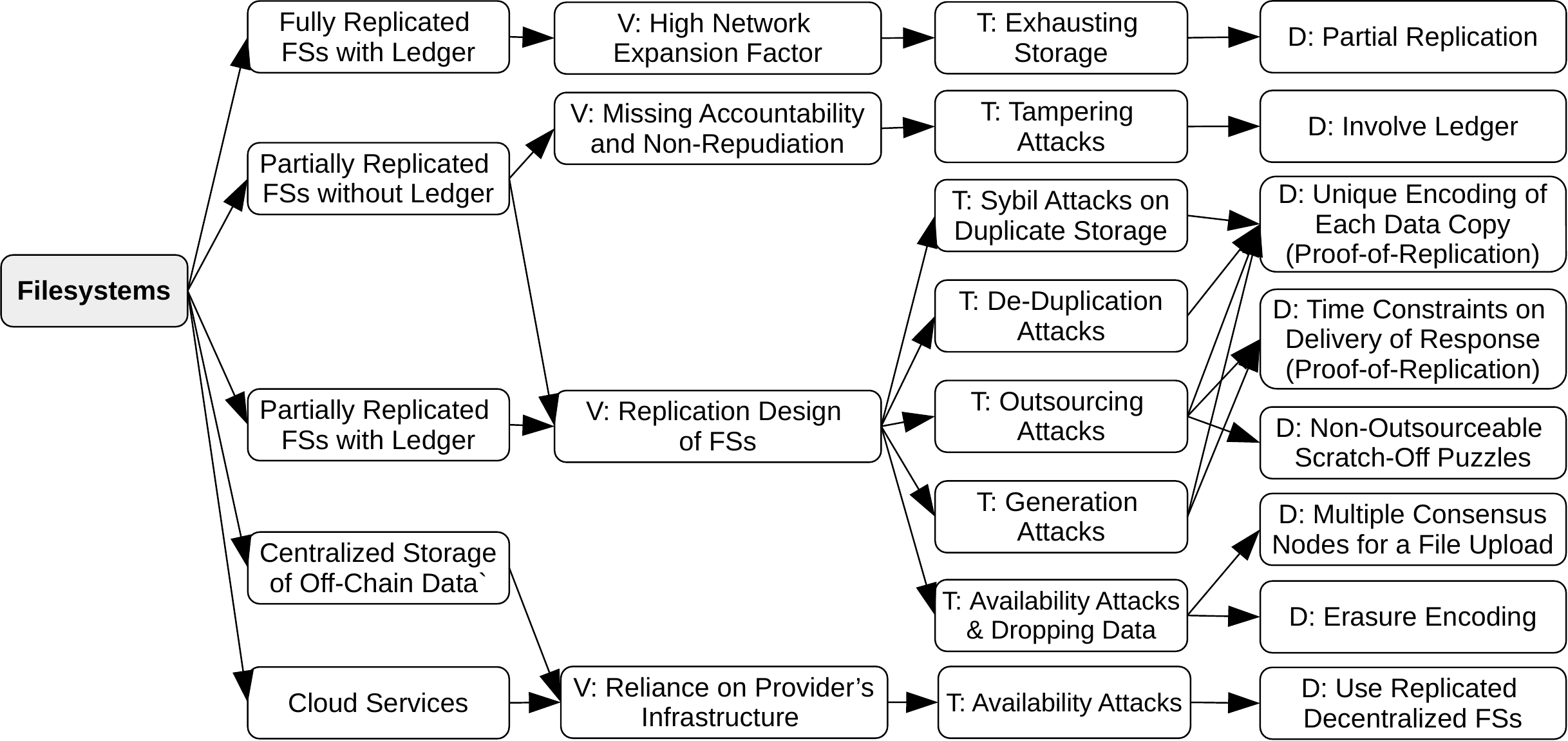} 		
		\caption{Vulnerabilities, threats, and defenses of the filesystems category (application layer).}
		\label{fig:attacks-APP-DFs}
	\end{center}	
\end{figure*}

\subsection{Filesystems}
\label{sec:filesystems}
Filesystems (FS) serve as a distributed data storage infrastructure that borrows ideas from peer-to-peer storage systems, while additionally incentivizing data preservation by tokens. 

\paragraph{Fully Replicated FS with Ledger}
A naive approach is to store the full content of data at the blockchain, and thus achieve full data replication, extremely high \textit{data durability} (i.e., availability of data) as well as \textit{network expansion factor} (i.e., storage overhead). 
An example is storing data using the instruction $OP\_RETURN$ in Bitcoin or storing data as key:value pairs within Namecoin.\footnote{\url{https://bit.namecoin.org/}} 
However, such an approach results in high storage overheads required for full replication of the data among the consensus nodes.

\paragraph{Partially Replicated FS with Ledger}
To decrease the costs while preserving reasonable durability, partial replication of the data with erasure encoding is often used (e.g., Permacoin~\cite{miller2014permacoin}, Storj~\cite{wilkinson2014storj}, and KopperCoin~\cite{kopp2016koppercoin}).
In erasure encoding, the data block is encoded using two numbers $(k, n)$, where $n$ represents the number of total erasure shares and $k$ represents the minimum number of shares required for data recovery. 
Permacoin incorporates Proof-of-Retrievability in the consensus layer, where consensus nodes store large segments of data provided by an authoritative file dealer. 
KopperCoin follows a similar approach, but it does not need the trusted dealer for the initial distribution of data files since files are uploaded by the users.
Storj uses a 3rd party distributed ledger for storage of metadata, such as file hash, network locations of copies, and Merkle roots of data. 
Permacoin, KopperCoin, and Storj enable probabilistic audits using Proof-of-Retrievability, which proves that a node stores certain data at the time of the challenge.
Filecoin is an incentive mechanism of any distributed FS (e.g., IPFS), and in contrast to previous works, it can guarantee the data possession over a certain time range in a setting of Proof-of-Spacetime.
Moreover, Filecoin uses Proof-of-Replication~\cite{benet2017proof}, which guarantees physically unique copies of data for each node.

\paragraph{Partially Replicated FS without Ledger}
IPFS and Swarm\footnote{\url{https://swarm-guide.readthedocs.io/en/latest/}} utilize the concept of distributed hash tables (DHT). 
DHT provides a decentralized data lookup service with key:data mappings, in which the set of nodes storing the data is unambiguously determined by the key  associated with the data (i.e., its hash).
Since the lookup service and data storage are (partially) distributed, a change in the set of participants causes only a negligible disruption of availability. 
IPFS does not contain any incentive mechanism and the availability of the data is dependent on its popularity. 
Although IPFS does not involve a blockchain, it may achieve its properties.
In particular, nodes may optionally store a BitSwap ledger that logs data transfers with other nodes.

\paragraph{Centralized Storage of Off-Chain Data}
Alternatively to decentralized filesystems, decoupling of the data from the blockchain itself through the storage of on-chain integrity proofs (see \autoref{sec:secure-timestamping}) and off-chain data is also an option; however, it introduces a single-point-of-failure and thus may not provide sufficient availability.
Besides centralized storage of data, cloud services are promising approaches for decentralized yet manageable data storage, for which integrity and consistency proofs are stored on some blockchain.

\mysubsubsection{Security Threats and Mitigations}
We present a taxonomy of vulnerabilities, threats, and defenses related to the filesystems category in \autoref{fig:attacks-APP-DFs}.
While fully replicated and partially replicated decentralized filesystems handle \textbf{availability} using decentralized infrastructure, centralized storage with integrity proofs and cloud services relies on a centralized provider.
In a \textbf{Sybil attack} on a replicated FS, a malicious node claims the storage of multiple copies of the same data.
Similarly, in a \textbf{de-duplication} attack, more consensus nodes may collude to claim that each of them is storing an independent copy of the data, while only one of the nodes stores the data.
These attacks can be prevented by a unique encoding of each data copy proposed in Proof-of-Replication~\cite{benet2017proof}.
In an \textbf{outsourcing attack}, a malicious consensus node claims the storage of more data than it can physically store while relying on data retrieval from outsourced data providers.
In a \textbf{generation attack}, a malicious node can re-generate the previously uploaded data upon request using some algorithm, which may increase its chances to be rewarded: a node might commit to storing of a huge volume of generated data.

On top of the unique encoding of each data copy, a mitigation technique for outsourcing and generation attacks is to put time constraints on the delivery of the response by a prover, as proposed in  Proof-of-Replication~\cite{benet2017proof}.
In detail, the function used for the encoding of data replicas must not be parallelizable (e.g., symmetric encryption in CBC mode) to mitigate generation attacks.
In the case of outsourcing attacks, the time constraints must distinguish whether cloud access was made or not.
Similar mitigation for outsourcing attacks is the use of non-outsourceable scratch-off puzzles~\cite{miller2014permacoin,miller2015nonoutsourceable}, in which the computation of the puzzle requires access to the storage in random order; hence, many round-trips are incurred during one attempt of solving a puzzle.
Another attack might target the reputation of a network by dropping data and its redundant copies.
A simple mitigation technique is to use multiple consensus nodes for a file upload, which diminishes the chance of the attack being successful.
Another mitigation is to increase the durability by erasure encoding.

\mysubsubsection{Side Effects and Implications of Countermeasures}
Although unique encoding of each data copy thwarts several attacks, on the other hand, it imposes higher overheads for file distribution on clients, which might negatively influence the throughput of data.
Similarly, additional overheads on the client during the file upload is imposed by the use of multiple consensus nodes for file upload.
Although erasure encoding aims to decrease costs, it must be viewed in a trade-off with the availability of data that is negatively affected by it.

\subsection{Identity Management}\label{sec:identity-management}
Identity management refers to binding identities of entities to their public keys.
This concept is also referred to as Public Key Infrastructure (PKI), and it has
a few security goals~\cite{fromknecht2014decentralized}: 
\begin{compactitem}
	\item[\textbf{Accurate Registration:}] The user must be unable to register an identity that she does not own.
	\item[\textbf{Identity Retention:}] The user must be unable to impersonate an identity already registered.
	\item[\textbf{Censorship Resistance:}] The user must be able to register any identity that she owns.	
\end{compactitem}
In computer science, some have conjectured that it is highly unlikely to design an identity management system in which identifiers would be selected in a \textit{distributed} fashion while remaining \textit{secure} and \textit{human-readable}.
These three properties are often referred to as Zooko's triangle~\cite{wilcox2003names}.
However, this situation has been changed with the invention of blockchains; in particular, their immutability feature (Appendix \ref{sec:background-features}).

Namecoin\footnote{\url{https://www.namecoin.org/}} is a native blockchain  that facilitates identity management since it allows for the unique registration of key:value mappings.
However, searching for a value associated with a key requires full storage and traversal of the blockchain, which is costly.
Blockstack~\cite{ali2016bootstrapping} is a similar approach as Namecoin, providing decentralized DNS, but in contrast to Namecoin, it off-chains the data storage of domain name mappings and keeps only references to hashes of zone-files in its blockchain.
Zone-files (and their referred DNS entries) are stored off-chain. 
Certcoin~\cite{fromknecht2014decentralized} is built on top of the Namecoin blockchain, where entities publish their public keys (PK) by posting an identity-PK pair to the Namecoin blockchain.
Certcoin utilizes cryptographic accumulators, which represent a space-efficient data structure that supports the verification of set membership within the $\{ID, PK\}$ domain, imposing only a logarithmic time complexity in the number of registered users (in contrast to linear time complexity of Namecoin).
Furthermore, to speed up the PK lookup queries, Certcoin leverages the concept of DHT (see \autoref{sec:filesystems}), which enables it to achieve a constant lookup time complexity. 
Ethereum Name Service\footnote{\url{https://ens.domains/}} (ENS) maps human-readable domain names to Ethereum addresses in a similar fashion as in DNS. 
The root domain of ENS is maintained by a multi-signature smart contract owned by trustworthy individuals from the Ethereum community. 
Similarly, uPort~\cite{Lundkvist2016} utilizes smart contracts to keep a registry that maintains a mapping of the user addresses to hashes of claims\footnote{Claims as such belong to the notaries category (see \autoref{sec:notaries}).} that are stored off-chain.
In contrast to ENS, uPort does not provide human-readable user identifiers and discusses the possibility of using ENS as a naming layer.
Smart contracts for managing identities of humans, groups, objects, and machines are also used in the ERC 725 standard,\footnote{\url{https://erc725alliance.org/}} in which, identity is associated with several keys serving various purposes. 
ShoCard~\cite{Shocard2017} is another example that builds on top of existing public blockchains, but in contrast to the previous examples, it builds a sidechain containing encrypted identity-specific data such as biometric data, scans of IDs, etc.
The user may then decide to whom she will reveal the encrypted data.
The Sovrin~\cite{Tobin2016} is an example providing a public permissioned blockchain that consists of consensus nodes approved by Sovrin.
It focuses on high throughput and low operational costs. 
Decentralized Identifiers (DIDs)~\cite{did-w3c} represent a new type of universally unique identifiers whose control is decentralized since all roots of trust are contained in the blockchain and each entity might create its own root of trust.
DID employs the same hierarchical scheme for globally unique strings as URI, and it maps strings to DID documents containing data such as PKs, endpoints of the entity, or links to off-chain data.
Only the owner can create, manage, and prove ownership of her DID entries.

\begin{figure}[t]
	\begin{center}			
		\includegraphics[width=0.48\textwidth]{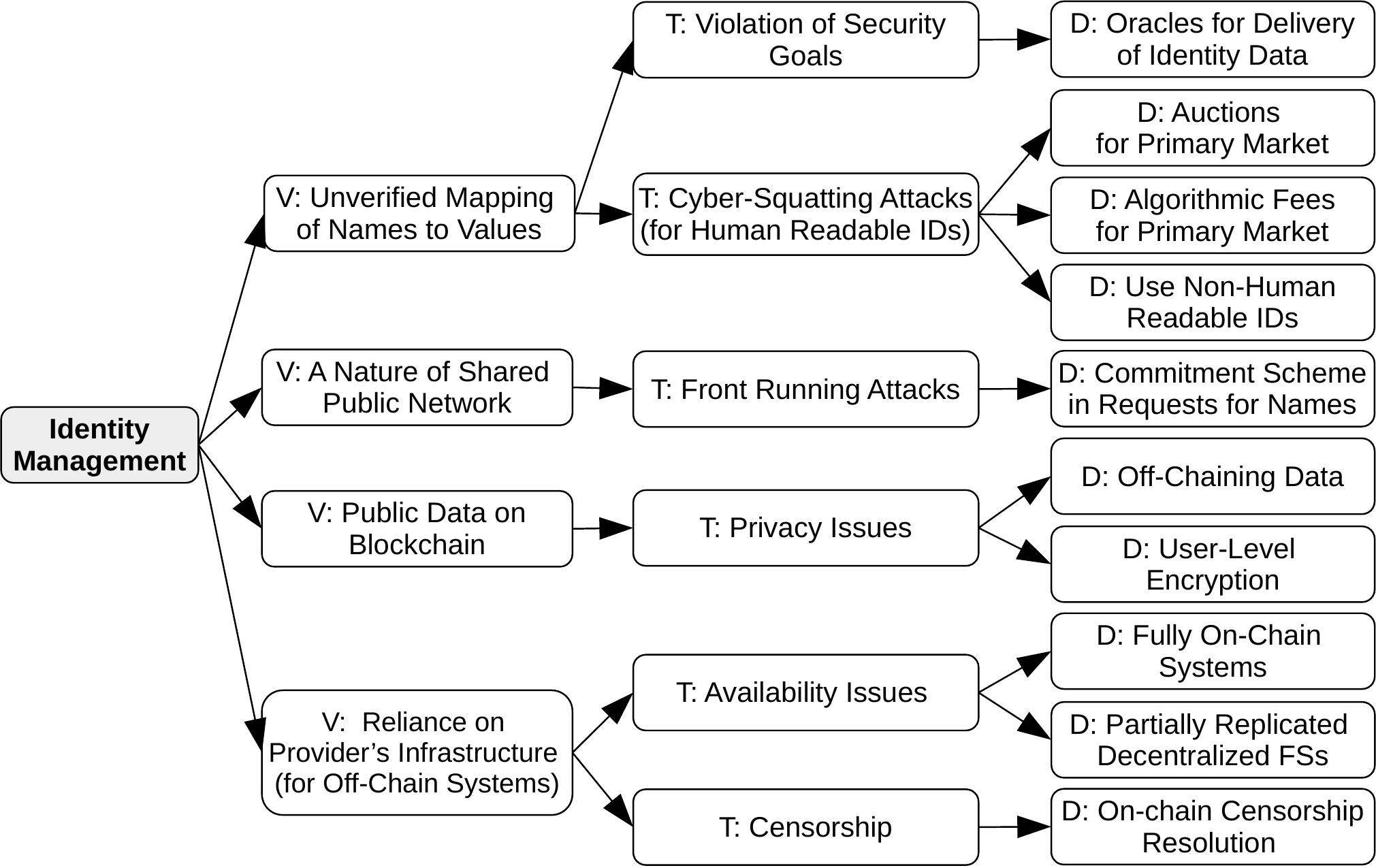} 		
		\caption{Vulnerabilities, threats, and defenses of the identity management category (application layer).}
		\label{fig:attacks-APP-identity}
	\end{center}	
\end{figure}

\mysubsubsection{Security Threats and Mitigations}
We present a taxonomy of vulnerabilities, threats, and defenses related to the identity management category in \autoref{fig:attacks-APP-identity}.
As mentioned by Kalodner et al.~\cite{kalodner2015empirical}, most of the solutions address the problem of mapping names to values but for identity management, it is essential to build a mapping of entities (i.e., persons, companies) to values.
However, to establish such a mapping in a trustworthy way, a human arbitration or a trusted party is needed.
For this purpose, oracles (see \autoref{sec:oracles}) might deliver verified data about the identity of users. 

Since the space of human-readable IDs is scarce, they hold some market value in contrast to an almost infinite number of non-human-readable IDs, such as hashes.
This opens a door to cyber-squatting attacks, in which anybody might seize an ID that does not belong to her and then sell the ID in the secondary market at an inflated price.
Kalodner et al.~\cite{kalodner2015empirical} found in 2015 that from around $120,000$ registered names in Namecoin, only $28$ were not squatted and had human-readable content.
They discussed two strategies to prevent such attacks within the primary market, in which names are issued for the first time. 
These two strategies stand for \textit{auctions} and \textit{algorithmic fees}, both having their respective cons.
Auctions are problematic since they may be initiated at any time, and some potentially interested bidders might not be available. 
An improvement is to specify a fixed time when auctions start.
Another approach for coping with cyber-squatting is an algorithmic fee solution, which assigns the price based on the deterministic observables, such as length of the name, a rank of the domain, occurrence of human-readable words, etc. 
Nevertheless, this approach may require a data feed provider (see \autoref{sec:oracles}).
In  contrast to human-readable identifiers, non-human readable identifiers, such as DIDs, do not suffer from the cyber-squatting threat. 

Another challenge is a front-running attack (i.e., a MITM attack), in which an adversary may intercept and override the user's transaction with a malicious transaction containing the same domain name but a higher fee.
A prevention technique is a variant of the commitment scheme where the user first publishes a sealed (domain) name and public bid, while in the second step she submits the plain text of the name.

Further, identity-related user data that are published on the blockchain are subject to certain privacy issues. 
Mitigation is to keep only integrity information (such as hashes) on the blockchain, while data should be stored off-chain~\cite{Shocard2017,ali2016bootstrapping,Lundkvist2016} or encrypted by the user's private key~\cite{Shocard2017}.
Moreover, all the approaches that rely on off-chain storage and service provisioning are vulnerable to availability issues (e.g.,~\cite{Lundkvist2016},~\cite{did-w3c}) and censorship attacks (e.g.,~\cite{Tobin2016}).
Mitigation that provides censorship evidence is an on-chain censorship resolution~\cite{szalachowski2018blockchain}.

\mysubsubsection{Side Effects and Implications of Countermeasures}
Side effects of oracles depend on their category, and we mention them in \autoref{sec:oracles}.
If a beginning time of the auctions on the primary market is fixed, bidders can be DoS-ed and prevented from bidding. 
Anonymization networks and VPNs used by bidders might mitigate this problem.
While off-chaining of identity-related data helps to cope with privacy issues and decreases operational costs, on the other hand, it means dependency on centralized storage, which negatively impacts the availability of data.
Possible mitigations are partially replicated decentralized filesystems (see \autoref{sec:filesystems}).

\subsection{Secure Timestamping}\label{sec:secure-timestamping}
The role of secure timestamping is to prove that some data existed prior to some point in time -- also referred to as proof-of-existence.
In the decentralized setting of blockchains, the blockchain serves as a trusted notary that enables such proofs since it provides immutability of the history. 
Nevertheless, the blockchain ``does not understand'' the semantics of data that are timestamped, and thus it cannot vet or certify them.

The simple examples of secure timestamping are CommitCoin~\cite{clark2012commitcoin},  STAMPD,\footnote{\url{https://stampd.io/}}  Bitcoin.com Notary,\footnote{\url{https://notary.bitcoin.com/}} and OriginStamp~\cite{GippMG15}, all enabling to post a document's hash into a single blockchain transaction.
OpenTimestamps\footnote{\url{https://opentimestamps.org/}} and POEX.IO\footnote{\url{https://poex.io/}} are examples that define a set of operations for creating timestamps and their verification as part of a Merkle tree that aggregates hashes of timestamped objects. 
The root of the Merkle tree is then stored in the blockchain and later used for verification of timestamped data.

\begin{figure}[t]
	\begin{center}			
		\includegraphics[width=0.48\textwidth]{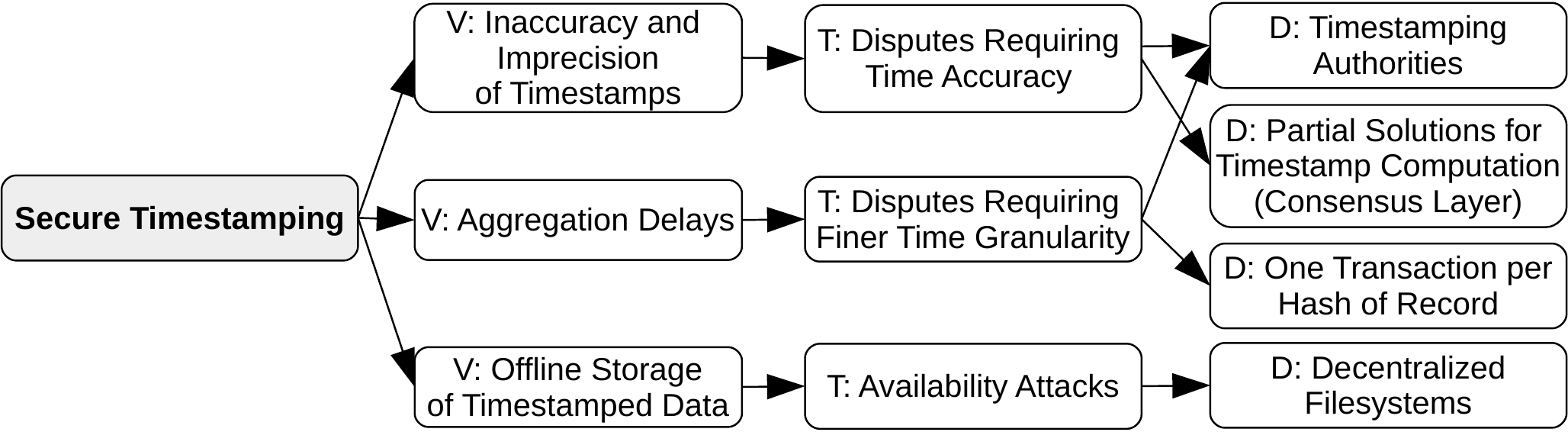} 		
		\caption{Vulnerabilities, threats, and defenses of the secure timestamping category (application layer).}
		\label{fig:attacks-APP-timestamps}
	\end{center}	
\end{figure}

\mysubsubsection{Security Threats and Mitigations}
\autoref{fig:attacks-APP-timestamps} depicts a taxonomy of vulnerabilities, threats, and defenses related to the secure timestamping systems.
Since these systems have a narrow principle of operation and provided functionality, their attack surface is very limited, too.
The main security threats stem from inaccuracy and imprecision of timestamps provided by blockchain network as well as aggregation delays of certain secure timestamping services.
As a consequence, the result of certain disputes might be influenced. 
A possible mitigation technique to improve the accuracy of timestamps is to use timestamping authorities~\cite{szalachowski2018short} or partial solutions for block timestamp computation~\cite{strongchain}.\footnote{Note that this is a countermeasure specific to the consensus layer.}
A mitigation technique for long aggregation delays is to employ timestamping authorities or use one transaction per hash of the timestamped record.
Another class of attacks concerns the availability of timestamped data, for which decentralized filesystems might be utilized as a mitigation technique while storing data in encrypted or plaintext form (depending on the use case).

\mysubsubsection{Side Effects and Implications of Countermeasures}
Although one transaction per hash of a timestamped record mitigates the impact of aggregation delays in solutions such as OpenTimestamps and POEX.IO, on the other hand, it requires a higher amount of data posted to blockchains, and thus it deteriorates the throughput and imposes higher costs. 
Therefore, the choice of either an aggregated solution or a single hash per record depends on a particular use case.

\section{Application Layer: Higher-Level Applications}
\label{sec:apps-applications}
In this section, we elaborate on more specific higher-level applications as opposed to ecosystem applications.
In detail, we deal with the following categories: 
(1) \textbf{e-voting}, (2) \textbf{reputation systems}, (3) \textbf{data provenance}, (4) \textbf{notaries}, (5) \textbf{direct trading}, (6) \textbf{escrows}, (7) \textbf{auctions}, and (8) \textbf{general application of blockchains}.
We describe each of the categories, present a few examples, and then summarize the potential security vulnerabilities, threats, and defenses.
For other detailed reviews of blockchain applications, we refer the reader to Casino et al.~\cite{casino2018systematic} and Zheng et al.~\cite{zheng2018blockchain}, which in contrast to our work follow domain-oriented classification.

\subsection{E-Voting}\label{sec:e-voting}
Kiayias et al.~\cite{Kiayias2002} and Groth~\cite{Groth2004} state several properties that are desirable in e-voting applications:
\begin{compactitem}
	
		\item [\textbf{Perfect Ballot Secrecy:}] 
		implies that finding partial results (i.e., partial tally) before the voting finishes is possible only if all voters are involved in its computation.	
		
		\item [\textbf{Fairness:}] the final tally can be computed only when all participants already had a chance to cast their vote.
		
		\item [\textbf{Public Verifiability:}] any public observer can verify the validity of all votes and final tally.
		This is achieved by using a public bulletin board (e.g., blockchain). 
		A consequence of the public verifiability is \textit{dispute-freeness}, i.e., the result of the voting is indisputable.

		\item [\textbf{Self-Tallying:}] once the voting stage has finished, anyone can compute the final tally. 
		This property together with fairness ensures that the last voter is unable to compute the tally before casting her vote.
		
		\item [\textbf{Fault Tolerance/Robustness:}] a voting protocol is able to recover from a fixed number of faulty voters who do not vote or whose vote is invalid.
		
		\item [\textbf{Receipt-Freeness:}] a participant is unable to supply a receipt of her vote after casting the vote.
		The goal is to prevent vote-selling and post-election coercion. 
\end{compactitem}

\noindent
E-voting~\cite{Chaum1981,Cohen1985} has tried to mimic many of the security properties provided by paper voting.
In decentralized e-voting, the protocol is carried out in phases and requires a multiparty computation (MPC)~\cite{Kiayias2002,Groth2004} executed by the voters.
Decentralized voting involves an interaction among participants and is less robust concerning fault tolerance -- i.e., if voters drop out midway, a recovery round has to be initiated.
The main advantages of using the blockchain for e-voting are its immutability,
public verifiability, 
enforcing protocol rules by the smart contract, and higher availability~\cite{boucher2016if}.

An example of a decentralized e-voting is the Open Voting Network (OVN)~\cite{McCorrySH17}, which for the first time utilizes blockchain as an instantiation of the public bulletin board.
OVN is implemented using Ethereum smart contracts, and it enables boardroom voting of up to $\sim$50 voters with support for two choices.
OVN requires the authority to initialize e-voting, compute multiparty keys from data submitted by voters, and reveal the result of the final tally.
However, the authority is unable to influence the outcome of voting or compromise the privacy of voters (i.e., cast votes).
Although OVN preserves the privacy of voters and provides self-tallying, it does not provide receipt-freeness. 
OVN uses deposit-based penalties to incentivize the authority and voters to actively participate.
Zhang et al.~\cite{Zhang18} present a distributed e-voting scheme that uses the blockchain, but the proposed protocol (like OVN) does not provide a fault tolerance -- a restart of voting is required if any voter does not cast her vote.
This enables sabotaging of the voting process by a single malicious voter.
Venugopalan et al.~\cite{venugopalan2020bbbvoting} propose a boardroom voting over Ethereum, which supports $n$ voting choices and provides a fault tolerance mechanism.
Li et al.~\cite{Li2019} propose an approach that assumes an interactive honest verifier for the zero-knowledge proof presented; however, the verifier can select a biased challenge, which enables collusion of the verifier and the authority.

\begin{figure}[t]
	\begin{center}			
		\includegraphics[width=0.48\textwidth]{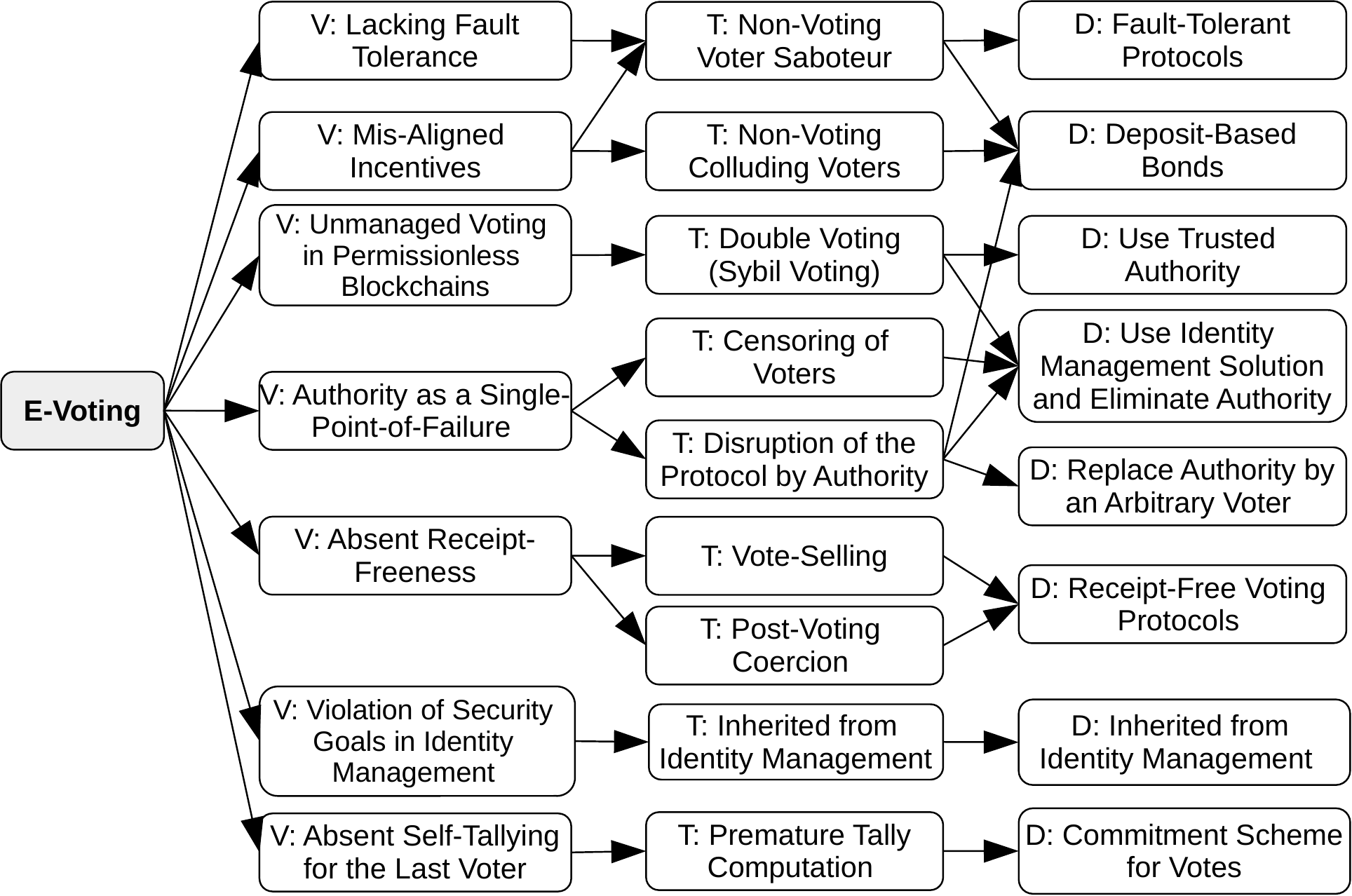} 		
		\caption{Vulnerabilities, threats, and defenses of the e-voting category (application layer).}
		\label{fig:attacks-APP-evoting}
	\end{center}	
\end{figure}

\mysubsubsection{Security Threats and Mitigations}
We present a taxonomy of vulnerabilities, threats, and defenses related to the decentralized e-voting category in \autoref{fig:attacks-APP-evoting}.
The first e-voting-specific group of threats represents vote-selling and post-voting coercion.
In vote-selling, the voter can prove to a briber that she voted as agreed, while in post-election coercion the voter is coerced to show her vote by decryption of the blinded vote.
Mitigation is to use receipt-free voting protocols that thwart both attacks~\cite{Benaloh1994,baudron2001practical}.
Existing solutions for achieving receipt-freeness assume a secret channel (bi-directional~\cite{Benaloh1994} or uni-directional~\cite{Sako95}), use deniable encryption~\cite{Canetti96,Canetti97}, or employ randomizers~\cite{Baudron2001}.

The next threat is double-voting (with Sybil accounts) in the case of unmanaged public voting in permissionless blockchains. 
To prevent double-voting and ensure that only eligible voters can vote, it is usually required that a voting authority permits voters to vote.\footnote{This is represented by know your client (KYC) compliance, and it is related to the principles of permissioned blockchains.}
Another issue is a possibility of voters not voting despite enrollment, resulting in the sabotage of the voting round (possible in~\cite{Groth2004,Zhang18,McCorrySH17}) or privacy issue related to more fine-grained inference of the actual votes of the remaining voters who voted.
Deposit-based bonds might be employed as penalties for saboteurs and disincentivize such behaviors~\cite{McCorrySH17}.
Another countermeasure for saboteurs is a fault-tolerant voting protocol (e.g.,~\cite{KhaderSRH12,venugopalan2020bbbvoting}), in which the remaining honest voters can recover the final tally even without votes of saboteurs.
Since e-voting might assume verified identities of all participants, the next group of threats that is worth noting is inherited from identity management (see \autoref{sec:identity-management}).

The last threat relates to the self-tallying property, which might not hold if no countermeasure is applied, as the last participant can compute the tally and only after that decide on her vote.
Simple mitigation is to use an additional ``dummy'' participant that is handled by the voting authority.
Another solution is to enforce participants to first commit to their votes and then cast the committed votes in the later stage~\cite{McCorrySH17}

\mysubsubsection{Side Effects and Implications of Countermeasures}
Receipt-free voting protocols can protect against vote-selling or coercion but on the other hand, they imply additional computational overhead for voters, which increases the cost of running a decentralized protocol. 
Although authority can prevent double voting, it is trusted with managing voters honestly and completing all actions required for progressing the protocol from one stage to the next stage (e.g.,~\cite{McCorrySH17,venugopalan2020bbbvoting}).
However, the authority represents a single-point-of-failure, since it might disrupt the execution of the protocol. 
Deposit-based bonds can be employed as a mitigation technique, or the authority can be replaced by an arbitrary voter for the purpose of execution of the voting protocol (but not for managing voters).
Regarding the management of voters, an important threat is the possibility that the authority might censor some eligible voters.
A solution for eliminating the authority (or a delegation of this problem to a different application type) is permitting voters to vote upon successful registration of their identities within a certain identity management application (see \autoref{sec:identity-management}).
Fault-tolerant voting protocols impose additional overheads, where all remaining voters must actively participate in the recovery round -- this implies additional costs on such voters, and it slows down the protocol.
Similarly, a commitment scheme for coping with violation of self-tallying in the case of the last voter implies additional overheads and costs for a dedicated commitment stage.

\subsection{Reputation Systems}\label{sec:reputation-systems}
Reputation systems commonly serve as a means to (1) measure the level of trust in particular entities that provide a certain service, (2) verify claims of user achievement or authenticity of issued counter-party/ownership tokens.
The reputation is usually quantified based on the voting of  parties/users that independently analyze the history of interactions/records produced by entities in a reputation query.

In reputation assessment, there are two options to determine the eligibility to rate. 
In the first option, an arbitrary legitimate participant can rate a product that she has bought or a service that she has consumed.
In the second option, only a limited number of selected participants can vote on the authenticity of individual records (e.g., accreditation).
In reputation-based systems, identity management is two-fold since the identity of both voters and the record owners/merchants/service providers needs to be verified. 

\paragraph{Rating by Arbitrary Participants}
A privacy-preserving reputation system for e-commerce was proposed in~\cite{schaub2016trustless}.
The authors utilize blinding signatures and merchant-issued tokens to achieve the privacy of reviews and avoid bad-mouthing and ballot-stuffing attacks.
A feedback-based reputation approach that utilizes the incentive-based scheme of the Bitcoin network is proposed in~\cite{Carboni2015}.
In this approach, any consumer might rate the service of the producer, while obtaining a voucher for the feedback.
Zhao et al.~\cite{zhao2019dynamic} propose a reputation management scheme that utilizes additive secret sharing to achieve the privacy of participants in reputation assessments.

\paragraph{Rating by Several Selected Participants}
An example of such a reputation-based application is related to accreditation of educational institutions by other higher-level institutions and organizations~\cite{grech2017blockchain}.

\begin{figure}[t]
	\begin{center}			
		\includegraphics[width=0.48\textwidth]{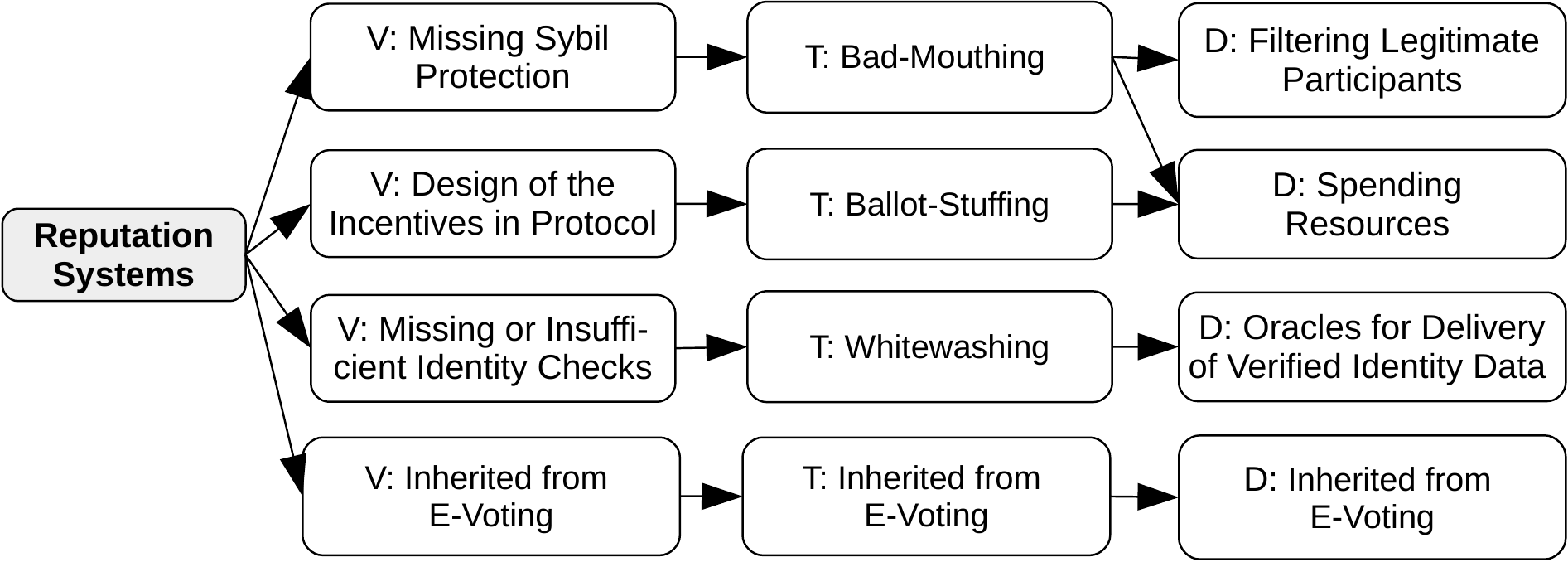} 		
		\caption{Vulnerabilities, threats, and defenses of the reputation systems category (application layer).}
		\label{fig:attacks-APP-reputation}		
	\end{center}	
\end{figure}

\mysubsubsection{Security Threats and Mitigations}
We present a taxonomy of vulnerabilities, threats, and defenses related to the reputation systems category in \autoref{fig:attacks-APP-reputation}.
Specific security threats to reputation systems with an arbitrary number of legitimate participants are \textbf{bad-mouthing}, \textbf{ballot-stuffing}, and \textbf{whitewashing} attacks~\cite{schaub2016trustless}.
In bad-mouthing attacks, the customer (e.g., competitor) lies about the product or service, while in the ballot-stuffing attacks, the service provider might increase her reputation by herself.
Bad-mouthing can be mitigated by filtering only authorized participants to submit reputation assessments (e.g., by review tokens~\cite{schaub2016trustless,Carboni2015}).
Although bad-mouthing cannot be completely prevented, it requires participants to spend resources (e.g., buying a product or paying transaction fees) to be eligible for rating a service provider.
Similarly to bad-mouthing, the ballot-stuffing attack cannot be eliminated but only mitigated by requiring to spend resources (i.e., tokens) for each rating entry.
If a service provider accumulates a significant negative reputation, it has an incentive for a whitewashing attack -- the service provider creates a new service with a neutral reputation, which is un-linkable to her previous service.
To mitigate this attack, oracles for obtaining verified data about identities of entities can be employed, possibly as part of the identity management system.

Since reputation systems resemble e-voting in general, concerning security threats are inherited from there (see \autoref{sec:e-voting}) and its dependency on identity management (\autoref{sec:identity-management}). 
In particular, only authorized participants are allowed to participate in the voting process and no duplicate votes are allowed (ensured by identity management), while the votes/ratings should remain blinded (ensured by e-voting) unless the particular use case requires otherwise.

\mysubsubsection{Side Effects of Countermeasures}
Since some countermeasures in reputation systems are inherited from the e-voting and oracles categories, the side effects are also inherited from these categories.

\subsection{Data Provenance}
Data provenance represents the ownership history of an arbitrary object.
However, in the cyber-world, objects are represented by data that can be changed, and thus the history
must account also for the modifications~\cite{buneman2001and}.
Data provenance with the use of blockchains has the potential to resolve various disputes and issues related to intellectual property, authorship, the validity of certificates or other issued documents, etc.
Data provenance assumes known verified identities of the involved parties. 

An application of data provenance is supply-chain management~\cite{KSHETRI201880,DeloittIntrie2017}, where the goal is to resolve potential issues in the traceability of goods and provenance of associated data~\cite{KimL2016a}.
Blockcloud is an approach that utilizes blockchains for data provenance in cloud computing~\cite{Shetty2017}.
The authors aim at the accountability of data creation and manipulation with the intention to detect malicious insiders and intrusions.
The idea of using the blockchain for tracking packages and mails as part of supply chain management was proposed in~\cite{jaag2017blockchain}.
ChainAnchor~\cite{hardjono2016cloud} is a  framework for the commissioning and decommissioning of IoT devices in a cloud-based ecosystem. 
In a commissioning procedure, devices prove their manufacturing provenance to a verifier in a privacy-preserving fashion without a need for interaction with the manufacturer. 
An additional goal of ChainAnchor is to reward owners of IoT devices for sharing data in a privacy-preserving manner.
A data provenance approach that focuses on the integrity of IoT-generated data is proposed in~\cite{Liu2017ICWS}.
A framework to achieve data provenance of multimedia objects such as artworks and books was proposed in~\cite{Bhowmik2017}. 
The authors use watermarking techniques to embed transaction metadata of objects into the objects themselves to prove data tampering.

Catena~\cite{Tomescu2017} guarantees a non-equivocation of its append-only log and relatively low storage overheads (i.e., storing all the blockchain headers).
The hash of each off-chain data block is added to the append-only log of Catena as a single transaction, which is bound to the previous Catena transaction. 
The same method of binding consecutive transactions in an append-only log was utilized in Contour~\cite{al2018contour}.
In contrast to general Catena, Contour focuses on non-equivocation during  the distribution of open-source software packages by a specified authority.
Grech and Camilleri~\cite{grech2017blockchain} elaborate on the usage of blockchain for the issuance of educational certificates by academic institutions.
Such certificates might be issued by institutions whose identities are verified (see \autoref{sec:identity-management}).

\begin{figure}[t]
	\begin{center}			
		\includegraphics[width=0.48\textwidth]{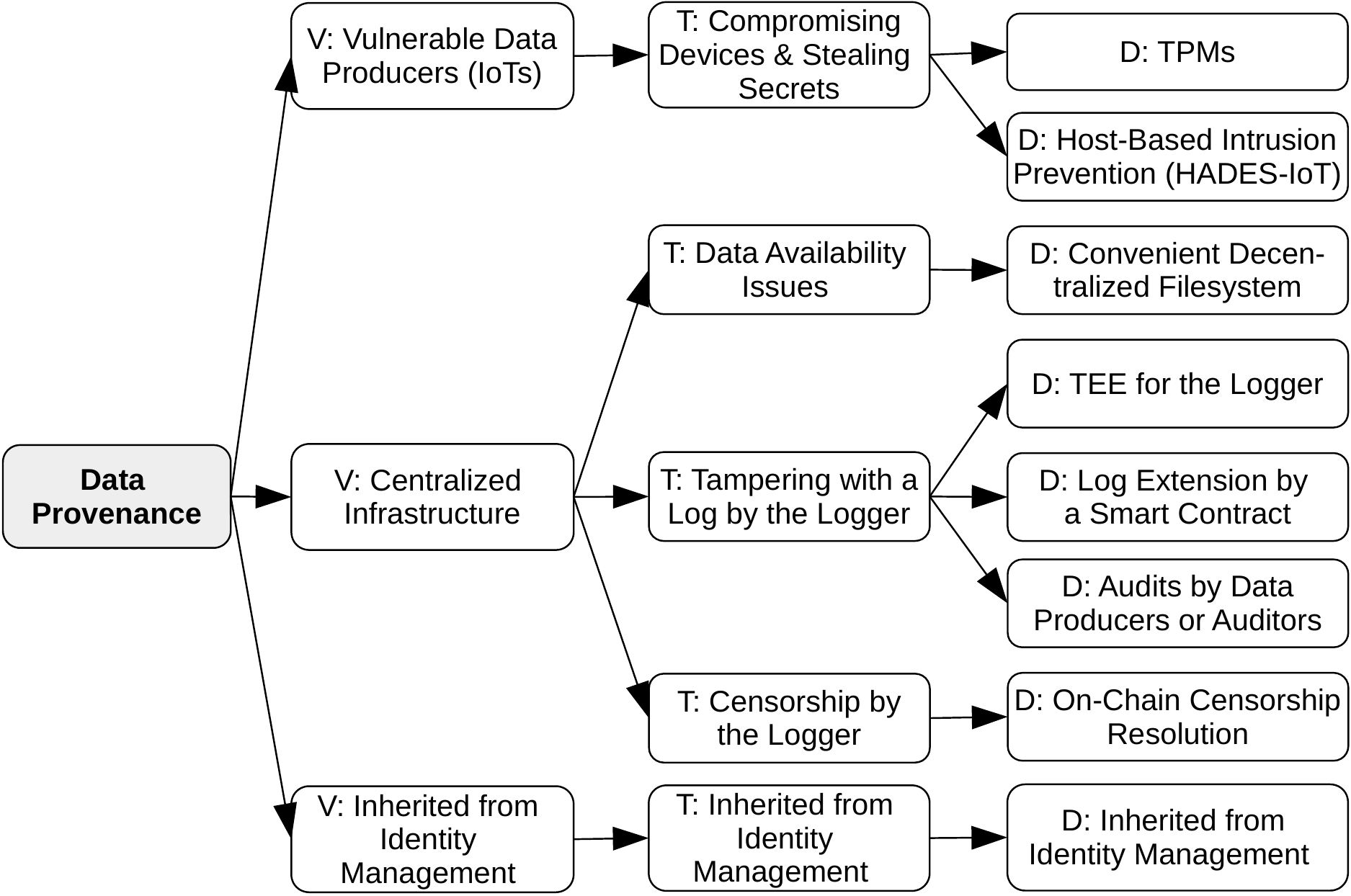} 		
		\caption{Vulnerabilities, threats, and defenses of the data provenance category (application layer).}
		\label{fig:attacks-APP-provenance}
	\end{center}	
\end{figure}

\mysubsubsection{Security Threats and Mitigations}
We present a taxonomy of vulnerabilities, threats, and defenses related to the data provenance category in \autoref{fig:attacks-APP-provenance}.
Since data provenance infrastructure might involve simple IoT devices and sensors that are often not updated nor physically protected, it is important to ensure that these devices are tamper-proof and no secret can be stolen from them.
As a countermeasure,  Trusted Platform Modules (TPMs) may be used if they are available. 
We note that the current trend is to involve TPMs in the hardware of contemporary IoT devices. 
However, TPMs are often not available in numerous legacy IoT devices. 
In such cases, it is possible to leverage kernel-space and user-space memory isolation  as part of the intrusion prevention system (e.g.,~\cite{breitenbacher2019hades}).
Another vulnerability originates from a possible centralized logger component of data provenance solutions 
(e.g.,~\cite{Tomescu2017,al2018contour,guarnizo2018pdfs,homoliak2020aquareum}). 
Hence, availability issues for data storage must be considered and possibly resolved by a convenient decentralized filesystem approach (see \autoref{sec:filesystems}). 
Another threat concerning the centralized infrastructure of loggers is the possibility of data tampering and censorship.
Data tampering can be detected by data producers  or auditors that do periodic audits~\cite{paccagnella2020custos}.
To cope with censorship, an on-chain smart contract-based censorship resolution can be utilized~\cite{szalachowski2018blockchain,homoliak2020aquareum}.

\mysubsubsection{Side Effects and Implications of Countermeasures}
Although auditors might detect tampering with data by a logger entity, their periodic activity implies high operational costs.
A possible option to save such costs is an auditor-free update of the log using a smart contract~\cite{guarnizo2018pdfs}, alternatively combined with TEE~\cite{homoliak2020aquareum} to enforce even stronger properties (i.e., the correctness of the execution by logger).
However, TEE has also its cons, which we discussed above. 
Moreover, since countermeasures of the data provenance category depend on the identity management and filesystems category, the side effects are inherited from them as well.

\subsection{Notaries}\label{sec:notaries}
In contrast to secure timestamping, the role of the notary system is not only to prove the existence of documents at certain points in time but also to vet and certify documents~\cite{notarization};
hence, notary systems assume known verified identities of involved parties who do the vetting. 
In addition to the above two functionalities, the definition of the notary system involves document preservation, which, however, in the context of the public blockchain is optional. 
The involved parties may decide whether to store vetted documents in a database of a notary service provider (e.g., PADVA~\cite{szalachowski2018blockchain}) or whether to keep it privately at the client-side (e.g., Blockusign\footnote{\url{https://blockusign.co/}}).

A blockchain-based notarization platform on Ethereum was proposed in the post~\cite{ethereum-notary}, where an arbitrary number of users/entities with verified identities may sign/approve the documents and their new versions, respectively.
The proposal assumes a certification authority that verifies the identities of involved entities, and as an example, the authors suggest the use of the ERC 725 standard~\cite{erc-725}.
ADVOCATE~\cite{rantos2018blockchain} is an approach for notarization of agreements about personal data processing in IoT between owners of IoT devices and data processing services -- both must co-sign an agreement.
Mizrahi~\cite{mizrahi2015blockchain} proposes a system for property ownership,
where, all ownership transfers can be executed without any trusted party, but the trusted party is required for introducing the initial ownership record to the blockchain.
The ownership register for vehicles was
proposed by Notheisen et al.~\cite{notheisen2017trading}, where trusted third parties, such as police departments and transport authorities provide and verify vehicle-specific information.
SilentNotary\footnote{\url{https://silentnotary.com/}} is a smart contract-based system for self-certifying of files produced by registered users.
PADVA~\cite{szalachowski2018blockchain} is a Transport Layer Security (TLS) notary service realized as a smart contract-based two-party agreement (i.e., Service Level Agreement). 
PADVA introduces notary entities that are obligated to periodically check the validity of PKs in a specified set of certificates.

\begin{figure}[t]
	\begin{center}			
		\includegraphics[width=0.48\textwidth]{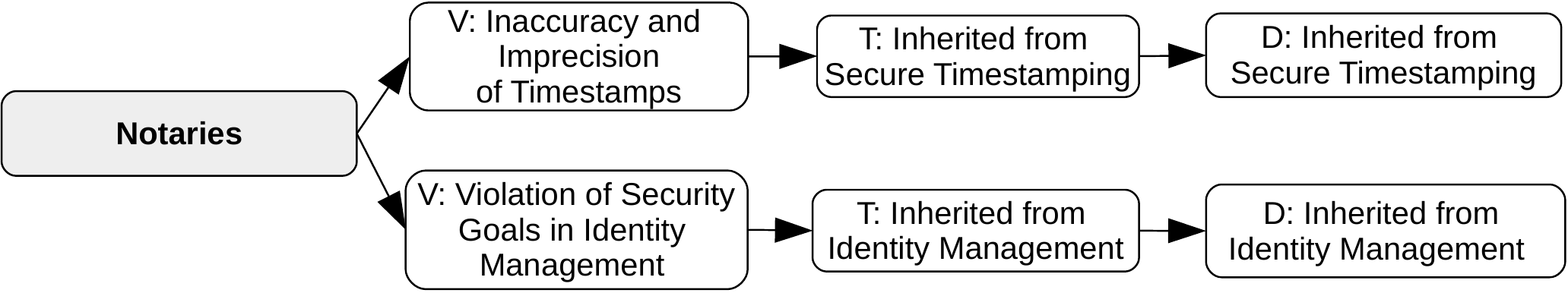} 		
		\caption{Vulnerabilities, threats, and defenses of the notaries category (application layer).}
		\label{fig:attacks-APP-notary}
	\end{center}	
\end{figure}

\mysubsubsection{Security Threats and Mitigations}
We present a taxonomy of vulnerabilities, threats, and defenses related to the notaries category in \autoref{fig:attacks-APP-notary}.
Notaries inherit security threats related to timestamp accuracy from the secure timestamping category (see \autoref{sec:secure-timestamping}).
In addition, since they assume verified identities of involved parties, they inherit security issues from the identity management category (see \autoref{sec:identity-management}).
In particular, many notary services assume a centralized identity management system, which might be subject to tampering or censorship issues.
Next, a very specific threat to PADVA is a cheating notary, who quietly does not  run the service periodically or runs it only sporadically.
Such cheating can be revealed by clients on an ad-hoc basis, who punish notaries by the smart contract logic that causes notaries to lose their deposit.

\mysubsubsection{Side Effects of Countermeasures}
Since security aspects in notaries are inherited from the secure timestamping and identity management categories, the side effects are also inherited from these categories.

\subsection{Direct Trading}\label{sec:trading}
While blockchain-based cryptocurrencies enable native secure transfers of crypto-tokens among their owners, a challenge arises when owners want to exchange crypto-tokens they hold for goods outside of the cryptocurrency blockchain.
This challenge is also referred to as the buyer/seller dilemma: ``\textit{Should the buyer trust the seller and pay her before receiving goods or should the seller trust the buyer and ship the goods before receiving the payment?}''
In the direct trading category, this problem is resolved directly between the buyer and seller, without the need for a mediator, under the assumption of a trusted seller with a verified identity.
For example in BIP-70~\cite{bip-70}, the buyer first verifies the authenticity of the seller using its X.509 certificate and then issues a payment transaction. 

\begin{figure}[t]
	\begin{center}			
		\includegraphics[width=0.48\textwidth]{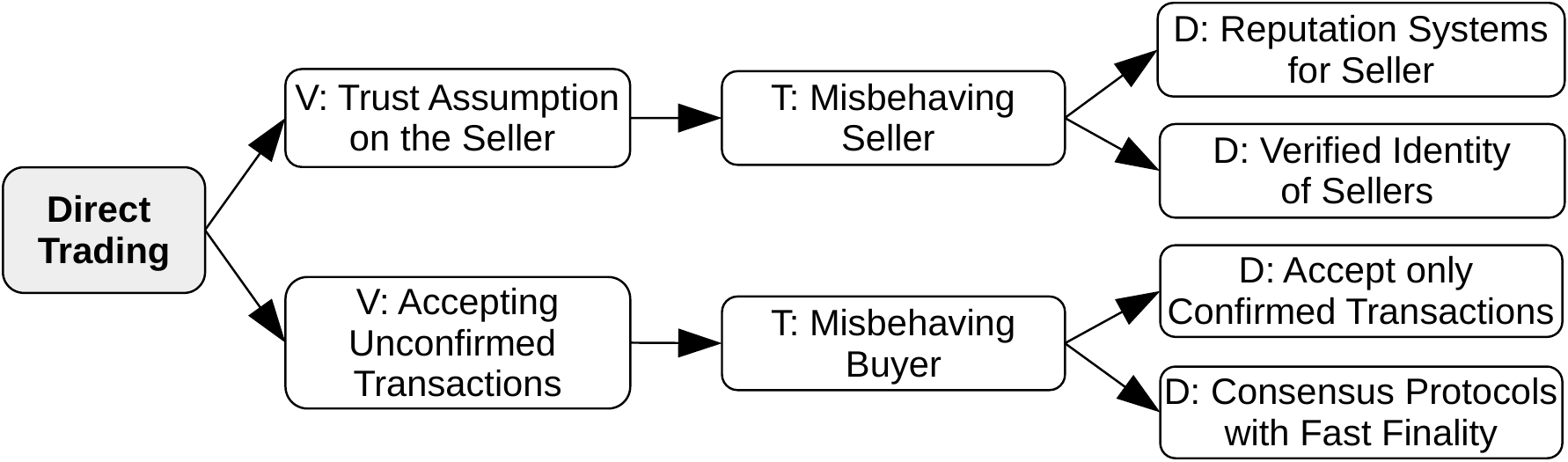} 		
		\caption{Vulnerabilities, threats, and defenses of the direct trading category (application layer).}
		\label{fig:attacks-APP-trading}
	\end{center}	
\end{figure}

\mysubsubsection{Security Threats and Mitigations}
We present a taxonomy of vulnerabilities, threats, and defenses related to the direct trading category in \autoref{fig:attacks-APP-trading}.
The first vulnerability represents the assumption of trust in the seller.
For example in BIP-70, the buyer might ask the seller to interrupt the request and get a refund but the seller may misbehave, and thus risk a reputation loss.
On the other hand, this might be tolerated if the seller spoofed her identity. 
A mitigation technique is to use a strong means for identity verification, including assessments from reputation systems.
Another attack on BIP-70 that is worth mentioning is the Silkroad trader attack~\cite{mccorry2016refund}, in which a malicious buyer might replace her refund address and then ask the seller for a refund. 
After a refund, the buyer might plausibly deny receipt of a refund (and ask for a refund again) due to missing authentication on the refund address. 
Another potential attack related to direct trading is double-spending performed as part of the selfish mining or 51\% attacks -- therefore,  it is important to wait for enough confirmations by the seller before releasing the goods or use consensus protocols having a fast finality (see \autoref{sec:general-consensus-attacks}).

\mysubsubsection{Side Effects and Implications of Countermeasures}
Enough confirmations by the seller imply a long waiting time for the buyer before the seller releases the goods and issues a receipt for it.
In particular, this might be problematic in the case of on-premise purchases.
The waiting time is dependent on the time to the finality of the underlying consensus protocol, and thus a consensus protocol with a low time to finality represents a solution.
However, it is the means of the consensus layer (see \autoref{sec:consensus}).

\subsection{Escrows}\label{sec:escrows}
Escrows address the same problem as direct trading but in contrast to direct trading, escrows do not assume a trusted seller; instead, escrows outsource the trust into the third party, referred to as a \textit{mediator}.
The mediator might actively participate in the escrow protocol or she might be involved only in the case of a dispute.
According to the decentralization of the mediator, escrow protocols can be split into \textit{single mediator} protocols and protocols with a \textit{group-based mediator}.
Goldfeder et al.~\cite{goldfeder2017escrow} propose a few escrow protocols from both categories, which we briefly review. 

\paragraph{Single Mediator}
Several proposed protocols contain a single mediator and involve 2-of-3 multi-signatures for splitting the control, threshold-based signatures for improving privacy, and protocols leveraging homomorphic properties of elliptic cryptography to achieve privacy (i.e., by blinding the mediator's next address) and non-interactiveness. 
Another protocol combines multi-signatures with bonds deposited by a mediator to avoid DoS by the mediator.
Note that blinding of the mediator's next address hides the execution of the protocol to the mediator only in the case when no dispute has arisen.

\paragraph{Group-Based Mediator}
In these protocols, disputes are resolved by a majority vote. 
DoS attack is thwarted as long as the majority of mediators is willing to finish the execution of the protocol.
The privacy of some protocols is preserved by blinding, similarly as in the case of single mediator protocols.

\smallskip
An example of a distributed marketplace was proposed in~\cite{kabi2018blockchain}, where a marketplace contract lists the products and an escrow agent contract serves for resolution of disputes by a mediator (viz. single or group-based mediator).
The authors discuss the integration of logistic parties with verified identities and reputation systems to assess these parties and mediators.
OpenBazaar\footnote{\url{https://openbazaar.org/}} is a distributed marketplace that uses smart contract-based escrows with 2-of-3 multi-signatures, where the mediator is agreed by the buyer and seller.
A similar example is Escaroo\footnote{\url{https://escaroo.com/}} but in contrast to Openbazaar, it has its own trusted mediator. 
Natmin\footnote{\url{https://www.natmin.io/}} is an escrow example that utilizes a public group of mediators for dispute resolution, while the reputation of the mediators is adjusted according to their votes and a result of the dispute.

\begin{figure}[t]
	\begin{center}			
		\includegraphics[width=0.48\textwidth]{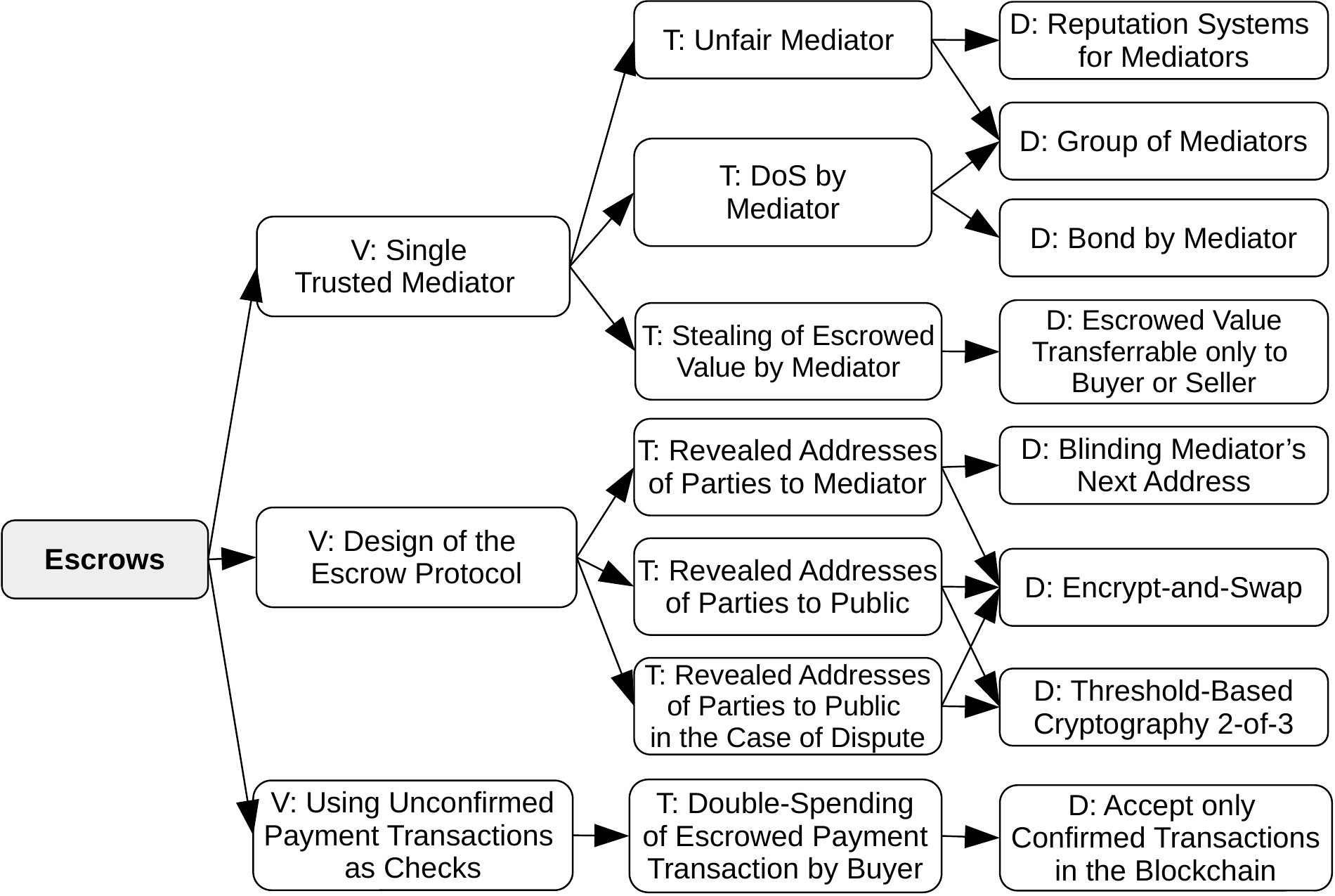} 		
		\caption{Vulnerabilities, threats, and defenses of the escrows category (application layer).}
		\label{fig:attacks-APP-escrow}
	\end{center}	
\end{figure}

\mysubsubsection{Security Threats and Mitigations}
We present a taxonomy of vulnerabilities, threats, and defenses related to the escrow category in \autoref{fig:attacks-APP-escrow}.
The first group of threats refers to a trusted mediator who represents a single-point-of-failure.
The mediator might disrupt the execution of the escrow protocol or decide unfairly in the case of a dispute. 
The existence of these threats depends on the design of an escrow protocol.
For example, in the Silk Road marketplace~\cite{christin2013traveling}, a mediator requests the sending of crypto-tokens to her address, while she is trusted to send the crypto-tokens to the seller upon delivery confirmation from the buyer.\footnote{Instead of a single mediator, the Silk Road marketplace utilized several intermediaries to increase the anonymity of the buyer and seller.} 
However, the mediator might refuse to do so and keep the value for herself.
The mitigation techniques are group-based mediators, requiring a consensus of the majority to decide on a case or requiring the mediator to put a bond into the escrow protocol.
Another mitigation technique is to use reputation systems for the assessment of a single mediator.

To avoid stealing of the escrowed value by the mediator, the protocol should by design allow releasing value only to the buyer or seller (using rules of a smart contract), while assuming a timeout.
For example, an early version of OpenBazaar utilized smart contracts for trading but without any timeout. 
As a result, many buyers did not release the funds to the seller upon successful delivery.
However, when a timeout is adopted, upon its expiration, a seller can unilaterally release and acquire the funds from the escrow.

The next class of threats is related to revealing the private information about running the protocol to the public or mediator (e.g., involved parties, occurrences of disputes). 
The countermeasures are blinding the mediator's address, threshold-based cryptography, including its special variant \textit{encrypt-and-swap}~\cite{goldfeder2017escrow}, which uses a 3-of-3 threshold signature protocol that assigns one private share to the buyer and one to the seller, while the third share is known to both parties.
Both parties reveal their private share to the mediator, who, upon dispute, provides the winning party with the missing share.

Another possible threat is the double-spending of an unconfirmed escrowed payment transaction by the buyer. 
For example, if a (naive) escrow protocol requires a mediator to escrow a signed transaction by the buyer and release it to the blockchain only upon delivery confirmation, the buyer might not confirm delivery and perform a double-spending attack (see \autoref{sec:consensus}).
As a prevention technique, unconfirmed transactions should not be accepted by the sellers at all.
Moreover, we highlight that some escrow protocols (similarly as atomic swaps) are sensitive to double-spending performed by the selfish mining or 51\% attacks -- therefore, in these protocols, it is important to wait for enough confirmations or use consensus protocols having a fast finality (see \autoref{sec:general-consensus-attacks}).

\mysubsubsection{Side Effects and Implications of Countermeasures}
Although group-based mediators are more robust to attacks misusing a trust in a single mediator, they are more expensive to run, requiring the interaction of enough mediators with the blockchain. 
This in turn slows down the throughput of the escrow protocol.
The extra operational overheads are also imposed by the reputation systems for single mediators.
Although encoding the escrow logic into a smart contract is supported in some implementations (e.g., OpenBazaar, Escaroo), they require the deployment of a new smart contract per each trade, which is a costly option.
Such logic can be implemented even within a single smart contract.
Similar to direct trading, not accepting unconfirmed transactions implies a long waiting time for the buyer; this is not the case for the consensus protocols with a fast finality.  
Finally, using reputation systems brings their security aspects.

\subsection{Auctions}\label{sec:auctions}
In auctions, sellers promote the sale of their goods through blockchain while buyers place bids for them.
Galal and Youssef~\cite{galal2018verifiable} specify several desired properties of auctions: 

\begin{compactitem}
	\item[\textbf{Privacy of bids}]  ensures that values of particular bids are not revealed to anybody before committing to them. 
	
	\item[\textbf{Posterior privacy}] ensures that all bids remain private  after the auction ends. 
	
	\item[\textbf{Publicly verifiable correctness}]  enables anybody to verify the results of the auction through the blockchain.
	
	\item[\textbf{Resistance against DoS}] ensures that no bidder or auctioneer can prematurely abort a protocol without being penalized.

\end{compactitem}

\smallskip \noindent
The authors of~\cite{galal2018verifiable} instantiate the auction as a smart contract, to which, bidders submit homomorphic commitments of their sealed bids and then reveal their commitments to the auctioneer via a PK encryption.
Afterward, the auctioneer deciphers the bids, determines the winner, and announces it to the public while providing zero-knowledge proofs of the correctness.
As part of their other contribution~\cite{galal2018succinctly}, the same authors improved on high costs intrinsic to their former work~\cite{galal2018verifiable} by using zk-SNARKs and its off-chain computation, which requires only a single on-chain proof verification of the whole auction process. 
However, in both approaches~\cite{galal2018succinctly,galal2018verifiable}, the auctioneer might compromise the privacy of all bids, which led the same authors to propose Trustee~\cite{galal2019trustee}, an approach based on TEE.
In Trustee, the bidders submit encrypted bids to the auctioneer's TEE, which
confidentially evaluates a winner and generates a blockchain transaction proving it.

Strain~\cite{blass2018strain} is an auction protocol that guarantees the privacy of bids against malicious bidders and, in contrast to~\cite{galal2018verifiable,galal2018succinctly} also against the auctioneer. 
Strain is executed in four rounds (i.e., four blocks), and it assumes a semi-honest (i.e., passive) auctioneer who acts as a judge verifying the correctness of zero-knowledge proofs. 
Neither the auctioneer nor a malicious bidder learns anything about bids of honest bidders; however, the order of the bids is leaked to the public.
Strain requires each bidder to commit publicly to her bid, while the proposed scheme enables a majority of honest bidders to open other bidders' commitments in the case that they abort the protocol.
For the sake of efficiency (i.e., a constant number of rounds), the authors provide weaker security properties in contrast to MPC protocols, where no semi-honest judge is required.
Finally, the authors propose an extension of their scheme to support the anonymity of all bidders by blinding RSA signatures and the Dining Cryptographers (DC) network. 

Another approach that preserves the privacy of the bid values (but not the privacy of their order) is proposed in~\cite{ma2019fully}.
The protocol requires off-chain interaction for two-party computation protocol
that performs a pairwise comparison of blinded bids among bidders and the
auctioneer. 

\begin{figure}[t]
	\begin{center}			
		\includegraphics[width=0.48\textwidth]{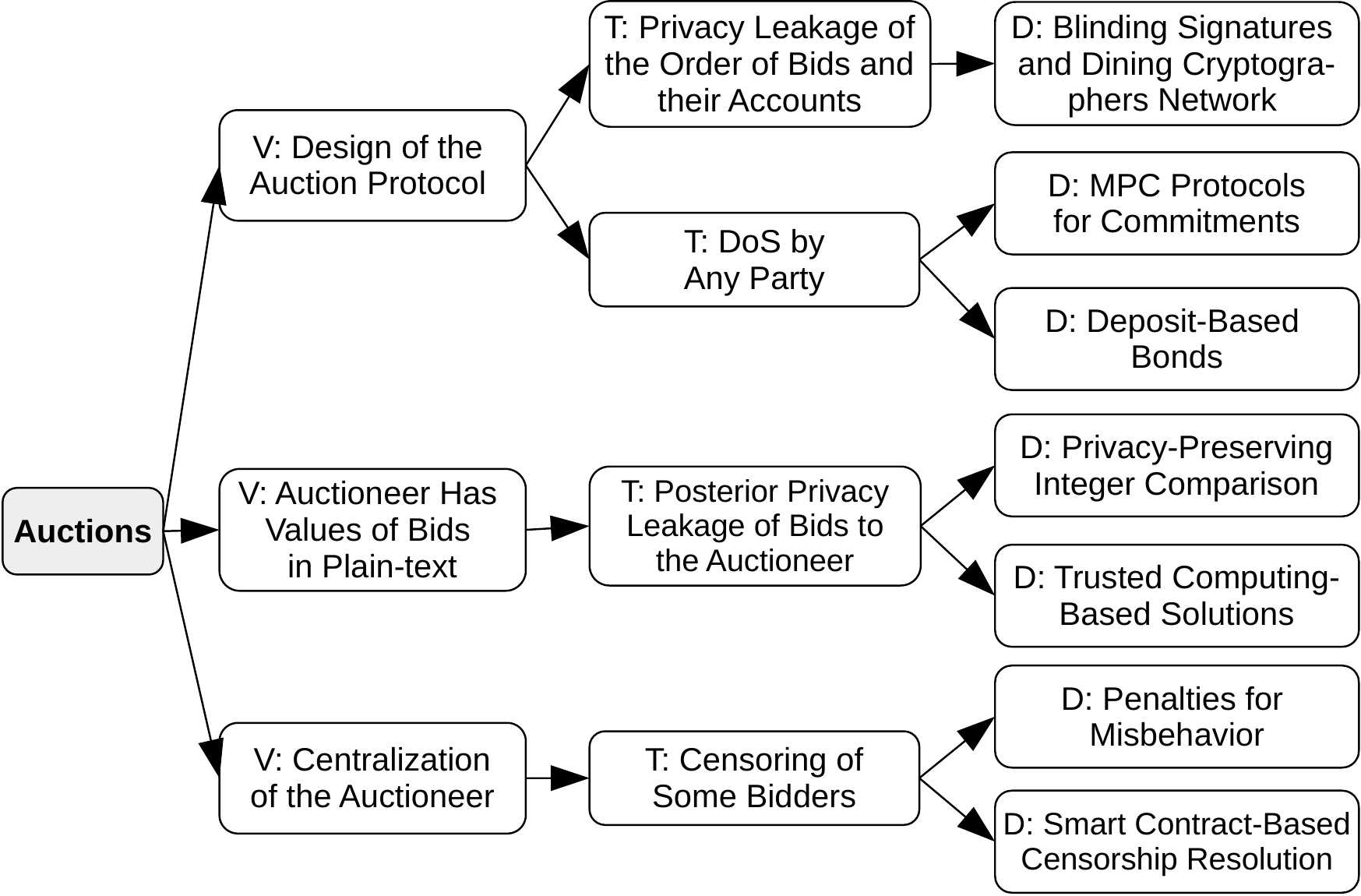} 		
		\caption{Vulnerabilities, threats, and defenses of the auctions category (application layer).}
		\label{fig:attacks-APP-auctions}
	\end{center}	
\end{figure}

\mysubsubsection{Security Threats and Mitigations}
We present a taxonomy of vulnerabilities, threats, and defenses related to the auctions category in \autoref{fig:attacks-APP-auctions}.
There are several possible issues related to privacy leakage in the auction protocols.
The first privacy issue stands for revealing addresses of bidders and/or order of their bids to the public.
For example, the authors of~\cite{galal2018verifiable} do not provide anonymity of bidders, since bidders use their existing Ethereum addresses to interact with the protocol.
A mitigation technique using blinding RSA signatures and the DC network was proposed in~\cite{blass2018strain}; however, network-level attacks on revealing locations/IP addresses of parties remain possible (see \autoref{sec:network}).

Since the auctioneer of some protocols (e.g.,~\cite{galal2018verifiable,galal2018succinctly}) sees the bidders and their bids in plain-text, she might either intentionally or accidentally (e.g., an external compromise) leak these data (and their corresponding proofs) that are attributable to particular bidders. 
The protection technique is to avoid the auctioneer from seeing the plain-text of the bids, and instead use privacy-preserving integer comparison (e.g.,~\cite{blass2018strain}) or trusted computing-based solutions (e.g.,~\cite{galal2019trustee}).

Another threat originates from a centralized auctioneer who might censor some bidders (e.g., due to collusion with another bidder) by claiming that their bids are invalid, i.e., claiming that a commitment does not open to the sealed bid.
To cope with this threat, the auction can utilize a smart contract-based resolution mechanism in which the bidder might prove the opposite by revealing her value of the bid, causing a penalization of the auctioneer~\cite{galal2018verifiable}. 
Deposit-based bonds and penalization of the involved parties can be used as a protection against abortion of the protocol (i.e., DoS) by any party.
For example, the protocol of~\cite{galal2018verifiable} splits penalties to honest participants in the case of abortion by some party.
Another option to cope with the abortion of the auction protocol is to use multiparty computation (MPC) for the commitment of sealed bids~\cite{blass2018strain}, which enables the opening of the commitment of the aborting party and thus to continue in the auction protocol.

\mysubsubsection{Side Effects and Implications of Countermeasures}
Although censoring of bids can be prevented by a smart contract-based resolution mechanism~\cite{galal2018verifiable}, it has privacy consequences since it leaks the value of the bid and its corresponding bidder.
Deposit-based bonds can disincentivize the abortion of the auction protocol but they require a restart of the round. 
In contrast to it, MPC protocols for commitments can recover the current auction round but with additional overheads and costs imposed on a smart contract platform.
In the case of using a TEE-based solution, the malicious auctioneer might perform censorship of some bidders due to collusion with other bidders. 
To avoid this threat, the authors of~\cite{galal2019trustee} propose a smart contract-based mechanism that verifies whether the set of sealed bids submitted to the smart contract corresponds to the list of bids in the proof generated by TEE.
Authors also discuss another option to cope with this attack by embedding an SPV client within the TEE that would evaluate the state of the blockchain; however, this solution would impose a high memory consumption of already constrained TEE and would expose TEE to vulnerabilities presented in SPV client.
Another implication of using TEE-based auction is the possibility of a local replay attack discussed in~\cite{galal2019trustee}, where the auctioneer might provide different instances of TEE with a different subset of the bids, and hence obtain the values of particular bids.
As described in~\cite{galal2019trustee}, such a privacy threat can be prevented by a TEE specific construct called \textit{hardware monotonic counter}, which cannot be reverted once incremented.
Finally, one has to consider that a vulnerability in trusted hardware may result in the unfairness of the auction process and compromise of its privacy.

\subsection{General Applications of Blockchains}\label{sec:app-general-application}
There are many use cases of applying blockchain to a particular domain that contains mutually untrusted participants:
these participants are represented by consensus nodes executing a consensus protocol. 
Applications from this category focus on leveraging inherent blockchain features and sometimes even on non-inherent features (see Appendix~\ref{sec:features-of-blockchain}).

An example that uses permissioned blockchain for the management of healthcare data is proposed in~\cite{esposito2018blockchain}, in which new data are included in the blockchain upon majority agreement.
The consensus nodes of this approach are represented by a patient, her family, and healthcare providers.
The authors discuss an issue regarding the right to delete personal data, which is contrary to the inherent features of the blockchain.
A data protection framework for energy grids and power systems was proposed in~\cite{liang2018distributed}. 
The authors suggest that smart meters act as consensus nodes and store the data of the blockchain in their memory.\footnote{These are unrealistic assumptions.}
DistBlockNet~\cite{sharma2017distblocknet} is the approach for ensuring state consistency among multiple SDN controllers with the utilization of dedicated blockchain.
In detail, flow rule tables of SDN controllers are managed by these controllers, while other entities in the system use blockchain as a reference point for downloading these tables.
A federated permissioned blockchain used as a cloud-based data storage requiring the consensus of all nodes\footnote{Note that such a proposal has very low fault tolerance.} is proposed in work~\cite{GaetaniABLMS17}.
The authors propose to use two tiers of blockchain.
The blockchain at the first tier is private and serves for consensus of participants, while the blockchain of the second tier is public PoW (e.g., Bitcoin) and serves for periodic publishing of integrity proofs of the first tier blockchain.
Two-tier blockchain was also applied in the domain of IoT~\cite{dorri2017towards}.
The authors of~\cite{dorri2017towards} deem the first tier blockchain as local to a group of IoT devices owned by a single party, while the second tier blockchain serves for sharing the data among multiple untrusted parties.
The authors demonstrate the applicability in a case study involving several smart homes.

\mysubsubsection{Security Threats and Mitigations}
Security threats of this general category vary case by case and usually concern the privacy of data shared among involved parties~\cite{esposito2018blockchain,liang2018distributed,dorri2017towards}.
Another common issue is an application of the blockchain with unrealistic assumptions about the target environment, e.g., low processing or storage performance of smart meters/IoT devices, no HW support for asymmetric cryptography, no tamper-proof means in devices producing transactions, etc. 
Some applications try to optimize throughput or finality of the blockchain by introducing their own consensus mechanisms (e.g., \cite{dorri2017towards,GaetaniABLMS17}); however, this might not be the best option since new attack vectors might be created. 
A general recommendation is to study and understand security issues and countermeasures of the state-of-the-art approaches of the consensus layer (see \autoref{sec:consensus}) as well as privacy concerns presented at the RSM layer (see \autoref{sec:smart_contracts}).

\section{Lessons Learned}
\label{sec:lessons}
In this section, we summarize lessons learned concerning the security reference architecture (SRA) and its practical utilization.
First, we describe the hierarchy of security dependencies among particular layers of the SRA.
Second, assuming such a hierarchy, we describe a security-oriented methodology for designers of blockchain platforms and applications.
Next, we summarize the design goals of particular blockchain types and discuss the security-specific features of the blockchains.
Finally, we analyze observations from the incidents that occurred in practice, limitations in the literature, and we propose future research directions.

\subsection{Hierarchy of Dependencies in the SRA}
In the proposed model of the SRA, we observe that consequences of
vulnerabilities presented at lower layers of the SRA are manifested in the same layers and/or at higher layers, especially at the application layer.
Therefore, we refer to \textit{security dependencies} of these layers on lower layers or the same layers, i.e., \textit{reflexive} and \textit{bottom-up} dependencies. 
We describe these two types of dependencies in the following.

\paragraph{Reflexive Dependencies}
If a layer of the SRA contains some assets, it also contains a reflexive security dependency on the countermeasures presented in the same layer.
It means that a countermeasure at a particular layer protects the assets presented in the same layer.
For example, in the case of the consensus layer whose protocols reward consensus nodes for participation,  the countermeasures against selfish mining attacks protect rewards (i.e., crypto-tokens) of consensus nodes.
In the case of the RSM layer, the privacy of user identities and data is protected by various countermeasures of this layer (e.g., blinding signatures, secure multiparty computations).
Another group of reflexive security dependencies is presented at the application layer. 
Although the application layer contains some bottom-up security dependencies (see \autoref{fig:dependencies}), we argue that with regard to the overall stacked model of the SRA they can be viewed as reflexive security dependencies of the application layer.

\paragraph{Bottom-Up Dependencies}
If a layer of the SRA contains some assets, besides reflexive security dependencies, it also contains bottom-up security dependencies on the countermeasures of all lower layers.
Hence, the consequences of vulnerabilities presented at lower layers of SRA might be manifested at the same layers (i.e., reflexive dependencies) but more importantly, they are manifested at higher layers, especially at the application layer.
For example, context-sensitive transactions and partial solutions as countermeasures of the consensus layer can protect against front-running attacks of intra-chain DEXes, which occur at the application layer.
Another example represents programming bugs in the RSM layer, which influence the correct functionality at the application layer.
The eclipse attack is an example that impacts the consensus layer from the network layer -- a victim consensus node operates over the attacker-controlled chain, and thus causes a loss of crypto-tokens by a consensus node and at the same time it decreases honest consensus power of the network.
In turn, this might simplify selfish-mining attacks at the consensus layer, which in turn might impact the correct functionality of a blockchain-based application at the application layer.
Bottom-up security dependencies are also presented in the context of the application layer, as we have already mentioned in \autoref{sec:apps}.

\setlength{\tabcolsep}{2.4pt}	
\begin{table*}[!h]
	\vspace{-0.4cm}
	\footnotesize
	\begin{tabular}{@{}ccclllll@{}}
		\toprule
		\textbf{Layer}                                                                                         & \multicolumn{3}{c}{\textbf{\begin{tabular}[c]{@{}c@{}}Category \\ in a Layer\end{tabular}}}                                                                                                                                                                                                                                   & \multicolumn{1}{c}{\textbf{Pros}}                                                                                                                                                                                                         & \multicolumn{1}{c}{\textbf{Cons}}                                                                                                                                                                                                           &  &  \\ 
		
		\Xhline{2\arrayrulewidth}\noalign{\smallskip}  
		
		\multirow{2}{*}{\textbf{\begin{tabular}[c]{@{}c@{}}\rotatebox[origin=l]{90}{Network Layer}\end{tabular}}}                      & \multicolumn{3}{c}{\textbf{\begin{tabular}[c]{@{}c@{}}Private\\ Networks\end{tabular}}}                                                                                                                                                                                                                                       & \begin{tabular}[c]{@{}l@{}}$\bullet$ low latency, high throughput\\ $\bullet$ centralized administration, ease of access control \\ $\bullet$ the privacy of data and identities\\ $\bullet$ meeting regulatory obligations\\ $\bullet$ resilience to external attacks\end{tabular} & \begin{tabular}[c]{@{}l@{}}$\bullet$ VPN is required for geographically spread\\ participants\\ $\bullet$ suitable only for permissioned blockchains\\ $\bullet$ insider threat and external attacks at nodes with\\administrative privileges\end{tabular}                                     &  &  \\
		
		\cmidrule(l){5-8} 
		
		& \multicolumn{3}{c}{\textbf{\begin{tabular}[c]{@{}c@{}}Public\\ Networks\end{tabular}}}                                                                                                                                                                                                                                        & \begin{tabular}[c]{@{}l@{}}$\bullet$ high decentralization\\ $\bullet$ high availability\\ $\bullet$ openness \& low entry barrier (low cost of \\ broadband connection, resistance to regulations)\end{tabular}                                                     & \begin{tabular}[c]{@{}l@{}}$\bullet$ high and non-uniform latency\\ $\bullet$ single-point-of-failure (DNS, IP, and ASes\\are managed by centralized parties)\\ $\bullet$ external adversaries (botnets, compromised\\BGP/DNS servers)\\ $\bullet$ stolen identities\end{tabular} &  &  \\
		
		\Xhline{2\arrayrulewidth}\noalign{\smallskip}  
		
		\multirow{6}{*}{\textbf{\begin{tabular}[c]{@{}c@{}}\rotatebox[origin=l]{90}{Consensus Layer~~~~~~~~~~~~~~~~~~}\end{tabular}}}                    & \multicolumn{3}{c}{\textbf{PoR}}                                                                                                                                                                                                                                                                                              & \begin{tabular}[c]{@{}l@{}}$\bullet$ high cost of overriding the history of blockchain\\ $\bullet$ high scalability\end{tabular}                                                                                                                            & \begin{tabular}[c]{@{}l@{}}$\bullet$ high operational costs\\ $\bullet$ low throughput\\ $\bullet$ low finality\end{tabular}                                                                                                                                           &  &  \\
		
		\cmidrule(l){5-8} 
		
		& \multicolumn{3}{c}{\textbf{BFT}}                                                                                                                                                                                                                                                                                              & \begin{tabular}[c]{@{}l@{}}$\bullet$ high throughput (with a small number of nodes)\\ $\bullet$ fast finality\end{tabular}                                                                                                                                                                 & \begin{tabular}[c]{@{}l@{}}$\bullet$ low scalability \\ $\bullet$ high communication complexity\\ $\bullet$ limited number of nodes (efficient use only\\in permissioned blockchains)\end{tabular}                                                                      &  &  \\
		
		\cmidrule(l){5-8} 
		
		& \multicolumn{3}{c}{\textbf{PoS}}                                                                                                                                                                                                                                                                                              & $\bullet$ energy efficiency                                                                                                                                                                                                                        & \begin{tabular}[c]{@{}l@{}}$\bullet$ PoS specific attacks and issues\\ $\bullet$ supports only semi-permissionless setting\\ $\bullet$ slow finality\end{tabular}                                                                                                      &  &  \\
		
		\cmidrule(l){5-8} 
		
		& \multicolumn{3}{c}{\textbf{PoS+BFT}}                                                                                                                                                                                                                                                                                          & \begin{tabular}[c]{@{}l@{}}$\bullet$ energy efficiency\\ $\bullet$ high scalability\\ $\bullet$ probabilistic security guarantees\\ $\bullet$ lower communication overheads than BFT\end{tabular}                                                                            & \begin{tabular}[c]{@{}l@{}}$\bullet$ some PoS specific attacks\\ $\bullet$ supports only semi-permissionless setting\end{tabular}                                                                                                                             &  &  \\
		
		\cmidrule(l){5-8} 		
		
		& \multicolumn{3}{c}{\textbf{PoR+BFT}}                                                                                                                                                                                                                                                                                          & \begin{tabular}[c]{@{}l@{}}$\bullet$ high scalability\\ $\bullet$ fast finality\end{tabular}                                                                                                                                                     & $\bullet$ spending some scarce resources                                                                                                                                                                                                             &  &  \\
		
		\cmidrule(l){5-8} 		
		
		& \multicolumn{3}{c}{\textbf{\begin{tabular}[c]{@{}c@{}}PoR+PoS \\ (i.e., PoA)\end{tabular}}}                                                                                                                                                                                                                                   & $\bullet$ high scalability                                                                                                                                                                                                                         & \begin{tabular}[c]{@{}l@{}}$\bullet$ spending some scarce resources\\ $\bullet$ some PoS specific attacks\\ $\bullet$ slow finality\end{tabular}                                                                                                                       &  &  \\

		\Xhline{2\arrayrulewidth}\noalign{\smallskip}  
		
		\multirow{8}{*}{\textbf{\begin{tabular}[c]{@{}c@{}}\rotatebox[origin=l]{90}{Replicated State Machine Layer~~~~~~~~~~~~~}\end{tabular}}} & \multirow{6}{*}{\rotatebox[origin=l]{90}{\textbf{Transactions~~~~~~~~~~~~~~~~~~~~~}}} & \multirow{4}{*}{\textbf{\begin{tabular}[c]{@{}c@{}}\rotatebox[origin=l]{90}{Protecting Identities~~~~}\end{tabular}}} & \textbf{Standard Approach}                                                                                                                 & \begin{tabular}[c]{@{}l@{}}$\bullet$ fast processing\\ $\bullet$ ease of verification\end{tabular}                                                                                                                                                          & \begin{tabular}[c]{@{}l@{}}$\bullet$ identities are only pseoudonymous and can\\be traced to IPs\\ $\bullet$ all data of transactions are publicly visible\end{tabular}                                                                                        &  &  \\
		
		\cmidrule(l){5-8} 
		
		&                                                                                     &                                                                                            & \textbf{\begin{tabular}[c]{@{}l@{}}Standard Approach\\ + Mixers\end{tabular}}                                                              & \begin{tabular}[c]{@{}l@{}}$\bullet$ privacy identity protection of users in  a group\\ $\bullet$ ease of verification\end{tabular}                                                                                                                         & \begin{tabular}[c]{@{}l@{}}$\bullet$ additional complexity, in some cases unlinkability\\ by the mixer or involved parties in a group\\ $\bullet$ all data of transactions are publicly visible\end{tabular}                                            &  &  \\
		
		\cmidrule(l){5-8} 
		
		&                                                                                     &                                                                                            & \textbf{\begin{tabular}[c]{@{}l@{}}NIZKs and \\ Ring-Signatures\end{tabular}}                                                              & \begin{tabular}[c]{@{}l@{}}$\bullet$ identities are anonymized to the extend of\\ the group\end{tabular}                                                                                                                                          & \specialcell{$\bullet$ additional computation overheads for running\\the schemes}                                                                                                                                                                                     &  &  \\
		
		\cmidrule(l){5-8} 
		
		&                                                                                     &                                                                                            & \textbf{\begin{tabular}[c]{@{}l@{}}MPC\\ Blinding Signatures, \\ Layered-Encryption\end{tabular}}                        & $\bullet$ unlinkability for all involved parties                                                                                                                                                                                                   & \specialcell{$\bullet$ additional computation overheads for running\\the schemes}                                                                                                                                                                                                                                                                                                                               &  &  \\
		
		\cmidrule(l){3-8} 
		&                                                                                     & \multirow{2}{*}{\textbf{\begin{tabular}[c]{@{}c@{}}\rotatebox[origin=l]{90}{Protecting Data}\end{tabular}}}       & \textbf{\begin{tabular}[c]{@{}l@{}}NIZKs, Blinding \\ Signatures, Homomor-\\phic Encryption\end{tabular}}                                 & \begin{tabular}[c]{@{}l@{}}$\bullet$ privacy of data in cryptocurrency \\ platforms\end{tabular}                                                                                                                                                   & \specialcell{$\bullet$ additional computation overheads for running\\the schemes}                                                                                                                                                                                                                                                                                                                                                                        &  &  \\
		
		\cmidrule(l){5-8} 
		
		&                                                                                     &                                                                                            & \textbf{\begin{tabular}[c]{@{}l@{}}Trusted Transaction \\ Managers,\\ Trusted Hardware,\\ MPC\end{tabular}} & \begin{tabular}[c]{@{}l@{}}$\bullet$ privacy of data in transactions of smart contract\\ platforms\end{tabular}                                                                                                                                    & \specialcell{$\bullet$ additional computation overheads for running\\the schemes}                                                                                                                                                                                                                                                                                                                                                                   &  &  \\
		
		\cmidrule(l){2-8} 
		
		& \multirow{2}{*}{\textbf{\rotatebox[origin=l]{90}{\begin{tabular}[c]{@{}c@{}}Smart\\ Contracts\end{tabular}}}} & \multicolumn{2}{c}{\textbf{\begin{tabular}[c]{@{}c@{}}Turing-Complete\\ Languages\end{tabular}}}                                                                                                                                     & \begin{tabular}[c]{@{}l@{}}$\bullet$ smart contracts may contain an arbitrary\\ programming logic\end{tabular}                                                                                                                                     & \specialcell{$\bullet$ wide surface for making the programming bugs\\that often results in vulnerabilities}                                                                                                                                                          &  &  \\
		
		\cmidrule(l){5-8} 
		
		&                                                                                     & \multicolumn{2}{c}{\textbf{\begin{tabular}[c]{@{}c@{}}Turing-Incomplete \\ Languages\end{tabular}}}                                                                                                                                  & $\bullet$ small attack surface and emphasis on safety                                                                                                                                                                                              & \specialcell{$\bullet$ the programming logic serves only for limited\\purposes}                                                                                                                                                                                    &  &  \\ 
		\bottomrule
		
	\end{tabular}
	\caption{Pros and cons of various categories within the first three layers of the stacked model.}\label{tab:pros-cons}
\end{table*}
\begin{figure*}[!h]
	\begin{center}		
		\includegraphics[width=0.70\textwidth]{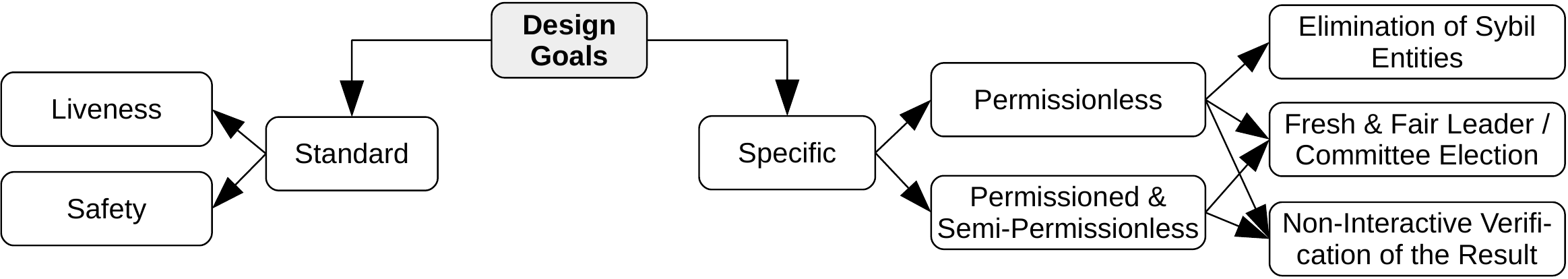} 		
		\caption{Standard and specific design goals of consensus protocols.}
		\label{fig:design-goals}
	\end{center}	
\end{figure*}

\setlength{\tabcolsep}{2.4pt}	
\begin{table*}[!h]
	\footnotesize
	\centering
	\begin{tabular}{@{}ccclllll@{}}
		\toprule
		\textbf{\specialcell{Application\\~~Category}}                                                                                         & \multicolumn{3}{c}{\textbf{\begin{tabular}[c]{@{}c@{}}Subcategory\end{tabular}}}                                                                                                                                                                                                                                   & \multicolumn{1}{c}{\textbf{Pros}}                                                                                                                                                                                                         & \multicolumn{1}{c}{\textbf{Cons}}                                                                                                                                                                                                           &  &  \\ 
		
		\Xhline{2\arrayrulewidth}\noalign{\smallskip}

		\multirow{3}{*}{\textbf{\begin{tabular}[c]{@{}c@{}}\rotatebox[origin=l]{90}{Wallets~~~~~~~}\end{tabular}}}                      & \multicolumn{3}{c}{\textbf{\begin{tabular}[c]{@{}c@{}}Server-Side\\Hosted\\ Wallets\end{tabular}}}                                                                                                                                                                                                                                       & \begin{tabular}[c]{@{}l@{}}$\bullet$ simplicity of control for end-users \\ $\bullet$ no storage requirements for end-users\\ \end{tabular} & \begin{tabular}[c]{@{}l@{}}$\bullet$ keys stored at the server, susceptibility to the theft of keys by external\\or internal attacks\\ $\bullet$ single-point-of-failure, availability attacks  \end{tabular}                                     &  &  \\
		
		\cmidrule(l){5-8} 
		
		& \multicolumn{3}{c}{\textbf{\begin{tabular}[c]{@{}c@{}}Client-Side\\Hosted\\Wallets\end{tabular}}}                                                                                                                                                                                                                                      & \begin{tabular}[c]{@{}l@{}}$\bullet$ simplicity of control for end-users \\ $\bullet$ no storage requirements for end-users\\ $\bullet$ keys stored locally \end{tabular}                                                     & \begin{tabular}[c]{@{}l@{}} $\bullet$ single-point-of-failure, availability attacks \\ $\bullet$ possibility of key theft by malware\\ $\bullet$ possibility of tampering attacks \end{tabular} &  &  \\
		
		\cmidrule(l){5-8} 
		
		& \multicolumn{3}{c}{\textbf{\begin{tabular}[c]{@{}c@{}}Self-Sovereign\\Wallets\end{tabular}}}                                                                                                                                                                                                                                      & \begin{tabular}[c]{@{}l@{}}$\bullet$ keys stored locally\\or in a dedicated hardware device \\ \end{tabular}                                                     & \begin{tabular}[c]{@{}l@{}} $\bullet$ moderate storage requirements for end-users \\ $\bullet$ more difficult control for end-users \\ $\bullet$ extra device to carry in the case of hardware wallet \end{tabular} &  &  \\
		
		\Xhline{2\arrayrulewidth}\noalign{\smallskip}

		\multirow{3}{*}{\textbf{\begin{tabular}[c]{@{}c@{}}\rotatebox[origin=l]{90}{Exchanges~~~~~~~~~~~~}\end{tabular}}}                      & \multicolumn{3}{c}{\textbf{\begin{tabular}[c]{@{}c@{}}Centralized\\ Exchange\end{tabular}}}                                                                                                                                                                                                                                       & \begin{tabular}[c]{@{}l@{}}$\bullet$ a high throughput and speed of operations \\ $\bullet$ the simplicity of control for end-users\\$\bullet$ low costs for exchange transactions\\ $\bullet$ trading of obscure crypto-tokens  \end{tabular} & \begin{tabular}[c]{@{}l@{}}$\bullet$ risk of insider threat due to centralization\\ $\bullet$ external threats to exchange infrastructure\\ $\bullet$ overheads for secure storage of secrets\\ $\bullet$ a fee specified by the operator \end{tabular}                                     &  &  \\
		
		\cmidrule(l){5-8} 
		
		& \multicolumn{3}{c}{\textbf{\begin{tabular}[c]{@{}c@{}}Direct\\Cross-Chain\\ Exchange\end{tabular}}}                                                                                                                                                                                                                                      & \begin{tabular}[c]{@{}l@{}}$\bullet$ fairness of the exchange \\ $\bullet$ no fee to any operator\\  \end{tabular}                                                     & \begin{tabular}[c]{@{}l@{}}$\bullet$ costs for 4 transactions of the atomic swap\\ $\bullet$ user has to find the counter-order on her own\\ $\bullet$ counter-orders might not exist\\ $\bullet$ a lower throughput than in a centralized exchange\\ $\bullet$ a higher complexity for end-users  \end{tabular} &  &  \\
		
		\cmidrule(l){5-8} 
		
		& \multicolumn{3}{c}{\textbf{\begin{tabular}[c]{@{}c@{}}Cross-Chain\\DEX\end{tabular}}}                                                                                                                                                                                                                                      & \begin{tabular}[c]{@{}l@{}}$\bullet$ fairness of the exchange \\ $\bullet$ order matching made by DEX \\ $\bullet$ trading of obscure crypto-tokens \end{tabular}                                                     & \begin{tabular}[c]{@{}l@{}}$\bullet$ costs for 4 or 6 transactions of the atomic swap\\ $\bullet$ a lower throughput than in a centralized exchange\\ $\bullet$ a fee specified by the operator\\ \end{tabular} &  &  \\
		
		\cmidrule(l){5-8} 
		
		& \multicolumn{3}{c}{\textbf{\begin{tabular}[c]{@{}c@{}}Intra-Chain\\ DEX\end{tabular}}}                                                                                                                                                                                                                                      & \begin{tabular}[c]{@{}l@{}} $\bullet$ fairness of the exchange\\ $\bullet$ uniform finality for every pair \\ $\bullet$ a high speed of operations \\  \end{tabular}                                                     & \begin{tabular}[c]{@{}l@{}}$\bullet$ a limited number of pairs that are specific to the target platform\\ $\bullet$ a fee specified by the operator\\ $\bullet$ costs for smart contract execution \end{tabular} &  &  \\
		
		\Xhline{2\arrayrulewidth}\noalign{\smallskip}  
		
		\multirow{3}{*}{\textbf{\begin{tabular}[c]{@{}c@{}}\rotatebox[origin=l]{90}{Oracles~~~~~~~}\end{tabular}}}                      & \multicolumn{3}{c}{\textbf{\begin{tabular}[c]{@{}c@{}}Prediction\\ Markets\end{tabular}}}                                                                                                                                                                                                                                       & \begin{tabular}[c]{@{}l@{}}$\bullet$ early (close to accurate) estimation\\~~of the future event's result \\ $\bullet$ decentralization \\ \end{tabular} & \begin{tabular}[c]{@{}l@{}}$\bullet$ possible conflict of interest\\ $\bullet$ a limited set of data specific to a few events \\ $\bullet$ a long time to obtain a final result, especially in\\~~the case of disputes \\ \end{tabular}                                     &  &  \\
		
		\cmidrule(l){5-8} 
		
		& \multicolumn{3}{c}{\textbf{\begin{tabular}[c]{@{}c@{}}Centralized\\ Data Feeds\end{tabular}}}                                                                                                                                                                                                                                      & \begin{tabular}[c]{@{}l@{}}$\bullet$ wide range of data \\ $\bullet$ fast provisioning time \\ $\bullet$ handling of private parameters of requests \\ $\bullet$ censorship evidence \\ \end{tabular}                                                     & \begin{tabular}[c]{@{}l@{}} $\bullet$ centralization (accidentally or intentionally wrong data) \\ $\bullet$ availability issues \\ \end{tabular} &  &  \\
		
		\cmidrule(l){5-8} 
		
		& \multicolumn{3}{c}{\textbf{\begin{tabular}[c]{@{}c@{}}Oracle\\ Networks\end{tabular}}}                                                                                                                                                                                                                                      & \begin{tabular}[c]{@{}l@{}}$\bullet$ decentralization \\ $\bullet$ wide range of data \\ $\bullet$ fast provisioning time \end{tabular}                                                     & \begin{tabular}[c]{@{}l@{}}$\bullet$ unsupported private parameters of requests\\ $\bullet$ publicly visible data and requests\\ \end{tabular} &  &  \\
		
		\Xhline{2\arrayrulewidth}\noalign{\smallskip}  
		
		\multirow{3}{*}{\textbf{\begin{tabular}[c]{@{}c@{}}\rotatebox[origin=l]{90}{Filesystems~~~~~~~~~~}\end{tabular}}}                      & \multicolumn{3}{c}{\textbf{\begin{tabular}[c]{@{}c@{}}Fully\\Replicated FSs\\with Ledger\end{tabular}}}                                                                                                                                                                                                                                       & \begin{tabular}[c]{@{}l@{}}$\bullet$ a high availability\\$\bullet$ accountability and auditability \end{tabular} & \begin{tabular}[c]{@{}l@{}}$\bullet$ a high storage overheads and operational costs\\ $\bullet$ a high price  \\ \end{tabular}                                     &  &  \\		
		\cmidrule(l){5-8} 
		
		& \multicolumn{3}{c}{\textbf{\begin{tabular}[c]{@{}c@{}}Partially\\Replicated FSs\\with Ledger\end{tabular}}}                                                                                                                                                                                                                                      & \begin{tabular}[c]{@{}l@{}}$\bullet$ reasonably high availability\\ $\bullet$ accountability and auditability\\ $\bullet$ a lower price than in a full replication \end{tabular}                                                     & \begin{tabular}[c]{@{}l@{}}$\bullet$ attack vectors specific to partial replication \end{tabular} &  &  \\
		
		\cmidrule(l){5-8} 
		
		& \multicolumn{3}{c}{\textbf{\begin{tabular}[c]{@{}c@{}}Partially\\Replicated FSs\\without Ledger\end{tabular}}}                                                                                                                                                                                                                                      & \begin{tabular}[c]{@{}l@{}} $\bullet$  reasonably high availability\\$\bullet$ a lower price than in a full replication\\  \end{tabular}                                                     & \begin{tabular}[c]{@{}l@{}}$\bullet$ a lack of native accountability and auditability\\$\bullet$ low durability due to a lack of incentives for storage \end{tabular} &  &  \\
		
		\cmidrule(l){5-8} 
		
		& \multicolumn{3}{c}{\textbf{\begin{tabular}[c]{@{}c@{}}Centralized\\Storage of\\Off-Chain Data\end{tabular}}}                                                                                                                                                                                                                                      & \begin{tabular}[c]{@{}l@{}}$\bullet$ a low price \\ $\bullet$ accountability and auditability   \end{tabular}                                                     & \begin{tabular}[c]{@{}l@{}}$\bullet$  a low availability \end{tabular} &  &  \\	
		
		\bottomrule
		
	\end{tabular}
	\caption{Pros and cons of some categories from the application layer.}\label{tab:pros-cons-2}
\end{table*}

\subsection{Methodology for Designers}\label{sec:designing-solutions}
A hierarchy of security dependencies in the SRA can be utilized during the design of new blockchain-based solutions.
When designing a new \textbf{blockchain platform} or a new \textbf{blockchain application}, we recommend designers to specify requirements on the blockchain features (see  Appendix~\ref{sec:features-of-blockchain}) and afterward analyze design options and their attack surfaces at the first three layers of the stacked model of SRA.
We briefly summarize the pros and cons of particular categories within the first three layers of SRA in \autoref{tab:pros-cons}, while security threats and mitigations are covered in \autoref{sec:network}, \autoref{sec:consensus}, and \autoref{sec:smart_contracts}.

On top of that, we recommend the designers of a new \textbf{blockchain application} to analyze particular options and their security implications at the application layer of SRA.
We list the pros and cons of a few categories from the application layer in \autoref{tab:pros-cons-2},\footnote{Note that the table contains only categories with specified sub-categorizations that represent the subject to a comparison.} while security threats and mitigation techniques of this layer are elaborated in \autoref{sec:apps} and \autoref{sec:apps-applications}.
During this process, we recommend the designers to follow security dependencies of the target category on other underlying categories (see \autoref{fig:dependencies}) if their decentralized variants are used (which is a preferable option from the security point-of-view).
For example, if one intends to design a decentralized reputation system, she is advised to study the security threats from the reputation system category and its recursive dependencies on e-voting, identity management, crypto-tokens \& wallets, and (optionally) filesystems.

\medskip
\paragraph{Divide and Conquer}
If a designer of the blockchain application is  also designing a blockchain platform, 
we recommend her to split the functionality of the solution with the divide-and-conquer approach respecting particular layers of our stacked model. 
In detail, if some functionality is specific to the application layer, then it should be implemented at that layer. 
Such an approach minimizes the attack surface of a solution and enables isolating the threats to specific layers, where they are easier to protect from and reviewed by the community.
A contra-example is to incorporate a part of application layer functionality/validation into the consensus layer. 
The consensus layer should deal only with the ordering of transactions, and it should be agnostic to the application. 

Nevertheless, it is worth noting that the divide-and-conquer approach might not be suitable for some very specific cases.
For example, some decentralized filesystems (see \autoref{sec:filesystems}) might combine data storage as an application-layer service with the proof-of-storage consensus algorithm, presented at the consensus layer. 
Therefore, the consensus layer also embeds a part of functionality from the application layer.
However, when filesystems are in security dependencies of the target application other than filesystems, one should realize that they are usually running on a different blockchain or infrastructure than the target application, and this exception is not a concern.

\subsection{Blockchain Types \& Design Goals}\label{sec:specific-design-goals}
We learned that the type of a blockchain (see \autoref{sec:types-of-blockchains}) implies the specific design goals of its consensus protocol (see \autoref{fig:design-goals}), which must be considered on top of the standard design goals (i.e., liveness and safety) and the inherent features (see Appendix~\ref{sec:background-features}) during the design of a particular blockchain platform and its consensus protocol.
In the following, we elaborate on such specific design goals.

\paragraph{Permissionless Type} 
The first design goal is to \textit{eliminate Sybil entities} -- such elimination can be done by requiring that some amount of scarce resources is spent for extension of the blockchain, and hence no Sybil entity can participate.
This implies that no pure PoS protocol can be permissionless since it never spends resources on running a consensus protocol.
The next design goal is \textit{a fresh and fair leader/committee election}, which ensures that each consensus node influences the result of a consensus commensurately to the number of scarce resources spent.
Moreover, freshness avoids the prediction of the selected nodes, and therefore elected nodes cannot become the subject of targeted DoS attacks. 
The last design goal is the \textit{non-interactive verification} of the consensus result by any node -- i.e., any node can verify the result of the consensus based on the data presented in the blockchain.

\paragraph{Permissioned and Semi-Permissionless Types}
These types of blockchains require \textit{fresh and fair leader/committee election} as well as \textit{non-interactive verification} of the result of the consensus.
However, in contrast to the permissionless blockchains, they do not require a means for the elimination of Sybil entities, as permission to enter the system is given by a centralized entity (i.e., permissioned type) or any existing consensus node (i.e., semi-permissionless type).

\paragraph{Blockchain Types and Incentives}
We observed that no application running on a \textit{public} (permissioned) blockchain has been able to work without introducing crypto-tokens (i.e., an incentive scheme), even if the use case is not financial in nature, e.g., e-voting, notaries, secure timestamping, or reputation systems. 
In these blockchains, incentive schemes serve as a means for the elimination of Sybil entities, besides other purposes.
The situation is different in the context of \textit{private} (permissioned) blockchains, which are usually provisioned by a single organization or a consortium and do not necessarily need crypto-tokens to operate. 
Misaligned incentives can cause consensus-level vulnerabilities, e.g., when it becomes profitable to drop blocks of other nodes to earn higher mining rewards~\cite{eyal2018majority} or transaction fees~\cite{instability_noReward}. 
The design of incentive mechanisms is a research field by itself and we refer the reader to the work of Leonardos et al.~\cite{leonardos2019presto}. 

\subsection{Security-Specific Features of Blockchains}\label{sec:lessons-finality}
We realized that consensus protocols are the target of most financially-oriented attacks on the decentralized infrastructure of blockchains, even if such attacks might originate from the network layer (e.g., routing and eclipse attacks).
The goal of these attacks is to overturn and re-order already ordered blocks while doing double-spending.
Hence, the finality is the most security-critical feature of the consensus layer.
The finality differs per various categories of the consensus layer. 
The best finality is achieved in the pure BFT protocols, and the worse finality is achieved in the single-leader-based PoR and PoS protocols.
On the other hand, combinations of the BFT with PoS protocols (i.e., introducing committees) slightly deteriorate the finality of BFT in a probabilistic ratio that is commensurate to the committee size.
In the case of PoR protocols with partial solutions, finality is improved as opposed to pure PoR protocols; however, it is also probabilistic, depending on the number of partial solutions.

\subsection{Incidents in Practice}
We list several incidents at each layer of the SRA in \autoref{tab:incidents-network},~\autoref{tab:incidents-cons},~\autoref{tab:smartcontractincidents}, and~\autoref{tab:incidents-app} of Appendix.
The observations about the number of different incident types vary layer by layer.
In the case of the network layer, many of the attacks described in the SRA occurred or were demonstrated with a proof-of-concept.
However, incidents that occurred at the consensus layer mostly contained 51\%
attacks with double-spending, while incidents that occurred on the application
layer were mostly caused by the exploitation of centralized components.
In the case of the RSM layer, most of the incidents occurred due to bugs in smart contracts.
Finally, we observed that most of the incidents that occurred at the application layer were caused due to a single-point-of-failure, e.g., centralized components or the insider threat.

Although the number of occurred incident types is low as compared to described vulnerabilities and threats, we argue that the adoption of blockchains for practical applications is still in its infancy, and thus we may expect that the number of different incident types observed in practice will grow.

\subsection{Limitations in the Literature and Practice}\label{sec:limitations}

\paragraph{Applications of Blockchains}
Although the literature contains surveys and overviews~\cite{casino2018systematic,zheng2018blockchain,wust2018you} of blockchain-based applications, these works introduce only domain-oriented categorizations (i.e., categories such financial, governance, security, education, supply chain, etc.) and they do not investigate the security aspects and functionalities that these applications leverage on and whether some of the applications do not belong to the same category from the security and functionality point-of-view.
To address this limitation, we provide a security-driven functionality-oriented categorization of blockchain-based applications (see \autoref{sec:apps}), which is agnostic to an application domain and thus can generalize different application scenarios.
Furthermore, our proposed categorization enables us to model security and functionality-based dependencies among particular categories, which is not possible with state-of-the-art categorizations.

\paragraph{Centralization}
Even though blockchains are meant to be fully decentralized, we have seen that this does not hold at some layers of the SRA -- the network and application layers, in particular.
In the network layer, some attacks are possible due to centralized DNS bootstrapping, while in the application layer a few categories utilize centralized components to ensure some functionality that cannot run on-chain or its provisioning would be too expensive and slow, which, however, forms the trade-off with the security.  
Some applications might depend on components from other application categories (e.g., identity management) but implementing these components in a centralized fashion, even though there exist some decentralized variants that are gaining popularity (e.g., DIDs~\cite{did-w3c} for identity management).

\subsection{Future Research Directions}

\paragraph{Fast Finality}
Although finality is the most security-critical feature of the consensus layer (see \autoref{sec:lessons-finality}), it forms the trade-off with scalability.  
Therefore, we believe that the future focus of the consensus research should be in a thorough evaluation of this trade-off across various consensus protocols.

\paragraph{Network-Layer Security}
We learned that a substantial body of security research in blockchains is focusing on the consensus and RSM layers since these layers are mostly identified with the blockchains.
As opposed to them (see \autoref{fig:category_totals}), the network layer is not so popular even though the serious threats originating from this layer might hurt the higher layers and their assets.
Therefore, a potential direction for future research lies in studying the
security aspects of network protocols, their suitability for a decentralized environment, and potential improvements.

\paragraph{Privacy Preservation \& Performance}
All cryptographic privacy preservation techniques (see \autoref{sec:privacy-threats}) bring additional computation overhead, and thus they negatively impact the throughput of the blockchains.
On the other hand, privacy-preserving solutions that are based on the trusted hardware might provide higher performance, but they rely on the manufacturer of trusted hardware and the assumption that it will not be compromised.
Therefore, we believe that optimizing the trade-off between performance and privacy-preservation is an important future research direction concerning the RSM layer.

\paragraph{Security Analysis of the Application Layer}
Although many references included in this study are presented in the application layer, only a very few of them analyze thoroughly security aspects of a particular application layer category or its instance. 
Therefore, as a future research direction, we recommend the authors of the blockchain-based applications to analyze the resistance of their applications to all known threats of a particular application category (e.g., with help of our work), while broadly think of new vulnerabilities and threats that might be specific to their application type.

\paragraph{Decentralization}
Since some blockchain applications utilize centralized components while their decentralized variants already exist \autoref{sec:limitations}, we suggest that a potential future direction for researchers and practitioners might be the concept of a fully decentralized blockchain ecosystem.
Such an ecosystem might consist of only decentralized (or partially decentralized) application types, for example, the ones that we reviewed in the application layer of the SRA.

\section{Discussion}
\label{sec:discussion}
In this section, we first discuss the versatility and modularity of the stacked model that is the proposed security reference architecture (SRA) based on. 
Then, we outline a few additional security aspects related to blockchains, which, for clarity and simplicity, we have not pursued throughout this paper or mentioned them only tangentially.
Finally, we discuss a few types of blockchain-oriented applications that directly inherit security aspects from already existing categories; therefore, we omitted such application types in our work.

\subsection{Stacked Model}
\paragraph{Versatility}
The hierarchical stacked model that is the SRA based on was already utilized in other domains before. 
A well-known example is the ISO/OSI model with seven layers or later derived TCP/IP model with four layers in the field of communication networks.
The stacked model was also applied in cloud computing, referred to as cloud stack~\cite{lenk2009s}, in which, each layer represents one service model in the model's hierarchy. 
Nevertheless, the stacked model was also applied in the field of blockchains~\cite{wang2018survey}.

The versatility of the stack model allows not only for modeling the hierarchy in a particular domain but also for partitioning the corresponding security issues and their countermeasures based on the layers of the model.
This was done for the ISO/OSI model~\cite{verschuren1993iso}, TCP/IP model~\cite{bellovin1989security}, and cloud computing~\cite{xiao2013cloud}, while in this work we focus on the security threats related to blockchains and propose the SRA.

\paragraph{Modularity}
The stack model of SRA enables extensions of particular categories within each layer by adding new vulnerabilities, threats, and their respective countermeasures.
Likewise, the new categories can be modularly added to each of the SRA layers.
Afterward, the security implications of a new category, threat, or a countermeasure within some layer should be studied with regard to particular categories in higher layers -- a new category might be beneficial or detrimental to them from a security point-of-view.
When introducing a new defense or mitigation technique, it is also important to evaluate its side effects and implications on the features of the blockchain that are manifested at the same and higher layers.

A general guideline for extending the SRA is to introduce only such categories that have unique features from the security point-of-view, while
in the case of the application layer, functionality point-of-view should be considered as well.

\subsection{Additional Security Aspects}

\paragraph{Secure Cryptography Primitives}
We emphasize that for each layer of our stacked model, we assume the use of secure cryptographic primitives with recommended key lengths\footnote{\url{https://www.keylength.com}} that are based on existing standards (e.g.,~\cite{nist-hash-stadard,nist-DSS-standard}). 
Examples include secure communication (i.e., network layer), the use of private keys for transaction signing (i.e., consensus layer), and password management for blockchain-based services (i.e., application layer).
Since the area of cryptographic primitives is standardized and extensively covered in existing research, we treat security incidents that break these primitives as out-of-scope in the current paper.

\paragraph{Semantic Bugs}
We deal with semantic bugs only at the level of the RSM layer as part of the smart contracts (see \autoref{sec:smart-contracts}).
However, we emphasize that semantic bugs in the blockchain infrastructure may occur at each of the proposed layers, whereas in the case of the RSM layer, besides smart contracts, they may occur in compilers, interpreters, etc.  
In this work, we assume that the software of the blockchain-related infrastructure does not contain any programming semantic bugs at each of the layers, and it provides the expected functionality. 
On the other hand, we emphasize that these semantic bugs had already accounted for several incidents in the past, e.g.,~\cite{Bitcoin-fork,Pigeoncoin-attack,Bitcoin-overflow-bug,Verge-Apr-Attack}.
To achieve safe and correct software at each of the layers, similar to the case of the RSM layer, developers and designers should utilize verification tools, testing, code reviews, audits, known design patterns, best practices, etc.

\subsection{Other Blockchain-Oriented Applications}
There are several other applications of blockchain that we do not mention in our work because their security aspects are inherited from one or more categories presented in \autoref{sec:apps} and \autoref{sec:apps-applications}.
For example, insurance applications running on smart contract platforms inherit security aspects from the oracles category, as they require data to be delivered into the blockchain from the outside world.
The next example is the trading of crypto-tokens within the same blockchain platform -- it inherits security aspects from the crypto-tokens and wallets category (see \autoref{sec:wallets}).
Another example is cross-chain communication, which is a generalization of the exchanges category, and it also inherits most of the security aspects from it.

\section{Related Work}
\label{sec:related}
The security reference architecture that has been presented in our work offers a comprehensive overview of blockchain-related security vulnerabilities, threats, and mitigation techniques. 
We adapted a custom version of the four-layer stacked model, initially presented in the work of Wang et al.~\cite{wang2018survey}. 
In the following, we present an overview of the state-of-the-art survey papers related to blockchain research while we highlight the differences in contrast to our work. 
We consider three groups of blockchain-oriented research: 
(1) papers that use a flat categorization of threats and vulnerabilities, 
(2) papers that use a stacked or other multi-layered models, and 
(3) papers that focus on incidents that belong to a single layer.

\paragraph{Research with Flat Categorization}
Bonneau et al.~\cite{2015-Bitcoin-SOK} present the first major survey of blockchain-specific security aspects, with a particular focus on Bitcoin and cryptocurrencies. 
The authors aim at the consensus-layer properties, although some network-layer aspects (e.g., DoS attacks) are discussed as well. 
Since smart contract functionality was in its early stages of development at that time, not much is said about RSM-layer properties, and little is said about applications beyond cryptocurrencies and data storage. 
Similarly, Tschorsch et al.~\cite{tschorsch2016bitcoin} and Yli-Huumo et al.~\cite{yli2016current} present early survey papers that focus mostly on consensus- and network-layer attacks, but they also deal with user privacy. 
The latter~\cite{yli2016current} has a particular focus on the publication details of blockchain research until 2016, e.g., the venues and the countries of the authors' institutions.
Li et al.~\cite{li2017survey} present a high-level overview of blockchain security threats and incidents, but the categorization is lacking. 
The authors deal with selfish mining, the DAO hack, BGP hijacking, and eclipse attacks, while all of them are mentioned as individual incidents. 
Conti et al.~\cite{conti2018survey} present an overview of consensus- and network-layer attacks inherent to the Bitcoin blockchain.
One interesting contribution is its overview of client-side attacks and attacks on exchange systems.
Many attacks presented in this work are supported by evidence of incidents. 
On the other hand, the authors spent only a little effort on the issues related to the RSM and application layers. 

\paragraph{Research with Layered or Stacked Categorization}
Wang et al.~\cite{wang2018survey} are the first to propose a 4-layer model denoted as ``\textit{a network implementation stack}.'' 
Despite proposing the stacked model, the authors do not focus on attacks and countermeasures concerning each of the layers.
The main focus of their work is on the consensus layer, where the authors dedicate most of their attention to PoR, PoS, and BFT protocols as well as improving blockchain performance by sharding, side-chains, and non-linear data organization.
In the application layer, the authors discuss a few types of emerging blockchain-based applications, such as general-purpose data storage and access control.
Saad et al.~\cite{saad2019exploring} identify three categories of attacks: blockchain structure attacks, peer-to-peer system attacks, and application-oriented attacks.
As compared to our work, it mostly holds that their \textit{peer-to-peer system attacks} encompass our network and consensus layer attacks, whereas their \textit{application-oriented attacks} include the RSM and application-layer attacks from our work.
In contrast to our work, Saad et al.~\cite{saad2019exploring} put double-spending attacks into application-oriented attacks, whereas in our case, they are part of the consensus layer since the means for their realization resides in this layer.
Moreover, the authors of this paper deal with crypto-jacking attacks, which are out-of-the-scope for our reference architecture, as they are not related to the infrastructure of the involved parties we consider (see \autoref{sec:involved-parties}).
Chen et al.~\cite{chen2019survey} propose a 4-layer model similar to ours, which is used to study vulnerabilities in Ethereum. The authors identify 44 vulnerabilities, 26 attacks, and
47 defenses in total.
In contrast to our model, the authors use the ``\textit{data}'' layer in place of our RSM layer.
This leads to a difference in interpretation between their framework and ours: e.g., they consider reentrancy bugs as an application layer vulnerability, whereas we treat them as an RSM-layer vulnerability. 
Since the authors focus on Ethereum, most vulnerabilities belong to the RSM layer; however, some of the other vulnerabilities (e.g., the BGP hijacking attack against MyEtherWallet~\cite{dnshijack2018,dnshijack2018b}) do not seem to be specific for Ethereum. 
In contrast, our work takes a broader view, and we do not constrain it to a single blockchain.
Another stack-based model was proposed by Zhang et al.~\cite{zhang2019security} and consists of six layers, where the layers stand for \textit{the application, contract, incentive, consensus, network, and a data layer}.
The works of Alkhalifah et al.~\cite{alkhalifah2019taxonomy} and Zhu et al.~\cite{zhu2018research} feature groupings consisting of five (\emph{network, consensus, mining pool, smart contract}, and \emph{client} vulnerabilities) and four (\emph{data privacy, data availability, data integrity}, and \emph{data controllability} attacks) categories, respectively.
Natoli et al.~\cite{natoli2019deconstructing} focus mainly on the consensus layer but include some network-layer attacks as well (e.g., eclipse attacks and BGP hijacking).

\paragraph{Research Focusing on a Particular Layer}
Finally, there is a number of survey papers that explicitly focus on specific layers: 
the network layer~\cite{neudecker2018network}, 
the RSM layer (in particular smart contracts) \cite{atzei2017survey,harz2018towards,diangelo2019survey}, 
protocols of the consensus layer~\cite{cachin2017blockchain,bano2017consensus,xiao2019survey,leonardos2019presto,garay2018sok,shahaab2019applicability}, 
incentives at the consensus layer~\cite{azouvi2019sok,huang2019survey}, and application layer from the general standpoint~\cite{casino2018systematic}, or with a focus on the IoT domain~\cite{panarello2018blockchain,ali2018applications,restuccia2019blockchain,Ferrag2019}.

\section{Conclusion}
\label{sec:conclusion}
In this paper, we focused on the systematization of the knowledge about security aspects of blockchain systems, while we aimed to create a standardized model for studying vulnerabilities and security threats.
We proposed a stack-modeled security reference architecture (SRA) consisting of four layers, and at each of the layers, we surveyed categories and options for their instantiation with their respective security implications and properties.
We modeled particular categories as vulnerability/threat/defense graphs, which we provided as a means for reasoning about imposed security aspects.
Next, we collected a sample of blockchain-related incidents that occurred in practice, which we further categorized using our proposed model. 
We observed that the number of incident types occurred in practice is substantially smaller than the number of described threats, especially in the consensus and application layer.
In the case of the application layer, most of the incidents occurred due to exploiting a centralized component by external or internal attackers, while in the case of the consensus layer, most of the incidents occurred due to temporary violation of protocol assumptions by 51\% attacks.
Finally, we presented a security-oriented methodology for designers of blockchains platforms and applications, respecting the proposed SRA.

\section*{Acknowledgment}
This work was supported by the NRF, Prime Minister's Office, Singapore, under its National Cybersecurity R\&D Programme (Award No. NRF2016NCR-NCR002-028) and administered by the National Cybersecurity R\&D Directorate.
Next, this work was supported by the project 20-07487S of the Czech Science Foundation, the H2020 ECSEL project VALU3S (876852), and the internal project of Brno University of Technology (FIT-S-20-6427).
We would also like to thank our colleagues Pieter Hartel, Stefanos Leonardos, and Mark van Staalduinen for their valuable feedback. 

\bibliographystyle{IEEEtran}

\bibliography{ref}

\appendix

\subsection{Features of Blockchains}\label{sec:features-of-blockchain}
Blockchains  were initially introduced as a means of coping with the centralization of monetary assets management, resulting in their most popular application -- a decentralized cryptocurrency with native crypto-tokens.
However, other blockchain applications have meanwhile started to proliferate as well, benefiting from features other than decentralization.
We summarize the inherent and non-inherent features of blockchains in the following. 

\smallskip
\mysubsubsection{Inherent Features}\label{sec:background-features}
\begin{compactitem}
	\item [\textbf{Decentralization:}] is achieved by a distributed consensus protocol -- the protocol ensures that each modification of the ledger is a result of interaction among participants. 
	In the consensus protocol, participants are equal, i.e., no single entity is designed as an authority.
	An important result of decentralization is resilience to node failures.
	
	\item [\textbf{Censorship Resistance:}] is achieved  due to decentralization, and it ensures that each valid transaction is processed and included in the blockchain.

	\item [\textbf{Immutability:}] means that the history of the ledger cannot be easily modified -- it requires a significant quorum of colluding nodes.
	The immutability of history is achieved by a cryptographic one-way function (i.e., a hash function) that creates integrity-preserving links between the previous record (i.e., block) and the current one. 
	In this way, integrity-preserving chains (e.g., blockchains) or graphs (e.g., direct acyclic graphs~\cite{sompolinsky2016spectre,rocket2018snowflake,popov2016tangle} or trees~\cite{sompolinsky2013accelerating}) are built in an append-only fashion.
	However, the immutability of new blocks is not immediate and depends on the time to the finality of a particular consensus protocol (see \autoref{sec:backgroun-design-goal}).
	
	\item [\textbf{Availability:}] although distributed ledgers are highly redundant in terms of data storage (i.e., full nodes store replicated data), the main advantage of such redundancy is paid off by the extremely high availability of the system.
	This feature may be of special interest to applications that cannot tolerate outages. 
	
	\item [\textbf{Auditability:}] correctness of each transaction and block recorded in the blockchain can be validated by any participating node, which is possible due to the publicly-known rules of a consensus protocol.

	\item [\textbf{Transparency:}] the transactions stored in the blockchain as well as the actions of protocol participants are visible to other participants and in most cases even to the public.	
	
\end{compactitem}

\smallskip
\mysubsubsection{Non-Inherent Features}
\hfill\\
Additionally to the inherent features, blockchains may be equipped with other features that aim to achieve extra goals.
Below we list a few examples of such non-inherent features.
\begin{compactitem}
	\item[\textbf{Energy Efficiency:}] running an open distributed ledger often means that scarce resources are wasted (e.g., Proof-of-Work).
	However, there are available consensus protocols that do not waste scarce resources, but instead emulate the consumption of scarce resources (i.e., Proof-of-Burn), or the interest rate on an investment (i.e., Proof-of-Stake). 
	See examples of these protocols in \autoref{sec:consensus}.
	
	\item[\textbf{Scalability:}] describes how the consensus protocol scales when the number of participants increases. 
	Protocols whose behavior is not negatively affected by an increasing number of participants have high scalability.
	
	\item[\textbf{Throughput:}] represents the number of transactions that can be processed per unit of time. 
	Some consensus protocols have only a small throughput (e.g., Proof-of-Work), while others are designed with the intention to maximize throughput (e.g., Byzantine Fault Tolerant (BFT) protocols with a small number of participants).
	See examples of BFT protocols in  \autoref{sec:BFT}.

	\item[\textbf{Privacy \& Anonymity:}] 
	by design, data recorded on a public blockchain is visible to all nodes or public, which may lead to privacy and anonymity issues. 
	Therefore, multiple solutions increasing anonymity (e.g., ring signatures~\cite{rivest2001leak} in Monero) and privacy (e.g., zk-SNARKs~\cite{ben2014succinct} in Zcash) were proposed in the context of cryptocurrencies, while other efforts have been made in privacy-preserving smart contract platforms~\cite{kosba2016hawk,cheng2018ekiden}.
	
	\item [\textbf{Accountability and Non-Repudiation:}] if blockchains or  applications running on top of them are designed in such a way that identities of nodes (or application users) are known and verified, accountability and non-repudiation of actions performed can be provided too.
	
\end{compactitem}

\subsection{Atomic Swap Protocols}\label{appendix:atomic-swaps}

\paragraph{Atomic Swap for Two Parties}\label{appendix:atomic-swaps-2}
Atomic swaps assume two parties $\mathbb{A}$ and $\mathbb{B}$ owning crypto-tokens in two different blockchains.
$\mathbb{A}$ and $\mathbb{B}$ wish to execute cross-chain exchange atomically and thus achieve a \textit{fairness} property, i.e., either both of the parties receive the agreed amount of crypto-tokens or neither of them.
First, this process involves an agreement on the amount and exchange rate, and second, the execution of the exchange itself.

In a centralized scenario~\cite{micali2003simple}, the approach is to utilize a trusted third party for the execution of the exchange.
In contrast to the centralized scenario, blockchains allow us to execute such an exchange without a requirement of the trusted party.
The atomic swap protocol~\cite{atomic-swap} enables conditional redemption of the funds in the first blockchain to $\mathbb{B}$ upon revealing of the hash pre-image (i.e., secret) that redeems the funds on the second blockchain to $\mathbb{A}$.
The atomic swap protocol is based on two Hashed Time-Lock Contracts (HTLC) that are deployed by both parties in both blockchains.

Although HTLCs can be realized by Turing-incomplete smart contracts with support for hash-locks and time-locks, for clarity, we provide a description assuming Turing-complete smart contracts, requiring four transactions:
\begin{compactenum}
	\item $\mathbb{A}$ chooses a random string $x$ (i.e., a secret) and computes its hash $h(x)$.
	Using $h(x)$, $\mathbb{A}$ deploys $HTLC_\mathbb{A}$ on the first blockchain and sends the agreed amount to it, which later enables anybody to do a conditional transfer of that amount to $\mathbb{B}$ upon calling a particular method of $HTLC_\mathbb{A}$ with $x = h(x)$ as an argument (i.e., hash-lock). 
	Moreover, $\mathbb{A}$ defines a time-lock, which, when expired, allows $\mathbb{A}$ to recover funds into her address by calling a dedicated method: this is to prevent aborting of the protocol by another party.
	
	\item When $\mathbb{B}$ notices that $HTLC_\mathbb{A}$ has been already deployed, she deploys $HTLC_\mathbb{B}$ on the second blockchain and sends the agreed amount there, enabling a conditional transfer of that amount to $\mathbb{A}$ upon revealing the correct pre-image of $h(x)$ ($h(x)$ is visible from already deployed $HTLC_\mathbb{A}$).
	$\mathbb{B}$ also defines a time-lock in $HTLC_\mathbb{B}$ to handle abortion by $\mathbb{A}$.
	
	\item Once $\mathbb{A}$ notices deployed $HTLC_\mathbb{B}$, she calls a method of $HTLC_\mathbb{B}$ with revealed $x$, and in turn, she obtains the funds on the second blockchain.
	
	\item Once $\mathbb{B}$ notices that $x$ was revealed by $\mathbb{A}$ on the second blockchain, she calls a method of $HTLC_\mathbb{A}$ with $x$ as an argument, and in turn, she obtains the funds on the first blockchain.
\end{compactenum}
If any of the parties aborts, the counter-party waits until the time-lock expires and redeems the funds.

\paragraph{Atomic Swap for Three Parties}\label{appendix:atomic-swaps-3}
In the following, we outline a three-way atomic swap protocol, where party $\mathbb{A}$ wishes to sell an asset $a$ for BTC, the party $\mathbb{B}$ wishes to buy $a$ for ETH, and DEX $\mathbb{E}$ is inter-mediating the asset transfer:
\begin{compactenum}
	\item $\mathbb{B}$ chooses a random string $x$ (i.e., a secret) and computes its hash $h(x)$.	
	Using $h(x)$, $\mathbb{B}$ deploys $HTLC_\mathbb{B}$ on the Ethereum blockchain and sends the agreed ETH amount there, which later enables anybody to do a conditional transfer of that amount to $\mathbb{E}$ upon calling a particular method of $HTLC_\mathbb{B}$ with $x = h(x)$ as an argument.
	Moreover, $\mathbb{B}$ defines a time-lock to handle abortion by any party.
	
	\item Once $\mathbb{E}$ notices that $HTLC_\mathbb{B}$ has been already deployed on the Ethereum blockchain, she deploys $HTLC_\mathbb{E}$ on the Bitcoin blockchain and sends the agreed BTC amount there, enabling a conditional transfer of that amount to $\mathbb{A}$ upon revealing the correct pre-image of $h(x)$ (which is visible in already deployed $HTLC_\mathbb{B}$).
	$\mathbb{E}$ also defines a time-lock in $HTLC_\mathbb{E}$.
	
	\item Once $\mathbb{A}$ notices that $HTLC_\mathbb{A}$ has been already deployed on the Bitcoin blockchain, she deploys $HTLC_\mathbb{A}$ on the asset blockchain and lock the asset $a$ there, enabling a conditional transfer of $a$ to $\mathbb{B}$ upon revealing the correct pre-image of $h(x)$ (which is visible in already deployed $HTLC_\mathbb{B}$ and $HTLC_\mathbb{E}$).
	$\mathbb{A}$ also defines a time-lock in $HTLC_\mathbb{A}$.
	
	\item When $\mathbb{B}$ notices that both $HTLC_\mathbb{E}$ and $HTLC_\mathbb{A}$ have been already correctly deployed, she reveals the secret $x$ as a part of the transaction sent to $HTLC_\mathbb{A}$. 
	This triggers a transfer of asset $a$ to $\mathbb{B}$.
	
	\item Once $\mathbb{A}$ notices that $x$ was revealed, she sends a transaction with $x$ to $HTLC_\mathbb{E}$, obtaining BTC from $\mathbb{E}$.
	
	\item Once $\mathbb{E}$ notices that $x$ was revealed, she sends a transaction with $x$ to $HTLC_\mathbb{B}$, obtaining ETH from $\mathbb{B}$.
\end{compactenum}

\setlength{\tabcolsep}{3.2pt}
\begin{table*}[b]
	\centering
	\begin{tabular}{@{}lp{3cm}cp{8.0cm}l@{}}
		\toprule
		\textbf{\specialcell{Acronym\\/ Incident}}& \textbf{Threat \newline / Vulnerability} & \textbf{Reference} & \textbf{Description} & \textbf{Impact (\$)} \\ \Xhline{2\arrayrulewidth}\noalign{\smallskip}  
		\specialcell{Gatecoin\\(May 2016)}  & Centralized server compromise (single-point-of-failure) & \cite{Gatecoin-attack}& The exchange was the victim of a man-in-the-middle attack. The malicious external party was involved in this breach, and it managed to alter Gatecoin's system so that deposit transfers bypassed the multisig cold storage and went to hot wallets, which were exploited due to OPSEC issues. & 25,160,000\\ \cmidrule(l){3-5} 
		\specialcell{Doge Vault\\(May 2014)}  & Centralized  server compromise (single-point-of-failure) & \cite{Dogecoin-vault-attack}  & The attacker gained access to the server where Doge Vault's virtual machines were running, providing him with full access to the systems. & 56,000 \\ \cmidrule(l){3-5} 
		\specialcell{BIPS.me\\(November 2013)} & DDoS + access subversion on centralized server (single-point-of-failure) &  \cite{BIPS-attack} & An initial DDoS attack caused vulnerability to the system, which has then enabled the attacker to gain access and compromise several wallets. & 554,260 \\ \cmidrule(l){3-5} 
		\specialcell{Bitfinex\\(August 2016)}  &  Centralized server (single-point-of-failure) & \cite{Bitfinex-attack} & Although Bitfinex used 2-of-3 multisig wallets, two of these keys were owned by Bitfinex (one stored in cold wallet), while the user owned only a single key. It is not clear whether this incident involved insider threat or it was conducted externally.  & 71,288,416\\ \cmidrule(l){3-5} 
		\specialcell{Bitcoin Central\\(April 2013)} & Centralized server compromise (single-point-of-failure) &  \cite{Bitcoin-central-attack} & Password was reset from the hosting provider's web interface, enabling the attacker to lock out of the interface and request a reboot of the machine into rescue mode. Next, the attacker has stolen private keys from the hot wallet. & --- \\ 
		\cmidrule(l){3-5}

		\specialcell{Cyber-squatting attacks\\in NameCoin} & Violation of accurate registration  & \cite{kalodner2015empirical}  & Cyber-squatters seized identities that do not belong to them nor represent themselves. & --- \\ 		\cmidrule(l){3-5} 
		\specialcell{Front-Running on\\Ethereum Exchanges} & Front-running in \newline Intra-Chain Exchanges  & \cite{daian2019flash}  & Arbitrage bots front-run user issued exchange transactions with the ones with the higher fees. Therefore, user issued transactions are discarded or traded with worse exchange rate as intended. & --- \\

		\bottomrule
	\end{tabular}
	\caption{Incidents that occurred at the application layer. }\label{tab:incidents-app}
	\vspace{2cm}
\end{table*}

\vfill\eject
\subsection{Examples of Incidents}\label{appendix:incident-tables}
In the current section, we list several incidents at each layer of the security reference architecture.
In detail, \autoref{tab:incidents-network} contains incidents of public networks at the network layer, \autoref{tab:incidents-cons} lists incidents of the consensus layer, \autoref{tab:smartcontractincidents} focuses on the RSM layer,  and~\autoref{tab:incidents-app} shows a few examples of incidents at the application layer.
\vspace{-15cm}

\begin{table*}
	\centering
	\footnotesize{
		\begin{tabular}{@{}lp{2.5cm}cp{6.0cm}lc@{}}
			\toprule
			\textbf{\specialcell{Acronym\\/ Incident}}& \textbf{Threat \newline / Vulnerability} & \textbf{Reference} & \textbf{Description} & \textbf{Impact (\$)} & \textbf{SWC Entry} \\ \Xhline{2\arrayrulewidth}\noalign{\smallskip}  
			\specialcell{DAO Attack\\(June 2016)} &Reentrancy & \cite{DAO} & A vulnerability in the code allowed a repeated withdrawal of ether, which could be exploited with a malicious fallback function.   & 70,000,000 & SWC-107  \\ \cmidrule(l){3-6} 
			\specialcell{King of the Ether Throne\\(February 2016)}&Unchecked return value & \cite{KoET}
			& An unchecked return value prevented the rightful compensation of the user.  & 300  & SWC-104  \\ \cmidrule(l){3-6} 
			RockPaperScissors&Storing secrets & \cite{RockPaperScissors} & A smart contract relied on a secret value that should not be visible to other users.  &  --- &  \\ \cmidrule(l){3-6} 
			&Integer overflow and underflow & \cite{overflow} & The user's input was not properly sanitized and overflow / underflow was possible. & & SWC-101 \\ \cmidrule(l){3-6} 
			&Unexpected value received & \cite{unexpectedEther}& A smart contract may contain conditions that rely on a certain balance, but the attacker is able to send an unexpected monetary value to the contract and disrupt the intended functionality. &  --- & SWC-132\\ \cmidrule(l){3-6} 
			&	Delegatecall & \cite{SoliditySecurityList} & Solidity has a feature called delegatecall that enables remote calls of other contracts. Such calls cause an execution of untrusted contracts in the context of the caller, hence all vulnerabilities of the remote code can be exploited. &  --- & SWC-112 \\ \cmidrule(l){3-6} 
			\specialcell{Parity multisig hack\\(July 2017)}&Default function visibility& \cite{multisig}& Solidity functions are public per default. This can be easily exploited if a developer forgets to make critical functions private.& 30,000,000 & SWC-100\\ \cmidrule(l){3-6} 
			\specialcell{TheRun\\(April 2016)}&Weak source of randomness& \cite{weakRandomness}& Relying on the block number or timestamp as a source for randomness is not safe, since it can allow a prediction of the next random number. &  --- & SWC-120\\ \cmidrule(l){3-6} 
			\specialcell{GovernMental\\(March 2016)}&DoS & \cite{ponziSchemes}&The gas limit or failing calls to external functions can lead to situations where a contract becomes unusable. & 11,000&\specialcell{SWC-113,\\ SWC-128} \\ \cmidrule(l){3-6} 
			\specialcell{EXTCODESIZE DoS attack\\ (September 2016)} & DoS & \cite{EXTCODESIZEDDOSAttack}& The attack exploited a mismatch between the computational cost of some operations and their gas cost. &  --- &\\ \cmidrule(l){3-6} 
			\specialcell{Rubixi\\(March 2016)}&Unprotected Ether withdrawal& \cite{SoliditySecurityList}& The access to administration or initialization functions had not been adjusted properly. Missing modifiers or an improperly named function might lead to unauthorized access.&  --- &\specialcell{SWC-105,\\ SWC-118} \\ \cmidrule(l){3-6} 
			\specialcell{Bancor\\(June 2017)}&Front-running / transaction order dependency& \cite{SoliditySecurityList}& Some contracts rely on the transaction order to make a decision. For example, a winner of a game is chosen from the first transaction with a correct answer. Since the transactions can be inspected by the consensus nodes before they are included, the attacker might steal the answer and make a transaction with a higher fee, which will be included as the first one. &  --- &SWC-114 \\ \cmidrule(l){3-6} 
			&Timestamp dependency& \cite{historicalVulnerabilities}& Some contracts might use the current timestamp to trigger certain events. However, timestamps can be adjusted by malicious consensus nodes. &  --- &SWC-116\\ \cmidrule(l){3-6} 
			&Write to an arbitrary storage location& \cite{arbitraryWrite}& By taking advantage of neighboring addresses in the storage and un-sanitized code,  the unauthorized attacker might write to sensitive storage locations. &  --- &SWC-124\\ \cmidrule(l){3-6} 
			\specialcell{Parity multisig hack 2\\(November 2017)}&Unprotected selfdestruct call& \cite{multisig2}&The unprotected self destruct functionality can be exploited to destroy contracts (or libraries), and potentially freeze funds in them. &150,000,000&SWC-106 \\ 
			\bottomrule
		\end{tabular}
	}
	\caption{Incidents and possible vulnerabilities at the RSM layer.}\label{tab:smartcontractincidents}
\end{table*}

\setlength{\tabcolsep}{3.8pt}	
\begin{table*}
	\centering
	\begin{tabular}{p{2.2cm}p{3.2cm}lp{8cm}l}
		\toprule
		\textbf{\specialcell{Acronym\\/ Incident}}& \textbf{Threat \newline / Vulnerability}  & \textbf{Reference} & \textbf{Description} & \textbf{Impact (\$)} \\ 
		\Xhline{2\arrayrulewidth}\noalign{\smallskip}  
		Ethereum Classic \newline (January 2019) & 51\% attack \& double spending (violation of assumptions) & \cite{Ethereum-Classic-attack} & A $51\%$ attack on Ethereum Classic led to a deep chain reorganization, which included double-spending attacks against Binance and Bitrue wallets. & 1,100,000 \\ \cmidrule(l){3-5} 
		Monacoin \newline (May 2018) & 51\% attack \& double spending (violation of assumptions) & \cite{Monacoin-attack} & A block reorganization on Monacoin included double-spending attacks that targeted several ``Western exchanges'' (particularly Livecoin). Unconfirmed reports on Reddit mention losses of 90,000 USD. & 90,000 \\ \cmidrule(l){3-5} 
		Bitcoin Gold \newline (May 2018) & 51\% attack \& double spending (violation of assumptions) & \cite{Bitcoin-Gold-attack} & A Block withholding attack on Bitcoin Gold led to 76 double-spent transactions, mostly targeting cryptocurrency exchanges. & 18,600,000\\ \cmidrule(l){3-5} 
		Litecoin Cash \newline (May 2018)& 51\% attack \& double spending (violation of assumptions) & \cite{Litecoin-Cash-attack} & A similar double-spending attack targeted Litecoin cash -- the lead developer ``Tanner'' stated that he believes the mining power may have been rented. & --- \\ \cmidrule(l){3-5} 
		Zencash \newline (June 2018) & 51\% attack \& double spending (violation of assumptions) & \cite{Zencash-attack} & A $51\%$ attack led to several block reorganizations, with the largest one that reversed 38 blocks. Only two transactions included double spending, but these were large enough to cause considerable losses.& 550,000 \\ \cmidrule(l){3-5} 
		Verge \newline (April 2018) &  51\% attack \& semantic bug \newline (violation of assumptions) & \cite{Verge-Apr-Attack} & A vulnerability in Verge's difficulty adjustment algorithm was used to amplify a $51\%$ attack. & 3,850,000\\ \cmidrule(l){3-5}
		Verge \newline (May 2018) & 51\% attack \& semantic bug  (violation of assumptions) & \cite{Verge-May-Attack} & An inadequate response by Verge developers to the previous attack led to it being repeated a month later. & 1,750,000\\  \cmidrule(l){3-5} 
		
		Checklocktime, CensorshipCon, GoldfingerCon, etc. & Bribery attacks & \cite{judmayer2019pay} & Various bribery attacks have been listed in the literature. In these attacks, it is profitable for consensus nodes to accept bribes for deviation from the protocol. & --- \\

		\bottomrule
		
	\end{tabular}
	\caption{Incidents that occurred at the consensus layer.}\label{tab:incidents-cons}
\end{table*}

\setlength{\tabcolsep}{4.4pt}
\begin{table*}[!ht]
	\centering
	\begin{tabular}{p{2.8cm}p{2.8cm}lp{8cm}l}
		\toprule
		\textbf{\specialcell{Acronym\\ / Incident}}& \textbf{Threat\newline / Vulnerability}  & \textbf{Reference} & \textbf{Description} & \textbf{Impact (\$)} \\ 
		\Xhline{2\arrayrulewidth}\noalign{\smallskip}  
		Canadian Bitcoin hijack \newline (February 2013) & BGP prefix hijacking (routing attack \& eclipse attack) & \cite{bgphijack2014} & Intercepted data between Bitcoin miners and Bitcoin mining pools. Tricked honest miners were mining on an attacker controlled pool. & 83,000 \\ \cmidrule(l){3-5} 
		Bitcoin deanonymization (March~2015) & Sybil attack (identity revealing attacks) & \cite{chainalysis2015} &  A large number of fake nodes were introduced to deanonymize client traffic. & --- \\ \cmidrule(l){3-5} 
		MyEtherWallet hijack \newline (April 2018) & BGP hijacking (routing attack) & \cite{dnshijack2018,dnshijack2018b} & A BGP hijack have been performed against Amazon DNS, leading to the wallet's server domain name being resolved to a Russian phishing site in several cities. & 152,000\\ \cmidrule(l){3-5} 
		Electrum DoS  \newline (August 2019)& Volumetric DoS & \cite{dosattack2019} & DoS of legitimate Electrum servers was conducted as an attempt to get connected vulnerable Electrum wallets to a malicious server, which resulted in a loss of wallet funds.  & In millions \\ \cmidrule(l){3-5} 
		Bitcoin partitioning \newline (proof-of-concept) & Prefix hijacking (routing attack) & \cite{mapostolwebsite} & An attack to partition Bitcoin network by hijacking IP prefixes.& --- \\ \cmidrule(l){3-5} 
		Erebus (proof-of-concept) & Malicious ISP attack & \cite{Erbus2019} & ISP uses its topological advantage to launch a stealthy partition attack. & --- \\ 
		\cmidrule(l){3-5} 
		Eclipse attack on Bitcoin (proof-of-concept) &  Un-authenticated and unreliable peers  & \cite{heilman2015eclipse} & A view of the network by eclipsed peers is fully under the attacker's control. & ---\\
		\cmidrule(l){3-5}  
		De-anonymization in Bitcoin with Tor \newline (proof-of-concept) &  Abusing Bitcoin DoS protection (identity revealing attack)  & \cite{BiryukovP14} & Bitcoin peers were forced to ban Tor exit nodes of attacker's choice by abusing Bitcoin DoS protection. Thus, the attacker was able to control all remaining Tor exit nodes and cause client traffic to pass through them. & ---\\		  
		\cmidrule(l){3-5} 
		Suspected Penny flooding on Bitcoin (December 2017) &  DoS on a mempool (attacking local resources)  & \cite{Saad2019mempool} & Flooding a mempool with low fee transactions resulted in clogged up memory of consensus nodes and increased transaction processing fees. & ---\\
		\cmidrule(l){3-5} 
		2X-Mempool-Attack on Bitcoin (suspected) &  Flooding by high-fee transactions (DoS on processing of legitimate transactions)   & \cite{penny-flooding} & Mempool is flooded by high-fee transactions, preventing regular-fee transactions being processed timely.  & ---\\		
		
		\bottomrule
		
	\end{tabular}
	\caption{Incidents that occurred at the network layer (public networks).}\label{tab:incidents-network}
\end{table*}

\end{document}